\documentclass{aa}  

\usepackage{graphicx}
\usepackage{txfonts}
\usepackage{epstopdf}
\usepackage{multicol}
\usepackage{multirow}
\usepackage{subfig}
\usepackage{rotating}
\usepackage{placeins}
\usepackage{xfrac}
\usepackage{hyperref}

\newcommand{\Teff}{$T_{\rm eff}$}
\newcommand{\logg}{$log~g$}
\newcommand{\vt}{V$_{t}$}

\begin{document}

   \title{Detailed chemical composition of classical Cepheids\\in the LMC cluster NGC~1866 and in the field of the SMC\thanks{Based on observations collected at the European Organisation for Astronomical Research in the Southern Hemisphere under ESO programme 082.D-0792(B).}}
   \author{B. Lemasle
          \inst{1}
          \and
          M. A. T. Groenewegen\inst{2}
          \and E. K. Grebel\inst{1}
          \and G. Bono\inst{3,4}
          \and G. Fiorentino\inst{5}
          \and P. Fran\c cois\inst{6,7}
          \and L. Inno\inst{8}          
          \and V. V. Kovtyukh\inst{9,10}
          \and N. Matsunaga\inst{11}
          \and S. Pedicelli \inst{3,12}
          \and F. Primas\inst{12}
          \and J. Pritchard\inst{12}
          \and M. Romaniello\inst{12,13}
          \and R. da Silva\inst{3,4,14}
    }

   \institute{Astronomisches Rechen-Institut, Zentrum für Astronomie der Universität Heidelberg, Mönchhofstr. 12-14, D-69120 Heidelberg, Germany~
              \email{lemasle@uni-heidelberg.de}
         \and
               Koninklijke Sterrenwacht van België, Ringlaan 3, 1180, Brussels, Belgium~
             \email{martin.groenewegen@oma.be}
         \and
               Dipartimento di Fisica, Università di Roma Tor Vergata, via della Ricerca Scientifica 1, 00133 Rome, Italy 
         \and
               INAF–Osservatorio Astronomico di Roma, via Frascati 33, 00078 Monte Porzio Catone, Rome, Italy 
         \and
               INAF-Osservatorio Astronomico di Bologna, via Gobetti 93/3, 40129, Bologna, Italy 
         \and
               GEPI, Observatoire de Paris, CNRS, Université Paris Diderot, Place Jules Janssen, 92190 Meudon, France 
         \and
               UPJV, Universit\'e de Picardie Jules Verne, 33 rue St. Leu, 80080 Amiens, France 
         \and 
               Max-Planck-Institut für Astronomie, D-69117, Heidelberg, Germany 
         \and 
               Astronomical Observatory, Odessa National University, Shevchenko Park, UA-65014 Odessa, Ukraine 
         \and 
               Isaac Newton Institute of Chile, Odessa Branch, Shevchenko Park, UA-65014 Odessa, Ukraine 
         \and 
               Department of Astronomy, School of Science, The University of Tokyo, 7-3-1 Hongo, Bunkyo-ku, Tokyo 113-0033, Japan 
         \and
               European Southern Observatory, Karl-Schwarzschild-Str. 2, 85748 Garching bei München, Germany 
         \and               
               Excellence Cluster Universe, Boltzmannstr. 2, 85748 Garching bei München, Germany 
         \and
               ASI Science Data Center, via del Politecnico snc, 00133 Rome, Italy 
}
   \date{Received September 15, 1996; accepted March 16, 1997}

 
  \abstract
   {Cepheids are excellent tracers of young stellar populations. They play a crucial role in astrophysics as standard candles. The chemistry of classical Cepheids in the Milky Way is now quite well-known. Despite a much larger sample, the chemical composition of Magellanic Cepheids has been only scarcely investigated.}
   {For the first time, we study the chemical composition of several Cepheids located in the same populous cluster: NGC 1866, in the Large Magellanic Cloud (LMC). To also investigate the chemical composition of Cepheids at lower metallicity, four targets are located in the Small Magellanic Cloud (SMC). Our sample allows us to increase the number of Cepheids with known metallicities in the LMC/SMC by 20\%/25\% and the number of Cepheids with detailed chemical composition in the LMC/SMC by 46\%/50\% .}
   {We use canonical spectroscopic analysis to determine the chemical composition of Cepheids and provide abundances for a good number of $\alpha$, iron-peak and neutron-capture elements.}
   {We find that six Cepheids in the LMC cluster NGC~1866 have a very homogeneous chemical composition, also consistent with red giant branch (RGB) stars in the cluster. Period--age relations that include no or average rotation indicate that all the Cepheids in NGC~1866 have a similar age and therefore belong to the same stellar population. Our results are in good agreement with theoretical models accounting for luminosity and radial velocity variations. Using distances based on period-luminosity relations in the near- or mid-infrared, we investigate for the first time the metallicity distribution of the young population in the SMC in the depth direction. Preliminary results show no metallicity gradient along the SMC main body, but our sample is small and does not contain Cepheids in the inner few degrees of the SMC.
   }
   {}

   \keywords{Stars: variables: Cepheids; Magellanic Clouds; Galaxies: star clusters: individual: NGC~1866}
   \titlerunning{Detailed chemical composition of classical Cepheids in the Magellanic Clouds}
   \authorrunning{B. Lemasle et al.} 
   \maketitle
%

\section{Introduction}

\par Classical Cepheids are the first step on the ladder of the extragalactic distance scale. Cepheid distances were first computed from period-luminosity (PL) relations in the optical bands, but the metallicity dependence of the optical PL-relations \citep[e.g.,][]{Roma2008} and the interstellar absorption led researchers to prefer period-luminosity or period-Wesenheit (PW) relations in the near-infrared \citep[e.g.,][]{Bono2010,Feast2012,Ripepi2012,Gie2013,Inno2013,Bhar2016} where the Wesenheit index is a reddening-free quantity \citep{Mado1982}. In the recent years, these relations have been extended to the mid-infrared \citep[e.g.,][]{Mon2012,Ngeow2012,Scow2013,Rich2014,Ngeow2015}. Most of them are tied to very accurate parallax measurements for the closest Cepheids \citep{Bene2007,vLeeuw2007}.\\

\par Independent distances to Cepheids can also be obtained with the Baade-Wesselink (BW) method, which combines the absolute variation of the radius of the star with the variation of its angular diameter. The former is obtained by integrating the pulsational velocity curve of the Cepheid that is derived from its radial velocity curve via the projection factor (p). The latter uses surface-brightness (SB) relations to transform variations of the color of the Cepheid to variations of its angular diameter. SB relations were first derived in the optical bands \citep[e.g.,][]{Wess1969,BE1976} and extended to the near-infrared by \cite{Welch1994}, and \citet{Fou1997}. Extremely accurate angular diameter variations can be obtained from interferometry \citep[e.g.,][]{Mou1997,Ker2004} but this technique is currently limited to the closest Cepheids.\\

\par Published values of the $p$-factor consistently cluster around $\sim$1.3. However, the exact value of the $p$-factor and its dependence on the pulsation period remain uncertain at the level of 5-10\% \citep{Ker2017}. In a series of papers, \cite{Storm2004a,Storm2004b,Gie2005,Fou2007}; and \citet{Storm2011a,Storm2011b} found that the $p$-factor strongly depends on the period. Similar conclusions were obtained independently by \cite{Groe2008,Groe2013}. Using hydrostatic, spherically-symmetric models of stellar atmospheres, \citet{Neil2012} indicate that the $p$-factor varies with the period, but the dependence derived is not compatible with the 
observational results of, e.g., \citet{Nar2014a} and \citet{Storm2011a,Storm2011b}. To overcome these issues, \citet{Mer2015} implemented a new flavor of the Baade-Wesselink method: they fit simultaneously all the photometric, interferometric and radial velocity measurements in order to obtain a global model of the stellar pulsation. Applying this method to the Cepheids for which trigonometric parallaxes are available, \citet{Breit2016} found a constant value of the $p$-factor, with no dependence on the pulsation period.\\

\par Among the aforementioned studies that include LMC/SMC Cepheids, only those of \citet{Groe2008,Groe2013} rely on abundance determinations for individual Cepheids while the others use either the (oxygen) abundances derived in nearby HII regions or a mean, global abundance for a given galaxy. Because the determination of nebular abundances is still affected by uncertainties as pointed out by \citet{Kew2008} \citep[but see, e.g.,][]{Pil2016}; and because the correlation between oxygen and iron varies from galaxy to galaxy, it is of crucial importance to have direct metallicity measurements in Cepheids. This task is now well achieved for Milky Way Cepheids \citep[see][and references therein]{Lem2007,Lem2008,Lem2013,Luck2011a,Luck2011b,Gen2013,Gen2014,Gen2015}. Despite the large number of Cepheids discovered in the Magellanic Clouds (3375/4630 in the LMC/SMC, respectively) by microlensing surveys such as OGLE \citep[the Optical Gravitational Lensing Experiment][]{Udal2015}, only a few dozens have been followed up via high-resolution spectroscopy in order to determine their metallicities \citep{Roma2005,Roma2008} or chemical composition \citep{Luck1992,Luck1998}. In this context, it is worth mentioning that by transforming a hydrodynamical model of $\delta$ Cephei into a consistent model of the same star in the LMC, \citet{Nar2011} found a weak dependence of the $p$-factor on metallicity (1.5\% difference between LMC and Solar metallicities).\\

\par NGC~1866 is of specific interest in that respect, as it is a young (age range of 100-200~Myr), massive cluster in the outskirts of the LMC that is known to harbor a large number (23) of Cepheids \citep[e.g.,][]{Welch1993}. Many studies investigated the pulsational and evolutionary properties of the intermediate-mass stars in NGC~1866 \citep[e.g.,][]{Bono1997,Fio2007,Mar2013,Mus2016} or the multiple stellar populations in LMC clusters \citep{Mil2017}. The focus on pulsating stars in NGC~1866 is obviously driven by the need to improve the extragalactic distance scale using either period-luminosity relations or the Baade-Wesselink methods \citep[e.g.,][]{Storm2011a,Storm2011b,Moli2012}.\\    

\par It is therefore quite surprising that the chemical composition of NGC~1866 stars has been investigated only in a few high-resolution spectroscopic studies: \cite{Hill2000} analyzed a few elements in three red giant branch (RGB) stars in NGC~1866 and report [Fe/H]=$-0.50\pm0.1$~dex. \cite{Mucc2011} derived the detailed chemical composition of 14 members of NGC~1866 and of 11 additional LMC field stars. They found an average [Fe/H]=$-0.43$~dex for NGC~1866. \cite{Colu2011,Colu2012a} determined the age and metallicity of NGC~1866 via high-resolution integrated light spectroscopy and extended their work to other elements in \cite{Colu2012b}. The study of \cite{Colu2012a} also includes three stellar targets in NGC~1866 for comparison purposes, with metallicities ranging from $-0.31$ to $-0.39$~dex.\\

\par In this paper, we focus on the chemical properties of six Cepheids in NGC~1866 and four field Cepheids in the SMC, and investigate what their chemical composition tells us about the stellar populations they belong to. Our sample increases the number of Cepheids with known metallicities in the LMC/SMC by 20\%/25\% and the number of Cepheids with known detailed chemical composition in the LMC/SMC by 46\%/50\%. The Baade-Wesselink analysis will be presented in a companion paper.


\section{Observations}

\par We selected stars for which both optical \& near-infrared light curves and radial velocity measurements of good quality are already available, but for which no direct determination of the metallicity exists. We selected six Cepheids in the LMC NGC~1866 cluster and four field Cepheids in the SMC. The LMC cluster stars were observed with the FLAMES/UVES high-resolution spectrograph \citep{Pas2002} while the SMC field stars were observed with the UVES high-resolution spectrograph \citep{Dek2000}. We used the red arm (CD \#3) standard template centered on 580~nm which offers a resolution of 47~000 and covers the 476--684 nm wavelength range with a 5~nm gap around the central wavelength. We used the ESO reflex pipeline \citep{Freu2013}\footnote[1]{ftp://ftp.eso.org/pub/dfs/pipelines/uves/uves-fibre-pipeline-manual-18.8.1.pdf\\ftp://ftp.eso.org/pub/dfs/pipelines/uves/uves-pipeline-manual-22.14.1.pdf} to perform the basic data reduction of the spectra. The heliocentric corrections of the radial velocities were computed with the IRAF task {\it rvcorrect}. The observing log is listed in Table~\ref{obslog}. For the FLAMES/UVES sample, the weather conditions deteriorated during the night. We therefore analyzed only the first three spectra of a series of six for each star, as they reached a higher S/N. The S/N values are listed in Table~\ref{atmparam}.\\

\par The phases were computed by adopting the period and the epoch of maximum light from OGLE IV \citep{Udal2015} as a zero point reference, except for HV~12202 for which no OGLE IV data are available. For this star, we used the values provided by \cite{Moli2012}. The computations were made using heliocentric Julian dates (HJD), i.  e., 0.5 days were added to the modified Julian dates (MJD) and the light travel time between the Earth and the Sun was taken into account. The HJDs were double-checked using the IRAF task {\it rvcorrect}.

\begin{table*}[!htbp]
\caption{Observing log. The first six lines are spectra taken with the FLAMES/UVES multi-object spectrograph. The other spectra were taken with the UVES spectrograph.}
\label{obslog}
\centering\begin{tabular}{ccccc}
\hline\hline
  Target &           Date          &      MJD       &  Airmass &  Exp time \\
         &                         &                &  (start) &    (s)    \\ 

\hline
NGC~1866 & 2008-12-06T00:35:23.385 & 54806.02457622 &  1.848   &  4800     \\
NGC~1866 & 2008-12-06T01:56:13.752 & 54806.08071473 &  1.541   &  4800     \\
NGC~1866 & 2008-12-06T03:17:03.638 & 54806.13684767 &  1.381   &  4800     \\
NGC~1866 & 2008-12-06T04:50:18.350 & 54806.20160128 &  1.320   &  4800     \\
NGC~1866 & 2008-12-06T06:11:08.646 & 54806.25773897 &  1.358   &  4800     \\
NGC~1866 & 2008-12-06T07:31:58.631 & 54806.31387305 &  1.490   &  3600     \\
HV~822   & 2008-11-15T00:57:58.745 & 54785.04026326 &  1.531   &  1000     \\
         & 2008-11-15T01:15:28.075 & 54785.05240828 &  1.523   &  1000     \\
         & 2008-11-15T01:32:57.446 & 54785.06455378 &  1.519   &  1000     \\
HV~1328  & 2008-11-15T00:11:13.084 & 54785.00779033 &  1.569   &   800     \\
         & 2008-11-15T00:25:22.587 & 54785.01762254 &  1.556   &   800     \\
         & 2008-11-15T00:39:31.570 & 54785.02744873 &  1.545   &   800     \\
HV~1333  & 2008-11-15T01:53:56.155 & 54785.07912217 &  1.531   &  1200     \\
         & 2008-11-15T02:14:45.620 & 54785.09358357 &  1.537   &  1200     \\     
         & 2008-11-15T02:35:34.996 & 54785.10804394 &  1.547   &  1200     \\
HV~1335  & 2008-11-15T03:03:49.091 & 54785.12765152 &  1.569   &  1300     \\
         & 2008-11-15T03:26:18.517 & 54785.14326987 &  1.594   &  1300     \\
         & 2008-11-15T03:48:48.862 & 54785.15889887 &  1.625   &  1300     \\
\hline
\end{tabular}
\end{table*}


\section{Chemical abundances}

\subsection{Data analysis}
\label{dao}
\par In our spectra, we measured the equivalent widths of the absorption lines with DAOSPEC \citep{Stet2008}: DAOSPEC fits lines with saturated Gaussians and all the lines detected are cross-correlated with a list of lines provided by the user. For each individual measurement of an equivalent width (EW), DAOSPEC provides the standard error $\sigma_{EW}$ on the measurement and a quality parameter Q that becomes higher in the regions where the quality of the spectrum decreases or for strong lines that deviate from a Gaussian profile. We selected only lines with $\sigma_{EW}$ $\leq$ 10~\% and Q~$\leq$ 1.25. For both the determination of the atmospheric parameters and the computation of the abundances, we considered only the lines with 20~$\leq$~EW~$\leq$~130~m\AA. 

\begin{figure}[!htbp]              
\centering                   
   \includegraphics[angle=-90,width=\columnwidth]{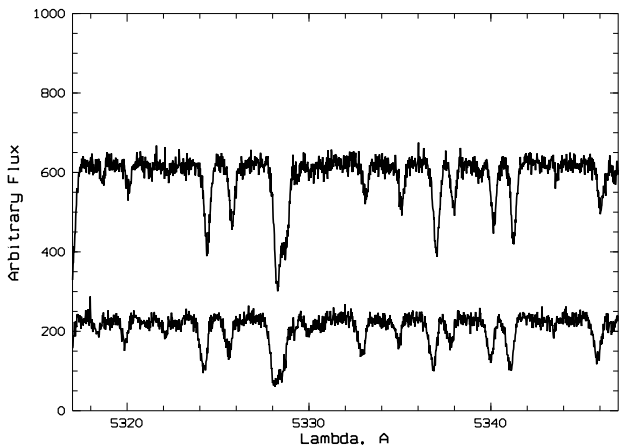}
   \caption{Excerpts of spectra covering the 5317--5347 \AA{} range. {\it Top:} HV1328 (SMC) at MJD=54785.01762254 (S/N$\approx$20).
{\it Bottom:} HV12198 (NGC~1866) at 54806.13684767 (S/N$\approx$30).}
\label{spec}
\end{figure}

The equivalent width method was favoured as it enables a more homogeneous continuum placement, especially for spectra with a relatively low S/N like ours (see examples in Fig.\ref{spec}). The hyperfine structure can therefore not be taken into account. Current studies indicate that the effects of hyperfine structure splitting (hfs) are negligible or small for Y, Zr, Nd, and Eu in Cepheids \citep{daSilva2016}, but not for Mn (Lemasle et al., in prep) or to a lesser extent La \citep{daSilva2016}. A more detailed discussion about the hfs is provided in Sect~\ref{hfs}.

\subsection{Radial velocities}

\par For the NGC 1866 sample, the accuracy of the radial velocity determined by DAOSPEC is in general better than $\pm$2~km~s$^{-1}$, with a mean error in the individual velocities measurement of 1.157~km~s$^{-1}$. Thanks to a higher S/N, the radial velocities for the SMC sample are even more accurate, with a mean error of 0.804~km~s$^{-1}$. Our measurements are listed in Table~\ref{Vr}. Comments in footnotes come from the OGLE-III database \citep{Sos2010a}. Note that in both cases the radial velocities obtained from the lower (L) and the upper (U) chip of the UVES red arm are in excellent agreement. The averaged radial velocities and the heliocentric corrections (computed with the IRAF task {\it rvcorrect}, with a negligible uncertainty of $\approx$0.005~km~s$^{-1}$) are also listed in Table~\ref{Vr}.

\begin{table*}[htbp!]
\centering
\caption{Radial velocities for our targets in the LMC cluster NGC~1866 and in the field of the SMC. The radial velocities derived for the lower (L) and upper (U) chips of the UVES red arm are listed in cols. 4 and 5. The averaged values are listed in col. 6, the barycentric corrections in col.7 and the final values for the radial velocity (after correction) in col. 8.}
\label{Vr}
\begin{tabular}{cccccccc}
\hline\hline
\multicolumn{8}{c}{{\bf Targets in the LMC cluster NGC~1866}}\\
\hline
Target & Period (P) & Phase $\phi$ & Vr$_{L}$\tablefootmark{a} & Vr$_{U}$\tablefootmark{b} & Vr (averaged) & Heliocentric correction & Vr corrected \\
       &    (d)     &              &       (km~s$^{-1}$)       &       (km~s$^{-1}$)       & (km~s$^{-1}$) &      (km~s$^{-1}$)      & (km~s$^{-1}$)\\ 
\hline  
HV~12197 & 3.1437642 & 0.081 & 283.399$\pm$1.919 & 282.260$\pm$3.202 & 283.098$\pm$1.646 &  -2.205 & 280.893$\pm$1.646 \\
         &           & 0.099 & 284.238$\pm$1.390 & 284.180$\pm$1.641 & 284.214$\pm$1.061 &  -2.242 & 281.972$\pm$1.061 \\                           
         &           & 0.117 & 285.289$\pm$1.567 & 284.997$\pm$1.378 & 285.124$\pm$1.035 &  -2.294 & 282.830$\pm$1.035 \\                           
HV~12198 & 3.5227781 & 0.643 & 315.940$\pm$1.384 & 315.921$\pm$1.552 & 315.932$\pm$1.033 &  -2.190 & 313.742$\pm$1.033 \\                           
         &           & 0.659 & 316.662$\pm$1.031 & 316.566$\pm$1.332 & 316.626$\pm$0.815 &  -2.227 & 314.399$\pm$0.815 \\                           
         &           & 0.675 & 316.960$\pm$1.305 & 316.943$\pm$1.097 & 316.950$\pm$0.840 &  -2.279 & 314.671$\pm$0.840 \\                           
HV~12199 & 2.6391571 & 0.928 & 289.626$\pm$2.129 & 289.346$\pm$4.279 & 289.570$\pm$1.906 &  -2.199 & 287.371$\pm$1.906 \\                           
         &           & 0.949 & 284.720$\pm$1.187 & 284.655$\pm$2.177 & 284.705$\pm$1.042 &  -2.236 & 282.469$\pm$1.042 \\                           
         &           & 0.970 & 281.507$\pm$1.528 & 281.312$\pm$1.867 & 281.429$\pm$1.182 &  -2.288 & 279.141$\pm$1.182 \\                           
HV~12202 & 3.101207  & 0.807 & 319.318$\pm$2.632 & 318.444$\pm$4.291 & 319.079$\pm$2.244 &  -2.180 & 316.899$\pm$2.244 \\                           
         &           & 0.825 & 316.907$\pm$2.040 & 316.018$\pm$2.053 & 316.465$\pm$1.447 &  -2.217 & 314.248$\pm$1.447 \\                           
         &           & 0.843 & 313.635$\pm$1.472 & 312.713$\pm$3.442 & 313.492$\pm$1.353 &  -2.269 & 311.223$\pm$1.353 \\                           
HV~12203 & 2.9541342 & 0.765 & 323.664$\pm$1.902 & 323.307$\pm$2.098 & 323.503$\pm$1.409 &  -2.180 & 321.323$\pm$1.409 \\                           
         &           & 0.784 & 322.905$\pm$1.775 & 322.400$\pm$1.448 & 322.602$\pm$1.122 &  -2.217 & 320.385$\pm$1.122 \\                           
         &           & 0.803 & 320.983$\pm$1.407 & 321.260$\pm$1.702 & 321.095$\pm$1.084 &  -2.269 & 318.826$\pm$1.084 \\                           
HV~12204 & 3.4387315 & 0.519 & 292.454$\pm$0.524 & 292.107$\pm$0.955 & 292.374$\pm$0.459 &  -2.163 & 290.211$\pm$0.459 \\                           
         &           & 0.535 & 293.664$\pm$0.919 & 293.164$\pm$0.751 & 293.364$\pm$0.582 &  -2.200 & 291.164$\pm$0.582 \\                           
         &           & 0.551 & 294.177$\pm$0.713 & 294.166$\pm$0.895 & 294.173$\pm$0.558 &  -2.252 & 291.921$\pm$0.558 \\                           
\hline                                                                                           
\multicolumn{8}{c}{{\bf Targets in the SMC}}\\
\hline
HV~822\tablefootmark{c}  & 16.7419693 & 0.998 & 101.288$\pm$1.736 & 101.438$\pm$2.309 & 101.342$\pm$1.388 & -12.692 &  88.650$\pm$1.388 \\
                         &            & 0.999 & 101.143$\pm$1.931 & 101.027$\pm$1.672 & 101.077$\pm$1.264 & -12.701 &  88.376$\pm$1.264 \\
                         &            & 0.999 & 101.450$\pm$0.939 & 100.845$\pm$1.543 & 101.286$\pm$0.802 & -12.709 &  88.577$\pm$0.802 \\
HV~1328\tablefootmark{d} & 15.8377104 & 0.883 & 121.800$\pm$1.013 & 121.771$\pm$1.605 & 121.792$\pm$0.857 & -12.808 & 108.984$\pm$0.857 \\
                         &            & 0.884 & 121.592$\pm$0.626 & 121.509$\pm$0.615 & 121.550$\pm$0.439 & -12.814 & 108.736$\pm$0.439 \\
                         &            & 0.884 & 121.402$\pm$1.069 & 121.716$\pm$0.876 & 121.590$\pm$0.678 & -12.821 & 108.769$\pm$0.678 \\
HV~1333                  & 16.2961015 & 0.659 & 179.553$\pm$1.333 & 179.786$\pm$0.735 & 179.732$\pm$0.644 & -12.742 & 166.990$\pm$0.644 \\
                         &            & 0.660 & 179.816$\pm$1.267 & 179.909$\pm$0.930 & 179.876$\pm$0.750 & -12.752 & 167.124$\pm$0.750 \\
                         &            & 0.661 & 179.486$\pm$1.555 & 179.972$\pm$0.827 & 179.865$\pm$0.730 & -12.761 & 167.104$\pm$0.730 \\
HV~1335                  & 14.3813503 & 0.318 & 163.310$\pm$0.763 & 163.376$\pm$1.360 & 163.326$\pm$0.665 & -12.750 & 150.576$\pm$0.665 \\
                         &            & 0.319 & 162.985$\pm$1.154 & 163.194$\pm$1.102 & 163.094$\pm$0.797 & -12.760 & 150.334$\pm$0.797 \\
                         &            & 0.320 & 163.352$\pm$1.466 & 163.198$\pm$0.877 & 163.239$\pm$0.753 & -12.769 & 150.470$\pm$0.753 \\
\hline                                                                                                
\end{tabular}                                                                                    
\tablefoot{
\tablefoottext{a}{Red arm lower chip.}
\tablefoottext{b}{Red arm upper chip.}
\tablefoottext{c}{secondary period of 1.28783d (OGLE-III database).}
\tablefoottext{d}{secondary period of 14.186d (OGLE-III database).}
}
\end{table*}

\par Because this was one of our target selection criteria, there is an extensive amount of radial velocity data available for the Cepheids in our sample. From these data it was possible to ascertain that our NGC~1866 Cepheids are indeed cluster members. Excluding variable stars, \citet{Mucc2011} report an average heliocentric velocity of v=298.5$\pm$0.4~km~s$^{-1}$ with a dispersion of $\sigma$=1.6~km~s$^{-1}$. For both the LMC and SMC targets our radial velocity measurements are in excellent agreement with the expected values at the given pulsation phase obtained from the radial velocity curves published in the literature \citep{Welch1991,Storm2004a,Storm2005,Moli2012,Mar2013,Mar2017}. 
\par Systematic shifts between different samples are generally attributed to the orbital motion in a binary system. Two stars in our sample (HV~12202 and HV~12204) were identified as spectroscopic binaries \citep{Welch1991,Storm2005}. As far as HV~12202 is concerned, our measurements are in good agreement with all the data compiled by \citet{Storm2005} except for their CTIO data and the latest part of the \citet{Welch1991} data, and therefore support the shifts of +18~km~s$^{-1}$ (respectively +21~km~s$^{-1}$) applied to these datasets in order to provide an homogeneous radial velocity curve. For the same purpose, the latest data from \citet{Welch1991} had to be shifted by +7~km~s$^{-1}$ and our measurements should be shifted by $\approx$ +15~km~s$^{-1}$ in the case of HV~12204. Binarity is a common feature for Milky Way Cepheids \citep[more than 50\% of them are binaries, see][]{Sza2003}, but there is a strong observational bias with distance and indeed the number of known binaries is much lower for the farther, fainter Cepheids in the Magellanic Clouds \citep{Sza2012}. It should be noted that \cite{Ander2014} found modulations in the radial velocity curves of four Galactic Cepheids. However, the order of magnitude of the effect ranges from several hundred m~s$^{-1}$ to a few km~s$^{-1}$ and cannot account for the differences reported here in the case of HV~12202 and HV~12204.\\

\subsection{Atmospheric parameters}
                  
\par As Cepheids are variable stars, simultaneous photometric and spectroscopic observations are in general not available and the atmospheric parameters 
are usually derived from the spectra only. \cite{KovGor2000} have developed an accurate method to derive the effective temperature \Teff{} from the depth ratio of carefully chosen pairs of lines that have been used extensively in Cepheids studies \citep{And2002a,Luck2011b}.
                  
As the red CCD detector of UVES is made of two chips side by side (lower: L; upper: U), there is a gap of $\approx$ 50~\AA{} around the central wavelength (580 nm in our case) and we could not use the lines falling in this spectral domain. Moreover, the line depth ratios have been calibrated for Milky Way Cepheids that are more metal-rich than the Magellanic Cepheids (especially in the case of the SMC)\footnote{The metallicity of Milky Way Cepheids continuously decreases from +0.4--+0.5 dex in the inner disk \citep[e.g., ][]{And2002b,Pedi2010,Martin2015,And2016} to $\approx$ $-0.4$~dex in the outer disk \citep[e.g., ][]{Luck2003, Lem2008}. Current high-resolution spectroscopic studies indicate that Cepheids have metallicities ranging from $-0.62$ to $-0.10$~dex \citep{Luck1992,Luck1998,Roma2008} in the LMC and from $-0.87$ to $-0.63$~dex in the SMC.}. It also turns out that several stars in our sample were observed at a phase where they reach higher \Teff{} ($>$6000 K) during the pulsation cycle. The combination of a high \Teff{} and a rather low metallicity made it very challenging to measure the depth of some lines, and in particular the weak line of the pairs. As a result, we could only use a limited number of line depth ratios (typically 5--10 out of 32) to determine \Teff. Moreover, for two stars (HV~12199 and HV~822), we were unable to determine \Teff{} from the line depth ratio as their temperature ($>$6400~K) at the time of the observations fell above the range of temperatures where most ratios are calibrated \footnote{Depending on the ratio, the upper limit varies between 6200 and 6700~K.}.\\
                  
\par To ensure the determination of \Teff, we double-checked that lines with both high and low $\chi_{\rm ex}$ values properly fit the curve of growth (See Appendix~\ref{CoG_newmarcs}) and that the Fe~I abundances are independent from the excitation potential of the lines. In a canonical spectroscopic analysis, we determined the surface gravity \logg{} and the microturbulent velocity \vt{} by imposing that the ionization balance between Fe~I and Fe~II is satisfied and that the Fe~I abundance is independent from the EW of the lines. On average we have at our disposal 42~Fe~I/7~Fe~II lines in the NGC~1866 Cepheids and 42~Fe~I/11~Fe~II lines in the SMC Cepheids. We note that the adopted \Teff{} values are in general in very good agreement with those derived from the line depth ratios. The atmospheric parameters are listed in Table~\ref{atmparam}. 
                  
\par As mentioned above, the Cepheids in our sample have rather high temperatures, two of them hot enough at the phase of the observations to prevent the use of line depth ratios to determine their temperature. It has been noted before \citep{Bro2004} that these stars are located in the color-magnitude diagram at the hot tip of the so-called "blue nose" experienced by core He-burning supergiants. During this evolutionary stage they cross the instability strip and start pulsating.

\par As we impose the ionization balance between Fe I and Fe II to derive \logg, NLTE effects affecting primarily Fe I could hamper an accurate determination of \logg{} \citep{LL1985}. There is currently no extensive study of NLTE effects in Cepheids, although NLTE abundances have been derived for some individual elements like O \citep[][and references therein]{Koro2014} or Ba \citep[][and references therein]{And2014}. It is beyond the scope of this paper to provide a full discussion of NLTE effects in Cepheids, and we refer the reader to the discussion in e.g., \citet{Kov1999} or \citet{Yong2006}. Several arguments have been brought forward to support the fact that NLTE effects may be limited in Cepheids. For instance, \citet{And2005} followed several Cepheids with 3d$<$P$<$6d throughout the entire period and found identical [Fe/H] and abundances ratios (within the uncertainties), although \Teff{} varies by $\approx$ 1000~K (the same holds for Cepheids with different period ranges studied in this series of papers). Also \citet[][]{Yong2006} found a mean difference [TiI/Fe]--[TiII/Fe]=0.07$\pm$0.02 ($\sigma$=0.11). As this difference falls within the measurement uncertainties, they concluded that the values of \logg{} obtained via the ionization equilibrium of FeI/FeII are satisfactory. All these arguments point toward the fact that a canonical spectroscopic analysis provides consistent, reliable results. However, the aforementioned studies deal with Milky Way Cepheids. As Magellanic Cepheids are slightly more metal-poor, NLTE effects should be a bit more pronounced than in the Galactic ones. In a study of 9 LMC F supergiants, \citet{Hill1995} introduced an overionization law and obtained higher (+0.6 dex) spectroscopic gravities that are in good agreement with those derived from photometry. They note that [Fe/H] becomes only +0.1 dex higher than in the LTE case and that the global abundance pattern remains unchanged, as already reported by, e.g., \cite{Spite1989}.

\subsection{Comparison with models}

\par \cite{Mar2013} have used non-linear convective pulsation models in order to reproduce simultaneously the lightcurves in several photometric bands and the radial velocity curves of a few Cepheids in NGC~1866. For HV~12197 they reached a good agreement between theory and observations and report a mean \Teff{} of 5850~K. They also plotted the temperature predicted by the model and for the phases 0.08--0.12 they found \Teff{} of the order of 6300~K and slightly below (their Fig.~8, bottom panel), in quite good agreement with our measurements that fall around 6150~K. For HV~12199, they report a mean \Teff{} of 6125~K but had to modify notably the projection factor to reach the best match with the radial velocity curve. They also mention that using the lightcurves only would lead to a hotter star (<~\Teff{}~> = 6200~K), but in this case an even lower (and unrealistic) value would be required for the projection factor in order to fit the radial velocity curve. The \Teff{} curve for HV~12199 (their Fig.~8, top panel) in the phases 0.93--0.97 shows a rapid rise of the temperature and the corresponding \Teff{} value of $\approx$ 6250--6300~K, somewhat below the values around 6600~K we determined for \Teff.   

\subsection{Abundance determinations}

\par Our abundance analysis is based on equivalent widths measured with DAOSPEC (see Sect.~\ref{dao}). We derived the abundances of 16 elements (several of them in two ionization states) for which absorption lines could be measured in the spectral domain covered by the UVES red arm (CD~ \#3, 580~nm) standard template. In a few cases we updated the linelists of \cite{Gen2013} and \cite{Lem2013} with oscillator strengths and excitation potentials from recent releases of the Vienna Atomic Lines Database \citep[VALD,][and references therein]{Kup1999} and from the Gaia-ESO survey linelist \citep{Heit2015}. We took the values tabulated by \citet{Anders1989} as Solar references, except for Fe and Ti for which we used $log~\epsilon_{Fe}$=7.48 and $log~\epsilon_{Ti}$=5.02. We used MARCS (1D LTE spherical) atmosphere models \citep{Gus2008} covering the parameter space of Magellanic Clouds Cepheids. Abundances were computed with {\it calrai}, a LTE spectrum synthesis code originally developed by \cite{Spite1967} and continuously updated since then. For a given element, the abundance derived from a single spectrum is estimated as the mean value of the abundances determined for each individual line of this element. The final abundance of a star is then obtained by computing the weighted mean (and standard deviation) for the three spectra analyzed, where the weight is the number of lines of a given element measured in each spectrum.  

\begin{table*}[!htbp]
\caption{Coordinates, properties and atmospheric parameters for the Cepheids in our sample. V magnitudes and periods are from OGLE IV, except for HV12202, for which they have been found in \citet{Moli2012} and \citet{Mus2016}. Col.~7 refers to the \Teff{} derived from the line depth ratio method \citep[LDR,][]{KovGor2000} while col.~8 is the \Teff{} derived from the excitation equilibrium. The last column lists the S/N around 5228 and 5928 \AA{} respectively.}
\label{atmparam}
\centering\small
\begin{tabular}{lccccccccccc}
\hline\hline
\multicolumn{12}{c}{{\bf Targets in the LMC cluster NGC~1866}}\\
\hline
 Cepheid & RA (J2000) & Dec (J2000) &   V    &      P     & $\phi$ & \Teff~(LDR)     & \Teff & \logg & \vt         & [Fe/H] &    S/N    \\
         &   (dms)    &   (dms)   & (mag)  &     (d)    &        &       (K)       &  (K)  & (dex) &(km~s$^{-1}$)& (dex)  & {\tiny(5228/5928\AA)} \\
\hline                                                      
HV~12197 & 05 13 13.0 & -65 30 48 & 16.116 &  3.1437642 & 0.081 & 6060$\pm$ 97  (3) & 6150  &  1.5  & 3.1 &  -0.35  & 16/15 \\ 
         &            &           &        &  3.1437642 & 0.099 &                   & 6150  &  1.5  & 3.2 &  -0.35  & 28/27 \\ 
         &            &           &        &  3.1437642 & 0.117 &                   & 6100  &  1.5  & 3.1 &  -0.35  & 27/25 \\ 
HV~12198 & 05 13 26.7 & -65 27 05 & 15.970 &  3.5227781 & 0.643 & 5634$\pm$ 85  (6) & 5625  &  1.4  & 3.4 &  -0.35  & 13/19 \\ 
         &            &           &        &  3.5227781 & 0.659 &                   & 5625  &  1.5  & 3.6 &  -0.35  & 20/26 \\ 
         &            &           &        &  3.5227781 & 0.675 &                   & 5625  &  1.4  & 3.6 &  -0.35  & 21/23 \\ 
HV~12199 & 05 13 19.0 & -65 29 30 & 16.283 &  2.6391571 & 0.928 &      --~~         & 6550  &  2.2  & 3.2 &  -0.30  & 15/14 \\ 
         &            &           &        &  2.6391571 & 0.949 &                   & 6600  &  2.1  & 3.0 &  -0.30  & 29/31 \\ 
         &            &           &        &  2.6391571 & 0.970 &                   & 6650  &  2.0  & 3.1 &  -0.35  & 26/32 \\ 
HV~12202 & 05 13 39.0 & -65 29 00 & 16.08  &  3.101207  & 0.807 & 5712$\pm$100  (6) & 5775  &  1.6  & 3.1 &  -0.40  & 17/14 \\ 
         &            &           &        &  3.101207  & 0.825 &                   & 5900  &  1.6  & 3.1 &  -0.40  & 20/25 \\ 
         &            &           &        &  3.101207  & 0.843 &                   & 5900  &  1.5  & 2.9 &  -0.40  & 20/24 \\ 
HV~12203 & 05 13 40.0 & -65 29 36 & 16.146 &  2.9541342 & 0.765 & 5856$\pm$117  (9) & 5850  &  1.7  & 3.5 &  -0.35  & 16/19 \\ 
         &            &           &        &  2.9541342 & 0.784 &                   & 5800  &  1.2  & 3.3 &  -0.35  & 17/26 \\ 
         &            &           &        &  2.9541342 & 0.803 &                   & 5800  &  1.6  & 3.4 &  -0.35  & 19/24 \\ 
HV~12204 & 05 13 58.0 & -65 28 48 & 15.715 &  3.4387315 & 0.519 & 5727$\pm$ 98 (11) & 5700  &  1.2  & 2.8 &  -0.35  & 19/23 \\ 
         &            &           &        &  3.4387315 & 0.535 &                   & 5725  &  1.3  & 2.9 &  -0.35  & 22/31 \\ 
         &            &           &        &  3.4387315 & 0.551 &                   & 5700  &  1.2  & 2.9 &  -0.35  & 21/28 \\ 
\hline                                                                                           
\multicolumn{12}{c}{{\bf Targets in the SMC}}\\
\hline
HV~822   & 00 41 55.5 & -73 32 23 & 14.524 & 16.7419693 & 0.998 &      --~~         & 6400  &  1.8  & 2.7 &  -0.75  & 33/48 \\
         &            &           &        & 16.7419693 & 0.999 &                   & 6400  &  1.8  & 2.7 &  -0.75  & 41/46 \\
         &            &           &        & 16.7419693 & 0.999 &                   & 6400  &  1.8  & 2.7 &  -0.75  & 35/42 \\
HV~1328  & 00 32 54.9 & -73 49 19 & 14.115 & 15.8377104 & 0.883 & 6325$\pm$ 98  (5) & 6100  &  1.9  & 2.6 &  -0.60  & 26/37 \\
         &            &           &        & 15.8377104 & 0.884 &                   & 6100  &  1.9  & 2.6 &  -0.60  & 31/37 \\
         &            &           &        & 15.8377104 & 0.884 &                   & 6100  &  1.9  & 2.6 &  -0.60  & 29/40 \\
HV~1333  & 00 36 03.5 & -73 55 58 & 14.729 & 16.2961015 & 0.659 & 5192$\pm$102  (8) & 5175  &  0.4  & 3.2 &  -0.90  & 18/26 \\
         &            &           &        & 16.2961015 & 0.660 &                   & 5200  &  0.4  & 2.8 &  -0.80  & 19/25 \\
         &            &           &        & 16.2961015 & 0.661 &                   & 5175  &  0.4  & 3.2 &  -0.90  & 16/30 \\
HV~1335  & 00 36 55.7 & -73 56 28 & 14.762 & 14.3813503 & 0.318 & 5566$\pm$156  (6) & 5600  &  0.6  & 2.6 &  -0.80  & 25/29 \\
         &            &           &        & 14.3813503 & 0.319 &                   & 5675  &  0.8  & 2.6 &  -0.75  & 28/29 \\
         &            &           &        & 14.3813503 & 0.320 &                   & 5600  &  0.6  & 2.7 &  -0.80  & 25/29 \\
\hline                                                     
\end{tabular}
\end{table*}

\subsection{Abundances}      
\label{Ab}                   
                             
\par We provide the abundances of one light element (Na), several $\alpha$-elements (Mg, Si, S, Ca, Ti), iron-peak elements (Sc, Cr, Mn, Fe, Ni), and neutron capture elements (Y, Zr, La, Nd, Eu). As already mentioned, we analyzed three individual (back to back) spectra for each star and the abundances derived are in most cases in excellent agreement. As expected, the size of the error bars is correlated to the number of lines analyzed. In contrast to our Cepheid studies in the Milky Way, where Si comes second after iron for the number of lines measured, the UVES red arm (CD \#3, 580~nm) spectral domain contains only a few Si lines with sufficient quality but a larger number of Ca lines and indeed 9--11 calcium lines were usually measured in our spectra. The individual abundances (per spectrum) are listed in Tables \ref{abund_12197}--\ref{abund_12204} for the NGC~1866 Cepheids and in Tables \ref{abund_822}--\ref{abund_1335} for the SMC Cepheids. The last two columns of these tables list the weighted means and standard deviations, adopted as the chemical composition of the star in the rest of the paper.

\par \cite{Moli2012} provide the metallicities for three Cepheids in NGC~1866, analyzed in the same way as the stars in \cite{Mucc2011}. Two of \cite{Moli2012} Cepheids are also included in our sample, namely HV~12197 and HV~12199: taking into account a tiny difference (0.02 dex) in the solar reference value for [Fe/H], the results agree very well: they report [Fe/H]=$-0.39\pm0.05$ for HV~12197 while we found $-0.33\pm0.07$~dex, and  [Fe/H]=$-0.38\pm0.06$ for HV~12199 while we found $-0.31\pm0.05$. 

\par For a good number of our spectra, several elements (Si, Ti, Cr) could be measured in two ionization states, in addition to the usual Fe~I~/~Fe~II. When the ionization equilibrium is reached for iron, it is usually also reached for the other elements as the abundances derived from the neutral and ionized species agree within the error bars, thus reinforcing our confidence in our atmospheric parameters, in particular \logg. In order to quantify how the results are affected by uncertainties in the atmospheric parameters, we computed the abundances with over- or underestimated values of \Teff ($\pm$150 K), \logg ($\pm$0.3 dex), \vt ($\pm$0.5 km~s$^{-1}$) for two spectra at different \Teff. Uncertainties in [Fe/H] leave the abundances unchanged and are therefore not considered in this exercise. The sum in quadrature of the differences in the computed abundances is adopted as the uncertainty in the abundances due to the uncertainties in the atmosphere parameters. The resulting values are listed in Table~\ref{err_budget_atmparam}. 

\begin{table}[htbp!]
\centering
\caption{Uncertainties in the final abundances due to uncertainties in the atmospheric parameters. Cols. 2, 3, 4 indicate respectively how the abundances are modified (mean values) when they are computed with over- or underestimated values of \Teff ($\pm$150 K), \logg ($\pm$0.3 dex), or \vt ($\pm$0.5 dex). The sum in quadrature of the differences is adopted as the uncertainty in the abundances due to the uncertainties in the atmosphere parameters}
\label{err_budget_atmparam}
\begin{tabular}{ccccc}
\hline\hline
\multicolumn{5}{c}{{Error budget for HV~12198, $\phi$=0.675 (MJD=54806.13684767)}}\\
\hline
Element & $\Delta$\Teff & $\Delta$\logg &  $\Delta$\vt  & Quadratic \\
        & {\small ($\pm$ 150 K)} & {\small ($\pm$ 0.3 dex)} & {\small ($\pm$ 0.5 km~s$^{-1}$)}  &     sum       \\
        &      (dex)    &      (dex)    &    (dex)      &    (dex)      \\        
\hline  
 [NaI/H] & 0.08 & 0.00 & 0.03 & 0.09\\{}
 [MgI/H] & 0.14 & 0.00 & 0.08 & 0.16\\{}
 [SiI/H] & 0.07 & 0.01 & 0.03 & 0.07\\{}
[SiII/H] & 0.12 & 0.12 & 0.11 & 0.19\\{}
  [SI/H] & 0.07 & 0.08 & 0.04 & 0.11\\{}
 [CaI/H] & 0.10 & 0.00 & 0.06 & 0.12\\{}
[ScII/H] & 0.03 & 0.12 & 0.08 & 0.15\\{}
 [TiI/H] & 0.15 & 0.00 & 0.03 & 0.15\\{}
[TiII/H] & 0.02 & 0.12 & 0.02 & 0.12\\{}
[CrII/H] & 0.02 & 0.11 & 0.07 & 0.13\\{}
 [MnI/H] & 0.12 & 0.00 & 0.03 & 0.12\\{}
 [FeI/H] & 0.13 & 0.00 & 0.05 & 0.13\\{}
[FeII/H] & 0.01 & 0.12 & 0.05 & 0.13\\{}
 [NiI/H] & 0.13 & 0.00 & 0.04 & 0.13\\{}
 [YII/H] & 0.04 & 0.12 & 0.05 & 0.13\\{}
[ZrII/H] & 0.03 & 0.11 & 0.03 & 0.12\\{}
[LaII/H] & 0.07 & 0.12 & 0.04 & 0.14\\{}
[NdII/H] & 0.07 & 0.22 & 0.05 & 0.23\\{}
[EuII/H] & 0.04 & 0.12 & 0.02 & 0.12\\  
\hline
\multicolumn{5}{c}{{Error budget for HV~1328, $\phi$=0.884 (MJD=54785.02744873)}}\\
\hline
Element & $\Delta$\Teff & $\Delta$\logg &  $\Delta$\vt  & Quadratic \\
        & {\small ($\pm$ 150 K)}   & {\small  ($\pm$ 0.3 dex)} & {\small($\pm$ 0.5 km~s$^{-1}$)}  &     sum       \\
        &      (dex)    &      (dex)     &    (dex)      &    (dex)      \\        
\hline  
 [NaI/H] & 0.10 & 0.04 & 0.02 & 0.11 \\{}
 [MgI/H] & 0.11 & 0.05 & 0.02 & 0.12 \\{}
 [SiI/H] & 0.18 & 0.07 & 0.02 & 0.19 \\{}
 [CaI/H] & 0.13 & 0.06 & 0.04 & 0.15 \\{}
[ScII/H] & 0.28 & 0.10 & 0.06 & 0.30 \\{}
 [TiI/H] & 0.09 & 0.11 & 0.03 & 0.15 \\{}
[TiII/H] & 0.12 & 0.03 & 0.07 & 0.14 \\{}
 [CrI/H] & 0.13 & 0.05 & 0.07 & 0.15 \\{}
[CrII/H] & 0.10 & 0.07 & 0.06 & 0.13 \\{}
 [FeI/H] & 0.11 & 0.06 & 0.08 & 0.15 \\{}
[FeII/H] & 0.22 & 0.13 & 0.09 & 0.26 \\{}
 [NiI/H] & 0.09 & 0.13 & 0.05 & 0.16 \\{}
 [YII/H] & 0.07 & 0.08 & 0.03 & 0.11 \\{}
[ZrII/H] & 0.31 & 0.12 & 0.02 & 0.33 \\{}
[LaII/H] & 0.13 & 0.05 & 0.03 & 0.14 \\{}
[NdII/H] & 0.10 & 0.15 & 0.03 & 0.18 \\{}
[EuII/H] & 0.28 & 0.08 & 0.03 & 0.29 \\
\hline                                                                                                
\end{tabular}                                                                                    
\end{table}

\label{hfs}
\par For the stars in our sample, NLTE effects are negligible for Na ($\leq$ 0.1 dex) as computed by \cite{Lind2011} for a range of atmospheric parameters including yellow supergiants and by \cite{Take2013} for Cepheids. Using DAOSPEC to automatically determine the EW of the lines (and the relatively low S/N of our spectra) made it impossible for us to take into account the contribution of the hyperfine structure splitting for iron-peak elements, and for neutron-capture elements as in \cite{daSilva2016}. Depending on the line considered, the latter authors estimated the hfs correction to range from negligible to $\approx$0.20 dex. In the case of the 6262.29 La~II line, it reaches $-0.211\pm0.178$~dex, where the quoted error represents the dispersion around the mean hfs correction for this line. In a forthcoming paper (Lemasle et al., in prep) we study the impact of the hfs on the Mn abundance in Milky Way Cepheids. The three Mn lines measured in our Magellanic Cepheids belong to the (6013,6016,6021\AA) triplet and as expected we find lower Mn abundances when the hfs is taken into account. For the 6013 \AA{} lines, we find a mean effect of $-0.18\pm0.21$~dex (max: $-0.65$~dex), while it is slightly lower for the 6016 \AA{} and 6021 \AA{} lines, with a mean effect of $-0.12\pm0.22$~dex (max: $-0.45$~dex) and $-0.12\pm0.16$~dex (max: $-0.40$~dex) respectively. It should be noted that the Milky Way Cepheids that are on average more metal-rich and somewhat cooler than the Magellanic Cepheids in our sample.   
                             
\section{Discussion}         
                             
\subsection{The chemical composition of Cepheids in NGC~1866}                                                                                                   
                             
\par
\begin{figure*}[!htbp]              
\centering                   
   \includegraphics[angle=-90,width=\textwidth]{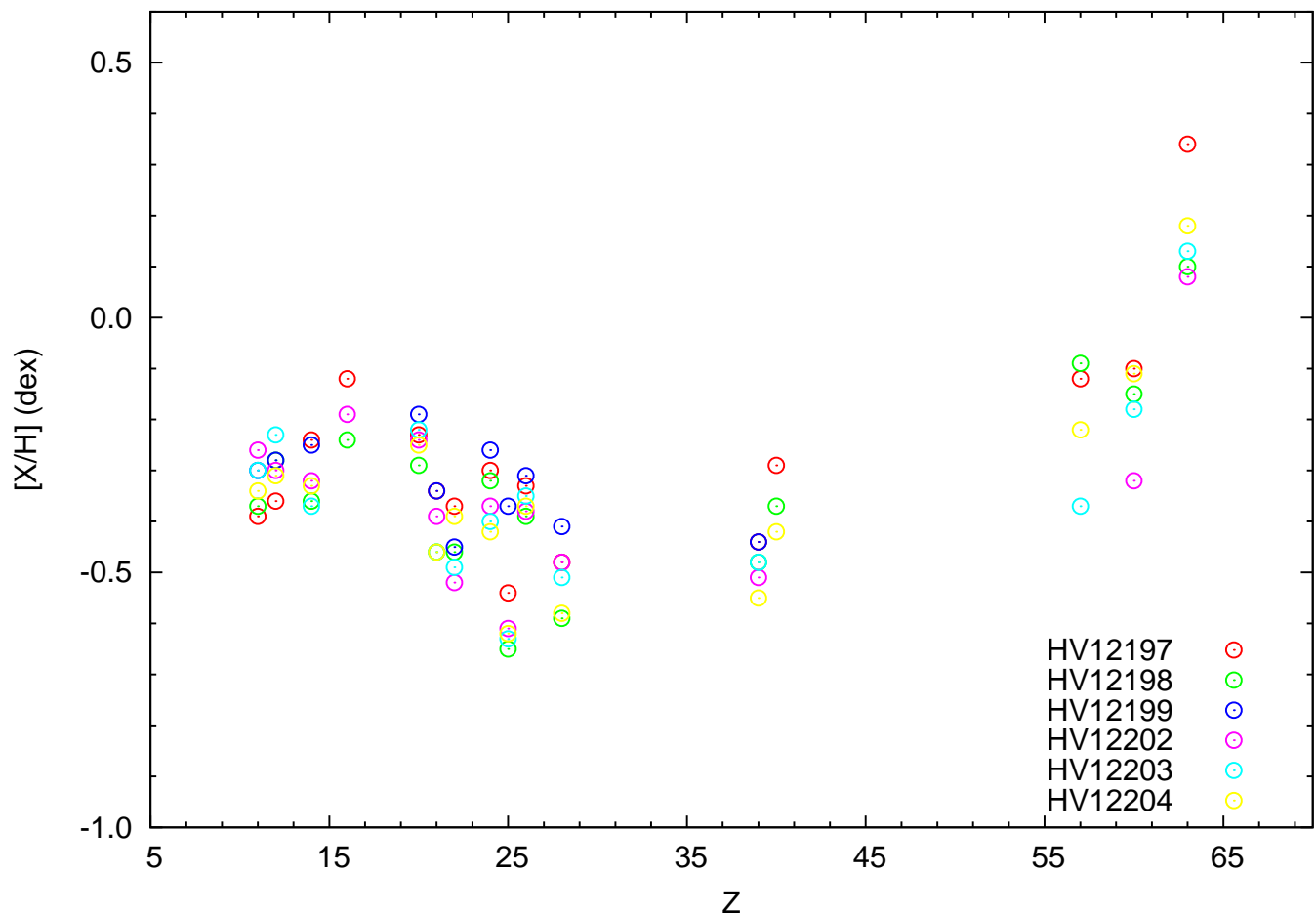}
   \caption{Abundance ratios ([X/H]) for our NGC~1866 Cepheids for different elements identified by their atomic number Z.}
\label{AbXH}
\end{figure*}

 The most striking feature of the abundance pattern of the NGC~1866 Cepheids is the very low star-to-star scatter (see Fig.~\ref{AbXH}): all the elements for which a good number of lines could be measured (e.g., Si, Ca, Fe) have abundances [X/H] that fall within $\approx$ 0.1 dex from each other. The same also applies for other elements (e.g., S, Sc, Ti, Ni, Y, Zr) where only a small number of lines could be measured, and even in the case of, e.g., Na or Mg, where only one line could be measured, the scatter remains smaller than 0.2 dex. In a few cases (mostly for neutron-capture elements), a star has a discrepant abundance for a given element, either because this element could be measured (probably poorly) in only one of the spectra (e.g., Mn in HV~12199, La in HV~12203) or because one of the spectra gives a discrepant value (e.g., Nd for HV~12202 or Eu for HV~12197). Ignoring the outliers, the star to star scatter is similar to the one observed for the other elements.
\par This small star-to-star scatter is a strong indication that the six Cepheids in our NGC~1866 sample are bone fide cluster members, sharing a very similar chemical composition as expected if they were born in the same place and at the same time. Indeed, they all have 2.64d~<~P~<~3.52d and it is well-known that classical Cepheids obey a period-age relation \citep[e.g.,][see also Sect.~\ref{age}]{Efre1978,Greb1998,Bono2005}.\\

\par With [Fe/H]~$\approx$~$-0.4$~dex, our NGC~1866 Cepheids can be compared to Cepheids located in the outer disc of the Milky Way, at Galactocentric distances R$_{G}$~>~10~kpc. A quick glance at the Cepheid abundances in, e.g., \cite{Lem2013}, and \citet{Gen2015} indicates that the [Na/Fe] and [$\alpha$/Fe] abundances in the NGC~1866 Cepheids fall slightly below those observed in the Milky Way Cepheids for the corresponding range of metallicities. The same comparison for neutron-capture elements \citep[in][]{daSilva2016} is less meaningful as the low S/N of our spectra prevented us from taking the hyperfine structure into account in the current study. The [Y/Fe] ratios appear to be similar, which is not surprising as the hfs corrections reported by \cite{daSilva2016} are small for the Y~II lines. The [La/Fe], [Nd/Fe], and [Eu/Fe] ratios appear to be higher than in the Milky Way Cepheids with similar metallicities. This is certainly partially due to the hfs corrections. Indeed \cite{daSilva2016} report that the abundances derived from some of the La~II lines can be smaller by up to $\approx$0.2~dex. On the other hand, their hfs corrections for the Eu lines are very small, and they did not apply any correction for Nd, which indicates that at least a fraction of the difference is intrinsic.\\ 
                             
\par Cepheids embedded in open clusters are extremely important: as the clusters' distances can be determined independently via main~sequence or isochrone fitting, their Cepheids can be used to calibrate the period-luminosity relations \citep[e.g.,][]{Tur2010}. Furthermore, they can be used to establish period-age relations since the ages of star clusters can be determined from their resolved color-magnitude diagrams. The search for Cepheids as members of open clusters or OB associations was conducted in a long term effort\footnote{\url{http://www.ap.smu.ca/~turner/cdlist.html}} by, e.g., \cite{Tur2012,Maj2013}, and references therein as well as by \cite{Ander2013} and \cite{Chen2015} in recent extensive studies. They combined spatial (position, distance) and kinematic data with additional information (age, [Fe/H]) about the stellar populations of the open clusters and found roughly 30 Cepheids associated with open clusters in the Milky Way. However the maximum number of Cepheids that belong to a given cluster is two, much lower than the 23 Cepheids found in NGC~1866 \citep[e.g.,][]{Welch1993}.\\
\indent Comparing the detailed chemical composition of the Cepheids with the one of the other cluster members, as done for the first time in this paper (see Sect.~\ref{compRGB}), speaks in favor of the Cepheid membership of the cluster and should be considered in the future as an important criterion when seeking to match Cepheids to open clusters. This argument holds only if the photospheric abundances in this evolutionary phase were not altered by stellar evolution. In the case of Cepheids, this is expected only for C, N (the first dredge-up alters the surface composition of C and N, and leaves O unaltered) and probably Na (the Ne--Na cycle brings Na-enriched material to the surface). As far as the Milky Way is concerned, the chemical composition of (RGB) stars in open clusters containing Cepheids is often missing, while the direct measurement of stellar abundances in more distant galaxies is out of reach for the current facilities, with the exception of bright, red supergiants \citep[RSGs, e.g., ][]{Davies2015,Pat2015,Gaz2015}. Obtaining detailed abundances from RSGs or cluster integrated light spectroscopy \citep{Colu2012a} for those extragalactic clusters harboring Cepheids would allow us to investigate the longstanding issue of a possible metallicity dependence of the period-luminosity relations that might affect the extragalactic distance scale \citep[e.g.,][]{Roma2008}.    
                             
\subsection{Comparison with giant stars in NGC~1866 and integrated light spectroscopy}
\label{compRGB}              
                             
\par
\begin{figure*}[!htbp]              
\centering                   
   \includegraphics[angle=-90,width=\textwidth]{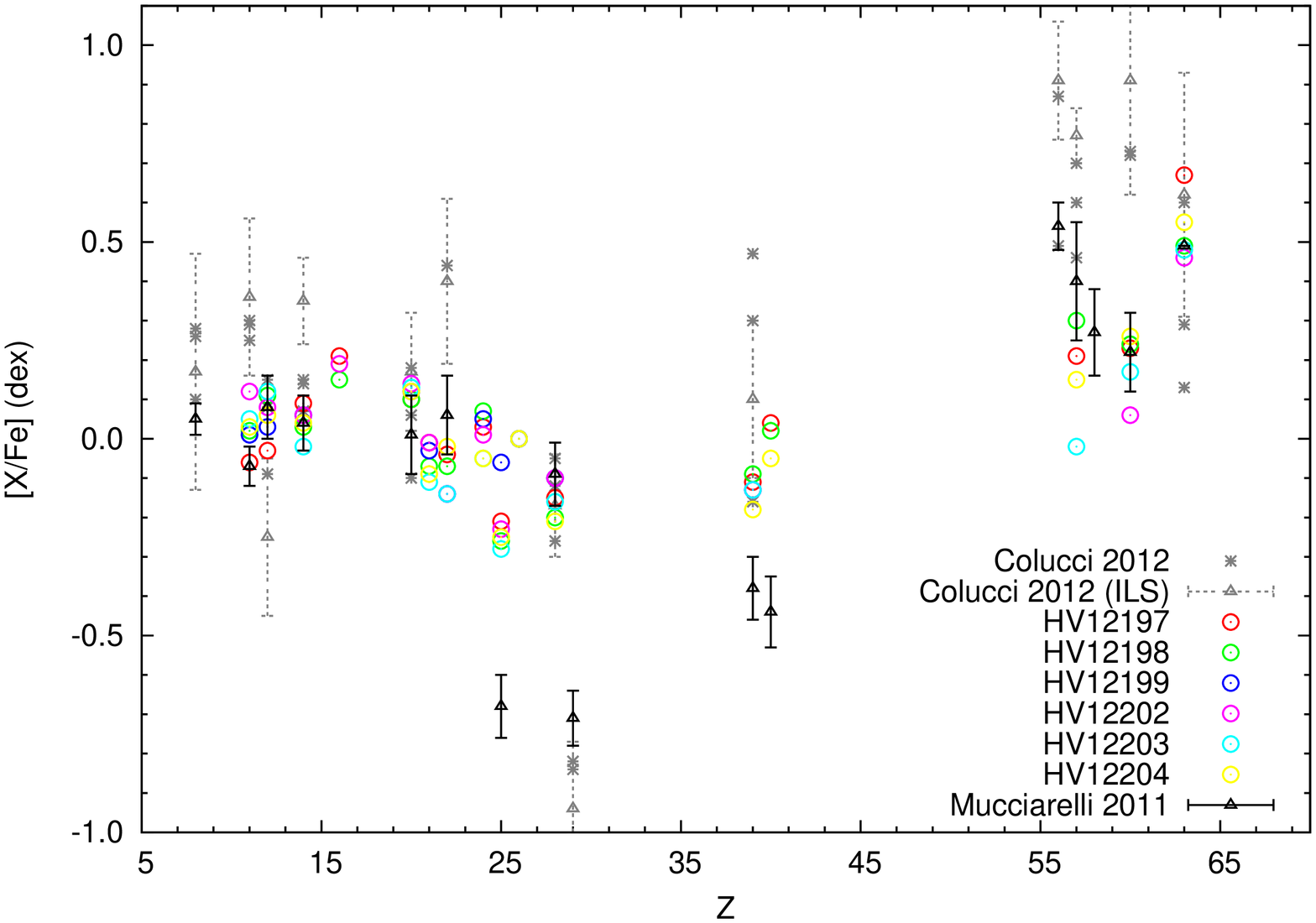}
   \caption{Abundance ratios ([X/Fe]) in NGC~1866 for different elements identified by their atomic number Z. Our Cepheids are the colored open circles, The mean value and dispersion for RGB stars in NGC~1866 from \citet{Mucc2011} are given by the black triangle and solid line. Individual stellar abundances in NGC~1866 by \citet{Colu2012b} are depicted by gray stars. The mean value and dispersion obtained by \citet{Colu2012b} via integrated light spectroscopy are indicated by the gray triangle and dotted line. All the abundance ratios have been rescaled to our Solar reference values.}
\label{Abratios}
\end{figure*}

 Fig.~\ref{Abratios} shows a comparison of the abundance ratios [X/Fe] between the six NGC~1866 Cepheids in our sample and other NGC~1866 stars: the 14 RGB stars of \cite{Mucc2011} for which we show the mean abundance ratios and dispersions and the three stars of \cite{Colu2012a} displayed individually. We also overplot the cluster mean abundance derived from integrated light spectroscopy by \cite{Colu2012a}. All the abundances have been rescaled to our solar reference values.\\
                             
\par Our Cepheids are slightly enriched in sodium with respect to the RGB stars of \cite{Mucc2011}. Similar Na overabundances have already been reported in the Milky Way \citep[e.g.,][]{Gen2015} when comparing Cepheids and field dwarfs in the thin and thick disc \citep{Sou2005}. Although this overabundance is probably partially due to NLTE effects (see Sect.~\ref{Ab}), it has been proposed that it may be caused by mixing events that dredge up material enriched in Na via the NeNa cycle into the surface of the Cepheids \citep{Sass1986,Deni1994,Take2013}. Similar Na overabundances have also been observed in RGB stars \citep[e.g.,][]{daSilva2015}, reinforcing this hypothesis. It is interesting to note that Na overabundances are quite homogeneous in Cepheids and do not depend on mass or period \citep{And2003,Kov2005b,Take2013,Gen2015}. In contrast, \citet{daSilva2015} report a positive trend with mass for [Na/Fe] for RGB stars (which cover a shorter mass range).
\par The agreement is excellent for the $\alpha$-elements Mg and Si, and to a lesser extent for Ca for which the Cepheid abundances are slightly larger than in the RGB stars. The agreement is good for Fe and excellent for Ni, the only two iron peak elements for which data are available for both RGB stars and Cepheids. For our 6 Cepheids we find a mean [Fe/H] = $-0.36$~dex with a dispersion of 0.03 dex. The 14 RGB stars in \cite{Mucc2011} have an average [Fe/H] of $-0.43$~dex (to which one should add 0.02 dex to take into account differences in the adopted Solar iron abundance) and a dispersion of 0.04 dex.
\par In contrast, the abundances of some neutron-capture elements are quite discrepant between the two studies: Y and Zr are found significantly more abundant (by 0.25/0.40 dex respectively) than in \citet{Mucc2011}. Our abundances of La agree only within the error bars whereas Nd and Eu abundances are in excellent agreement with those reported by these authors. The hfs corrections reported by \citet{daSilva2016} are negligible for Y and therefore cannot account for the difference. In contrast hfs corrections can reach $-0.2$~dex for several La lines, and a good agreement between both studies could be achieved if they were taken into account. A possible explanation for these discrepancies could be that the transitions used to derive the abundances of these elements are associated with different ionization stages. For instance, \citet[][their Fig.~13]{Allen2006} derived lower Zr abundances from Zr I lines than from Zr II lines in the Barium stars they analyzed, possibly because ionized lines are the dominant species and therefore less affected by departures from the LTE. In the end, their Zr II abundances span a range of 0.40 $\leq$ [Zr II/Fe] $\leq$ 1.60 while their Zr I abundances are found in the $-0.20$ $\leq$ [Zr I/Fe] $\leq$ 1.45 range. \citet{Mucc2011} do not provide their linelist but given the wavelength range of their spectra, it is likely that they used neutral lines. Unfortunately, neutral lines for these elements are too weak and/or blended in the spectra of Cepheids and therefore cannot be measured to test this hypothesis. Only the Zr I lines at 6134.58 and 6143.25 \AA, and the Y I line at 6435.05 \AA{} could possibly be measured in the most metal-rich Milky Way Cepheids, but they become too weak already at Solar metallicity.
\par The abundance ratios derived by \citet{Mucc2011} for NGC~1866 members are in very good agreement with the field RGB stars in the surroundings they analyzed. It is interesting to note that the [La/Fe] and [Eu/Fe] ratios derived in NGC~1866 by \citet{Mucc2011} are in good agreement with other LMC field RGB stars \citep[e.g.,][and references therein]{vdS2013}, while their [Y/Fe] and [Zr/Fe] ratios fall at the lower end of the LMC field stars distribution. 
\par Y and Zr belong to the first peak of the s-process, while La and Ce belong to the second peak of the s-process that is favored when metal-poor AGB stars dominate the chemical enrichment \citep[e.g.,][]{Crist2011}. The large values of [La/Fe] and [Ce/Fe] demonstrate that the enrichment in heavy elements is dominated by metal-poor AGB stars for both the Cepheids and RGB stars in NGC~1866. Cepheids show higher Y and Zr abundances than RGB stars. If this difference turns out to be real, it might hint that they experienced extra-enrichment in light s-process elements from more metal-rich AGB stars. 

\par Similar conclusions can be drawn when comparing the Cepheid abundances with the stellar abundances derived by \cite{Colu2012b}: the $\alpha$-elements (except Ti) and the iron-peak elements abundance ratios (with respect to iron) they obtained are very similar to those of the Cepheids, while their abundance ratios for the n-capture elements are higher than in the Cepheids, and even higher than those derived by \cite{Mucc2011}. \cite{Colu2012b} did not measure Mn in their NGC~1866 stellar sample. However, they found values ([Mn/Fe]$\approx-0.35$~dex) slightly lower than ours ([Mn/Fe]$\approx-0.25$~dex) in the stars belonging to other young LMC clusters. The Mn abundances reported by \citet{Mucc2011} are also (much) lower than ours. This is almost certainly due to the fact that both studies included hfs corrections for Mn, which are known to be very significant \citep[e.g.,][]{Pro2000}. Because these ratios are lower than in Milky Way stars of the same metallicity, they proposed that the type~Ia supernovae yields of Mn are metallicity-dependent, as reported/modeled in other environments by, e.g., \citet{McWil2003}, \citet{Ces2008}, and \citet{North2012}.\\ 

\par In contrast, the abundance ratios they derived from integrated light spectroscopy are almost always significantly larger than those obtained for RGB stars by \cite{Mucc2011} or for Cepheids (this study), or at least at the higher end. This might be due to the fact that Colucci et al.'s work based on integrated light includes contributions of many different stellar types (and possibly contaminating field populations). This is nevertheless surprising because the integrated flux originating from a young cluster such as NGC~1866 should be dominated by young supergiants, and one would therefore expect a better match between the Cepheids and the integrated light spectroscopy abundance ratios.    

\subsection{Multiple stellar populations in NGC 1866}   
\label{age}

\par In a recent paper, \citet{Mil2017} reported the discovery of a split main sequence (MS) and of an extended main sequence turn-off in NGC~1866. These intriguing features have already been reported in many of the intermediate-age clusters in the Magellanic Clouds as well as for some of their young clusters \citep[e.g.,][]{Bert2003,Glatt2008,Mil2013}, although there is no agreement whether this is indeed due to multiple stellar populations. The blue MS hosts roughly $\sfrac{1}{3}$ of the MS stars, the remaining \sfrac{2}{3} belonging to a spatially more concentrated red MS. \citet{Mil2017} rule out the possibility that age variations solely can be responsible for the split of the MS in NGC~1866. Instead, the red MS is consistent with a $\approx$200~Myr old population of extremely fast-rotating stars ($\omega$=0.9$\omega_{c}$) while the blue MS is consistent with non-rotating stars of similar age, including a small fraction of even older stars. However, according to \citet{Mil2017} the upper blue MS can only be reproduced by a somewhat younger population ($\approx$140~Myr old) accounting for roughly 15\% of the total MS stars.

\par As the age range of Cepheids is similar to the one of the NGC~1866 MS stars, it is natural to examine how they fit in the global picture of NGC~1866 drawn by \citet{Mil2017}. These authors clearly state in their conclusion that the above interpretation should only be considered as a working hypothesis and our only intent here is to examine if Cepheids can shed some light on this scenario. 
\par It is possible to compute individual ages for Cepheids with a period-age relation derived from pulsation models \citep[e.g.,][]{Bono2005}. Because rotation brings fresh material to the core during the MS hydrogen burning phase, fast-rotating stars of intermediate masses stay longer on the MS and therefore cross the instability strip later than a non-rotating star. Including rotation in models then increases the ages of Cepheids by 50 to 100\%, depending on the period, as computed by \citet{Ander2016}. Following the prescriptions of \citet{Ander2016} we derive ages for all the Cepheids known in NGC~1866: we use a period-age relation computed with models with average rotation ($\omega$=0.5$\omega_{c}$) and averaged over the second and third crossing of the instability strip. Periods are taken from \citet{Mus2016}. In the absence of further information, we assume that they are fundamental pulsators, except for V5, V6, and V8, as \citet{Mus2016} report that their periods and light curves are typical of first overtone pulsators. Even more importantly they lie on the PL relations of first overtones. For comparison, we also derive ages using the period-age relation from \citet{Bono2005}, which was computed using non-rotating pulsation models. Ages are listed in Table~\ref{ages}. 

\par We first notice that in both cases the age spread is very limited, thus reinforcing previous findings stating that there is no age variation within NGC1866, or at least that Cepheids all belong to the same sub-population. As expected, the ages calculated with the period-age relation from \citet{Bono2005} lead to younger Cepheids and therefore appear to be compatible only with the 140 Myr old stars populating the upper part of the blue main sequence. None of the period-age relations by \citet{Bono2005} and \citet{Ander2016} enables us to compute individual error bars. Uncertainties on the ages of the NGC~1866 Cepheids of the order of 25--30~Myr can be derived by using the standard deviation of the period-age relation by \citet{Bono2005} as the error. However, given quoted error bars of 50\% or more \citep{Ander2016}, an age of 200 Myr cannot be completely excluded. On the other hand, the ages computed with the period-age relation including rotation from \citet{Ander2016} correspond very well to the fast-rotating red MS population. However the reader should keep in mind that ages should be directly compared only when they are on the same scale, which requires that they were all calculated based on the same models.  

\par Using evolutionary tracks computed with either canonical (no overshooting) or non-canonical (moderate overshooting) assumptions (but no rotation), \citet{Mus2016} favor an age of 140 Myr. The location of the Cepheids, in between the theoretical blue loops computed in each case, does not allow us to discriminate the two overshooting hypotheses. Adopting a canonical overshooting and an older age of 180 Myr enables us to better fit the observed luminosities of the Cepheids, but the theoretical blue loops are then too short to reach the Cepheids' locus in the CMD. Finally, using high-resolution integrated light spectroscopy and CMD-fitting techniques, \citet{Colu2011} report a similar age of 130~Myr.

\par Ages of Cepheids, derived using period-age relations computed with either no rotation or an average rotation ($\omega$=0.5$\omega_{c}$), do not allow us to confirm or rule out the hypothesis of \citet{Mil2017}. Unfortunately, \citet{Ander2016} do not provide period-age relations for fast-rotators ($\omega$=0.9$\omega_{c}$). As far as Cepheid ages are concerned, it is interesting to note that the Cepheids in NGC~1866 match very well the peak of the age distribution for LMC field Cepheids, computed by \citet{Inno2015b} using new period-age relations (without rotation) at LMC metallicities. 

\begin{table}[!htbp]
\caption{Individual ages for Cepheids in NGC~1866 computed with the period-age relations of \citet{Bono2005} or \citet{Ander2016} for fundamental pulsators, and the periods listed in \citet{Mus2016}}
\label{ages}
\centering
\begin{tabular}{rccc}
\hline
\hline
 Cepheid & Period & Age\tablefootmark{a} & Age\tablefootmark{b}                 \\
         &        & (no rotation)        & (rotation: $\omega$=0.5$\omega_{c}$) \\  
         &   (d)  &(Myr)& (Myr)          \\ 
\hline
V6\tablefootmark{c} & 1.9442620 & 114.5 & 258.7 \\
V8\tablefootmark{c} & 2.0070000 & 111.7 & 252.0 \\
V5\tablefootmark{c} & 2.0390710 & 110.3 & 248.7 \\
           HV~12199 & 2.6391600 & 120.6 & 222.7 \\
           HV~12200 & 2.7249800 & 117.6 & 218.0 \\
               We~4 & 2.8603600 & 113.2 & 211.1 \\
               WS~5 & 2.8978000 & 112.1 & 209.3 \\
                New & 2.9429300 & 110.8 & 207.1 \\
           HV~12203 & 2.9541100 & 110.4 & 206.6 \\
               We~8 & 3.0398490 & 108.0 & 202.7 \\
               We~3 & 3.0490400 & 107.7 & 202.3 \\
              WS~11 & 3.0533000 & 107.6 & 202.1 \\
               We~2 & 3.0548500 & 107.6 & 202.1 \\
               WS~9 & 3.0694500 & 107.2 & 201.4 \\
                 V1 & 3.0845500 & 106.8 & 200.8 \\
           HV~12202 & 3.1012000 & 106.3 & 200.0 \\
           HV~12197 & 3.1437100 & 105.2 & 198.2 \\
               We~5 & 3.1745000 & 104.4 & 197.0 \\
               We~7 & 3.2322700 & 102.9 & 194.6 \\
               We~6 & 3.2899400 & 101.5 & 192.3 \\
                 V4 & 3.3180000 & 100.9 & 191.3 \\
           HV~12204 & 3.4388200 &  98.1 & 186.8 \\
                 V7 & 3.4520700 &  97.8 & 186.3 \\
           HV~12198 & 3.5228000 &  96.3 & 183.8 \\
\hline
\end{tabular}
\tablefoot{
\tablefoottext{a}{Period-age relation from \citet{Bono2005}.}
\tablefoottext{b}{Period-age relation from \citet{Ander2016}.}
\tablefoottext{c}{Ages computed using period-age relations for first overtone pulsators.}}
\end{table}

\subsection{The metallicity gradients from Cepheids in the SMC}   

\par The existence of a metallicity gradient across the SMC is a long-debated issue. Using large numbers of RGB stars, \citet{Carr2008b,Dobb2014b}, and \citet{Pari2016} report a radial metallicity gradient ($-0.075\pm0.011$~dex.deg$^{-1}$ vs. $-0.08\pm0.02$~dex.deg$^{-1}$ in the two latter studies) in the inner few degrees of the SMC. In both cases, this effect is attributed to the increasing fraction of younger, more metal-rich stars towards the SMC center. However, the presence of such a gradient was not confirmed by C- and M- type AGB stars \citep{Cioni2009}, populous clusters \citep[e.g.,][and references therein]{Pari2015,Pari2016}, or RR Lyrae studies \citep[e.g.,][and references therein]{Has2012a,Deb2015,Sko2016}.
\par
\begin{figure}[!htbp]              
\centering                   
   \includegraphics[width=\columnwidth]{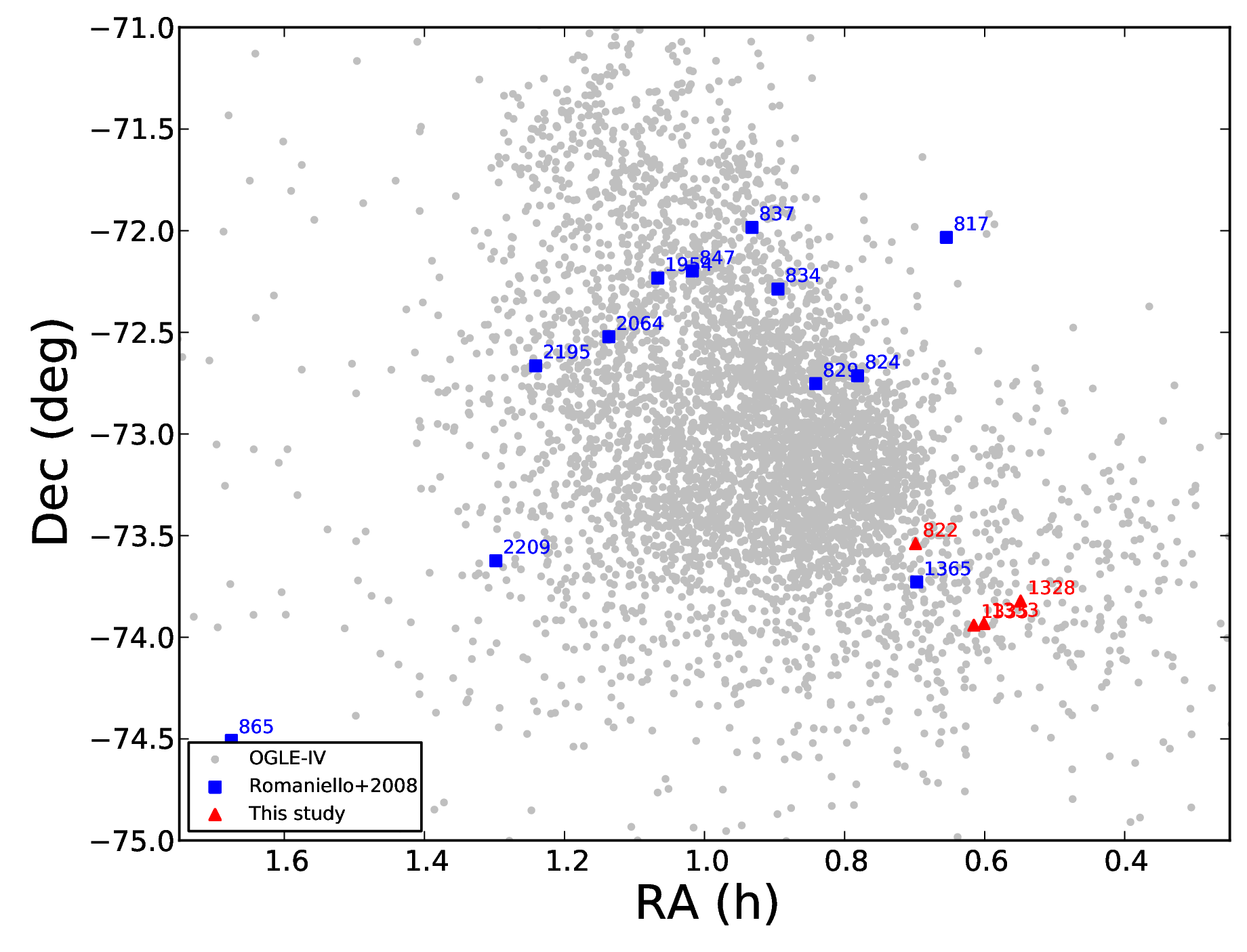} 
   \caption{SMC Cepheids with known metallicities (red: this study; blue: \citet{Roma2008}. SMC Cepheids in the OGLE-IV database are shown as gray dots).}
\label{SMC_sample}
\end{figure}

 The SMC is very elongated and tilted by more than 20$^\circ$ \citep[e.g.,][]{Has2012b,Subra2012,Nide2013}. Moreover, old and young stellar populations have significantly different spatial distributions and orientations \citep[e.g.,][]{Has2012b,Jac2017}. Recent studies using mid-infrared Spitzer data \citep{Scow2016} or optical data from the OGLE IV experiment \citep{Jac2016} clearly confirmed this complex shape. Our Cepheid abundances combined with those found in the literature \citep{Roma2008}, and the possibility to derive accurate distances thanks to the period-luminosity relations allow us to shed new light on the SMC metallicity distribution. For the first time, we are able to probe the metallicity gradient in the SMC' young population in the "depth" direction. As Cepheids are young stars, it should be noted that our study only concerns the present-day abundance gradient, and as such, the metal-rich end of the metallicity distribution function ([Fe/H]$>-0.90$~dex). Moreover, our sample is small (17 stars) and does not contain stars in the inner few degrees of the SMC in an on-sky projection (see Fig. \ref{SMC_sample}). For old populations traced by RR Lyrae stars, no significant metallicity gradient was found in the "depth" direction \citep{Has2012a}.

\par Individual distance moduli for SMC Cepheids were computed using the [3.6] $\mu$m mean-light magnitudes tabulated by \cite{Scow2016} and the corresponding PL-relation in the mid-infrared (MIR) established by the same authors. Combining the extinction law of \citet{Inde2005} with that of \citet{Card1989}, \citet{Mon2012} reported a total-to-selective extinction ratio of A$_{[3.6]}$/E(B-V)=0.203. We adopted the average color excess found by \citet{Scow2016} for the SMC: E(B-V)=0.071$\pm$0.004 mag, which leads to A$_{[3.6]}$=0.014$\pm$0.001 mag. There is no MIR photometry available for HV822 and HV823. For HV822, we use the distance of 67441.4 pc derived by \cite{Groe2013} via the Baade-Wesselink method. The typical uncertainty on the individual MIR distances is of the order of $\pm$3~kpc \citep{Scow2016}. 

\par
\begin{figure}[!htbp]              
\centering                   
   \includegraphics[width=\columnwidth]{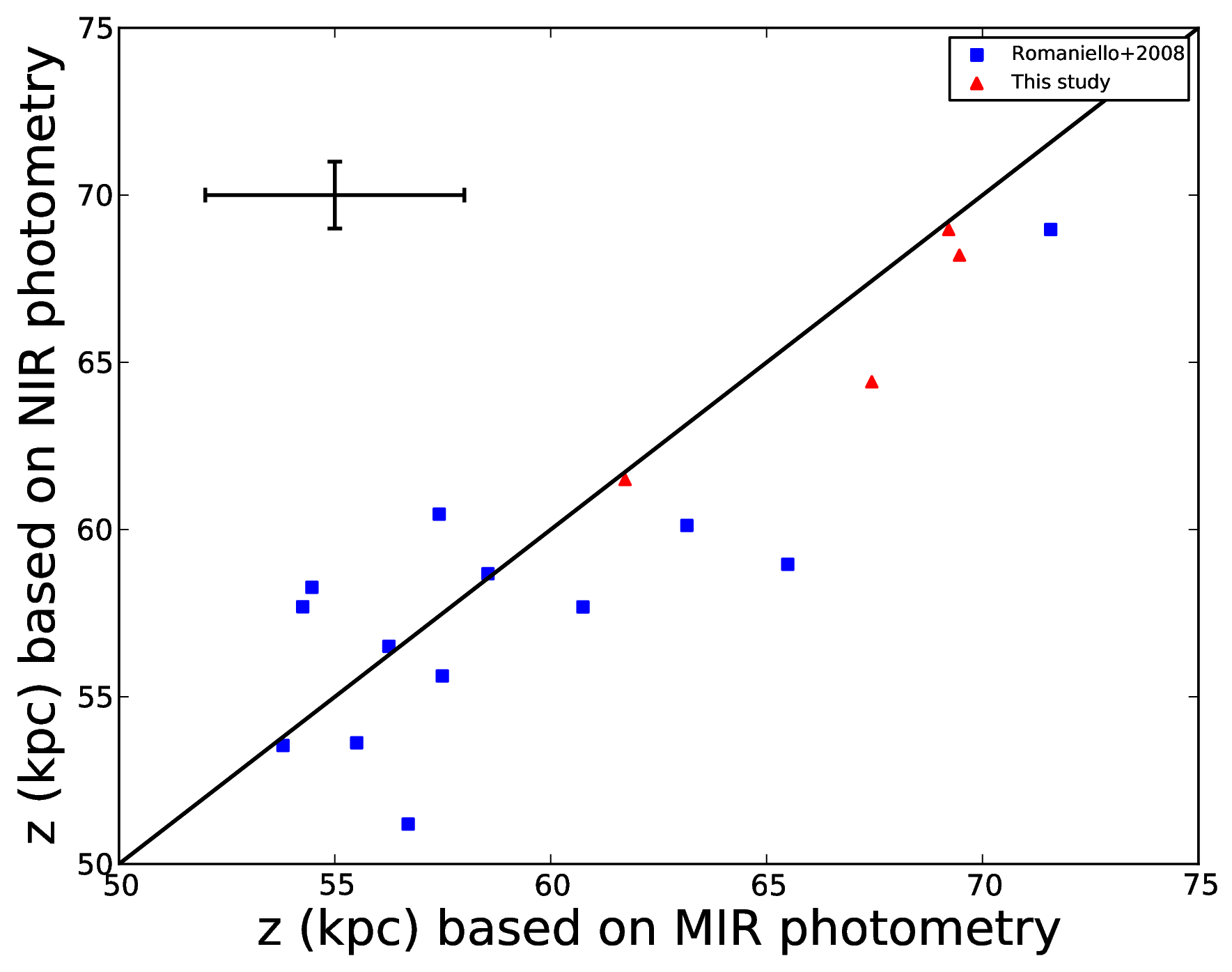} 
   \caption{Comparison of distances derived either from near-infrared or from mid-infrared photometry (red: this study; blue: \citet{Roma2008}. Typical uncertainties are shown in the top left corner.}
\label{comp_dist}
\end{figure}

 For comparison purposes, we also computed distances based on NIR photometry. For the Cepheids in the OGLE-IV database, we used near-infrared J, H and K$_{\rm{S}}$ magnitudes from the IRSF/SIRIUS catalog \citep{Kato2007} that were derived by using the near-infrared light-curve templates of \citet{Inno2015a}. Distances were computed using period-Wesenheit (PW) relations calibrated on the entire SMC sample of fundamental mode Cepheids \citep[$>$2200 stars][in prep]{Inno2017}. Wesenheit indices are reddening-free quantities by construction \citep{Mado1982}. We used the W$_{HJK}$ index as defined by \citet{Inno2016}: W$_{HJK}$~=~$H$--1.046~$\times$~($J-K_{\rm{S}}$) which is minimally affected by the uncertainty in the reddening law \citep{Inno2016}. For stars that are not in the OGLE-IV database, the same procedure was adopted, except that the distances are derived from 2MASS \citep{Skru2006} single epoch data (with no template applied). Individual uncertainties on distances are listed in Table~\ref{SMC_grad}. The typical uncertainty, computed as the average of the individual uncertainties is 993$\pm$41~pc and can be rounded to 1~kpc. It is beyond the scope of this paper to compare both sets of distances. We simply mention here that they are in very good agreement despite some star-to-star scatter (see Fig.~\ref{comp_dist}).

\par To investigate the metallicity gradient in the SMC, we combine our [Fe/H] abundances with those of \cite{Roma2008}, to which we added 0.03 dex to take into account differences in the Solar reference values. The Cepheids were placed in a Cartesian coordinate system using the transformations of \citet{vdM2001} and \citet{Wein2001}. We adopted the value tabulated in SIMBAD for the center of the SMC: $\alpha_{0}$=00h52m38.0s, $\delta_{0}$=-72d48m01.00s (J2000). For the SMC distance modulus, we adopted the value reported by \cite{Gra2014} using eclipsing binaries: 18.965$\pm$0.025 (stat.) $\pm$ 0.048 (syst.) mag which translates into a distance of 62.1$\pm$1.9 kpc. Individual distances and abundances can be found in Table~\ref{SMC_grad}, as well as ages derived with the period-age relation of \citet{Bono2005}.\\

\par
\begin{figure*}[!htbp]              
\centering                   
   \includegraphics[width=\textwidth]{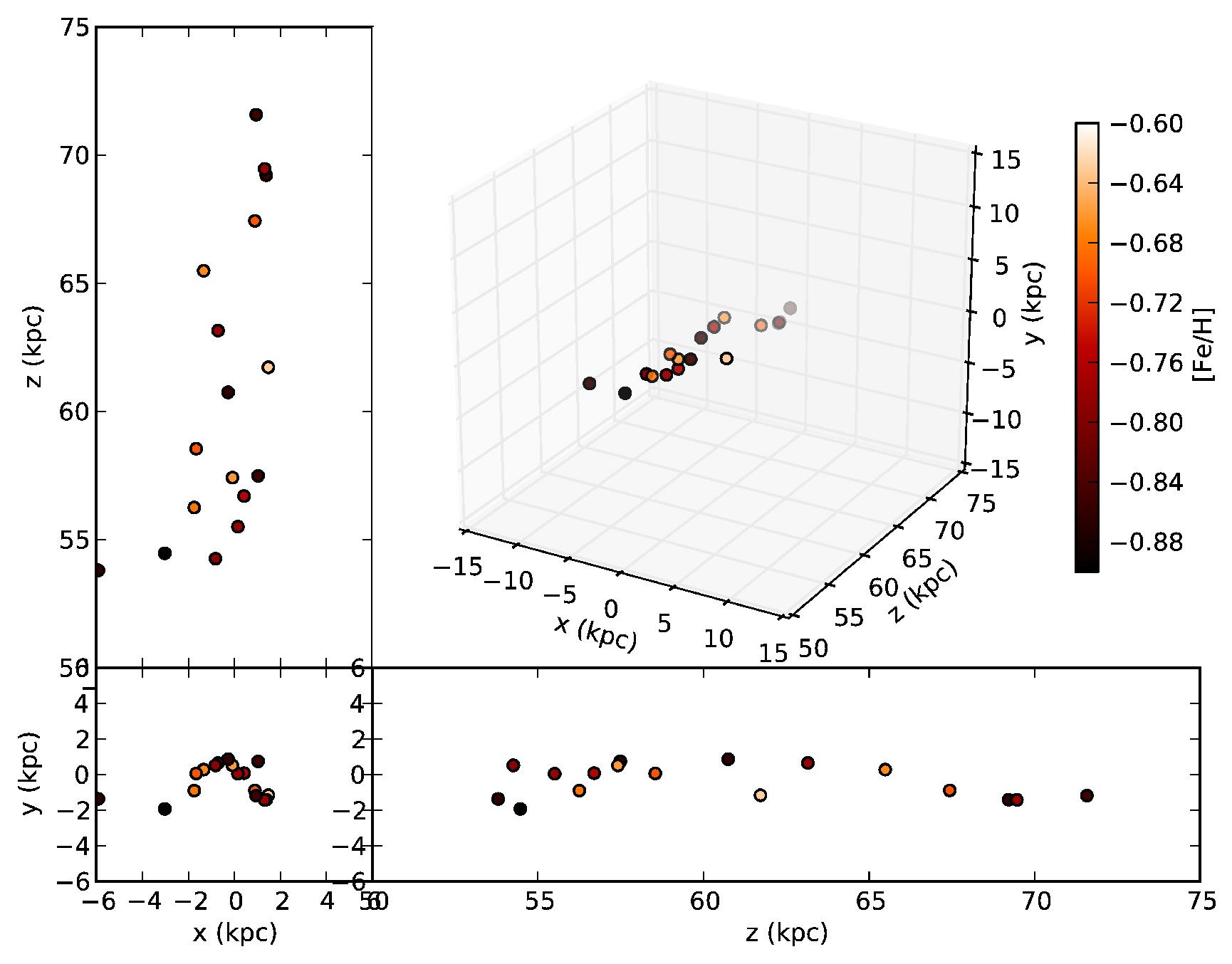} 
   \caption{Metallicity distribution of SMC Cepheids in Cartesian coordinates. Distances are based on mid-infrared photometry.}
\label{met_dist_SMC}
\end{figure*}

\begin{figure}[!htbp]              
\centering
   \includegraphics[width=\columnwidth]{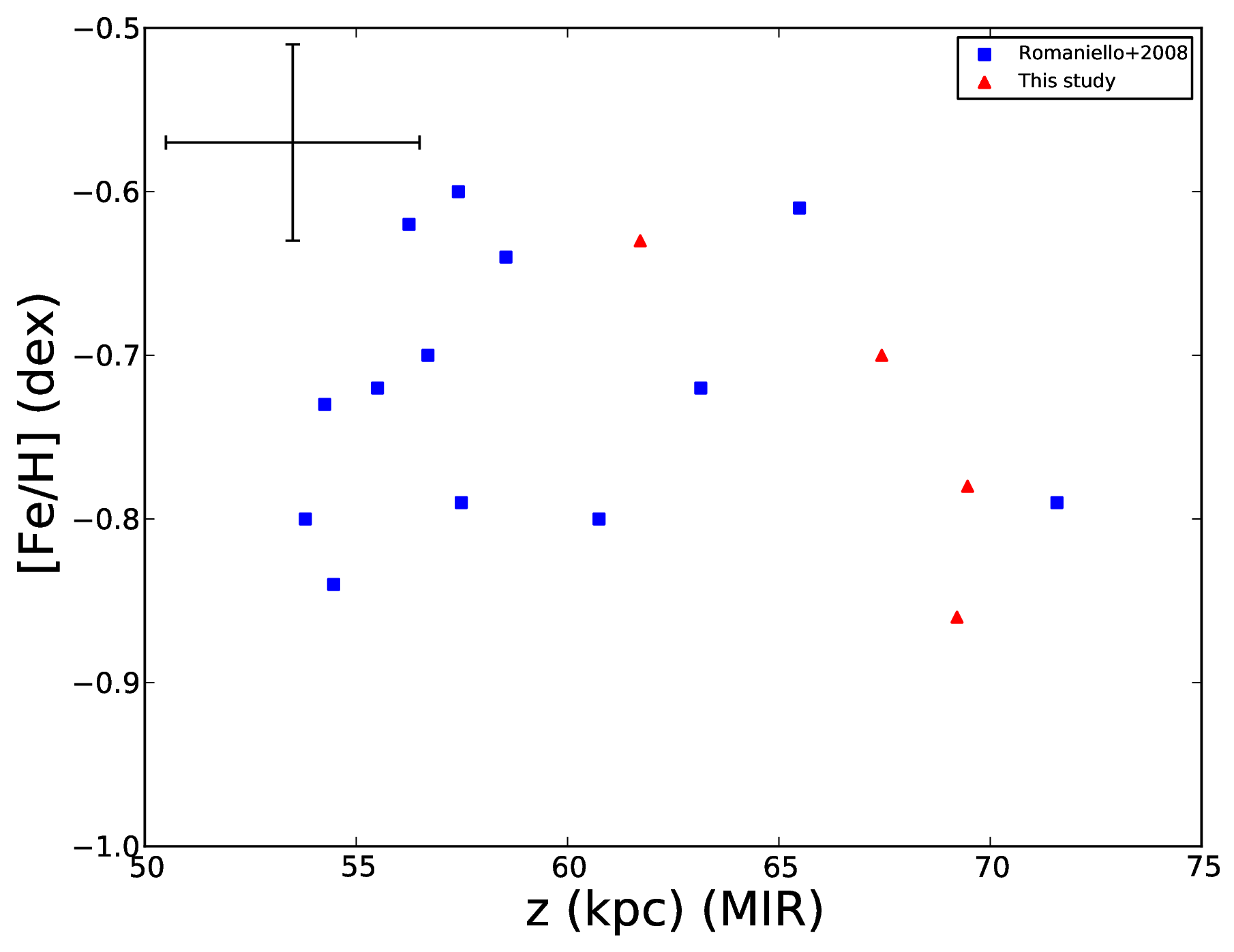} 
   \includegraphics[width=\columnwidth]{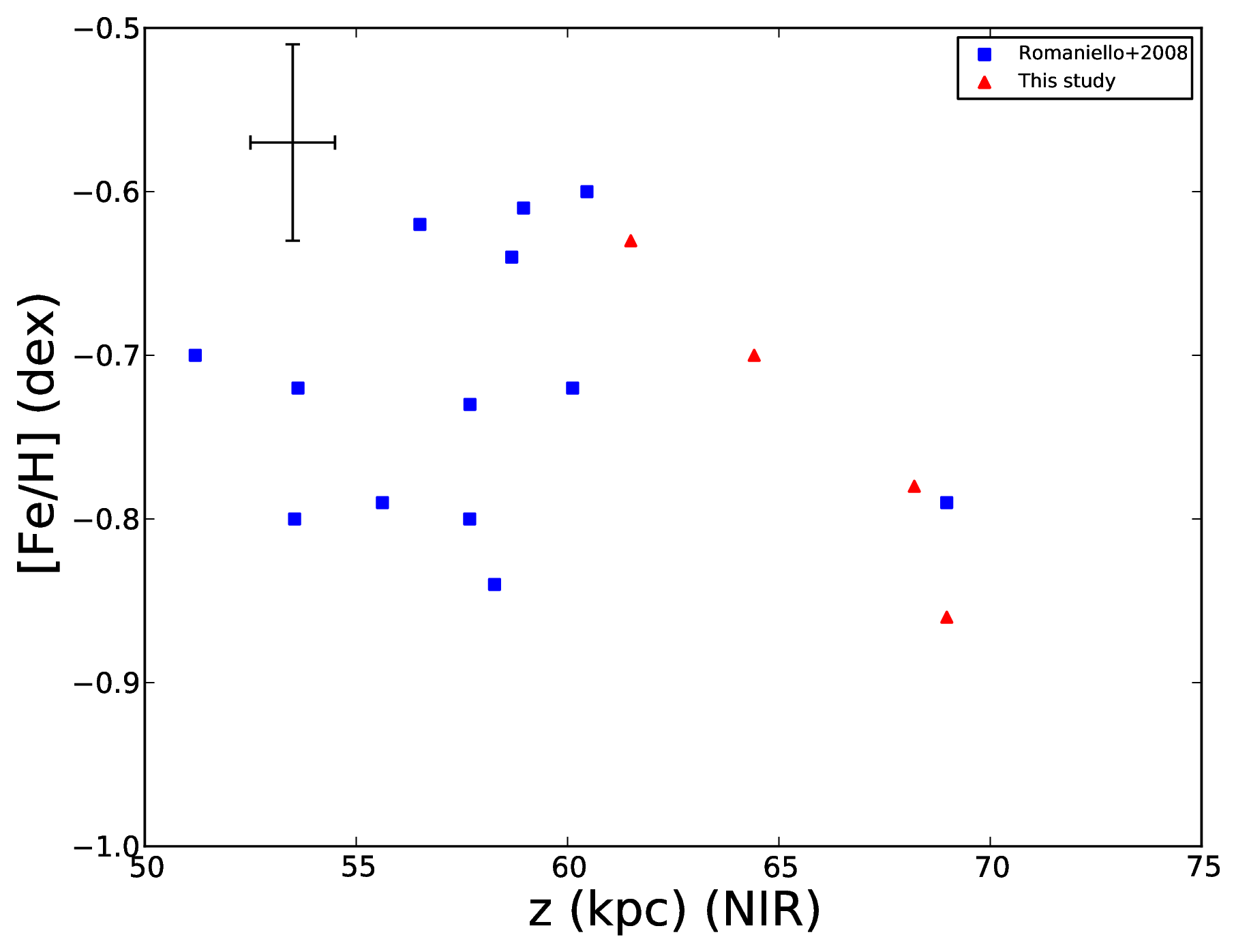}
   \caption{SMC metallicity distribution from Cepheids in the z (depth) direction. {\it Top panel:} distances derived from mid-infrared photometry; {\it Bottom panel:} distances derived from near-infrared photometry. Typical error bars are shown in the top left corner.}  
\label{grad_SMC}
\end{figure}

 A first glance at Fig.~\ref{met_dist_SMC} shows that the (x,y) plane is not very relevant because it does not reflect the depth of the SMC. This fact is reinforced in the case of Cepheids as they are bright stars that can be easily identified and analyzed, even at very large distances. More interesting are the (x,z) and especially the (y,z) plane, as they allow us to study for the first time the metallicity distribution of Cepheids along the SMC main component. Our 17 Cepheids adequately sample the z direction, but the reader should keep in mind that most of our targets are located above the main body of the SMC \citep[see Fig.~\ref{SMC_sample} or][their Fig.~16]{Jac2016}. Fig~\ref{met_dist_SMC} and Fig~\ref{grad_SMC}, where [Fe/H] is plotted as a function of z, show no evidence of a metallicity gradient along the main axis of the SMC. The metallicity spread barely reaches 0.3 dex, but both ends of the z-axis seem to be slightly more metal poor that the inner regions as they miss the more metal-rich Cepheids. The age range spans only 100~Myr and we see no correlation between age and metallicity or distance. These interesting findings should nevertheless be considered only as preliminary results, given the small size of our sample and the location of our Cepheids outside the main body of the SMC.     

\begin{table*}[!htbp]
\caption{Individual distances, ages, and metallicities for SMC Cepheids.
Metallicities from \citet{Roma2008} have been put on the same metallicity scale (by adding 0.03 dex to them) as our data.}
\label{SMC_grad}
\centering
\begin{tabular}{rcccccc}
\hline
\hline
 Cepheid & log P & Age\tablefootmark{a} &   Distance  &   Distance & Uncertainty on     &  [Fe/H] \\
         &  (d)  & (Myr)                &  (MIR) (pc) & (NIR) (pc) & NIR distances (pc) &  (dex)  \\ 
\hline
   HV817 & 1.277 & 212.4 & 57502 & 55636\tablefootmark{c}      &  1136      & -0.79 \\
   HV823 & 1.504 & 186.9 &    -  & 60770\tablefootmark{b}      &   964      & -0.77 \\
   HV824 & 1.819 & 161.2 & 56700 & 51195\tablefootmark{c}      &   957      & -0.70 \\
   HV829 & 1.926 & 154.2 & 55506 & 53625\tablefootmark{c}      &   964      & -0.73 \\
   HV834 & 1.867 & 157.9 & 57420 & 60463\tablefootmark{c}      &  1168      & -0.60 \\
   HV837 & 1.631 & 175.5 & 60752 & 57692\tablefootmark{c}      &  1033      & -0.80 \\
   HV847 & 1.433 & 194.2 & 63160 & 60129\tablefootmark{b}      &  1175      & -0.72 \\
   HV865 & 1.523 & 185.2 & 54586 & 58401\tablefootmark{c}      &  1018      & -0.84 \\
  HV1365 & 1.094 & 239.7 & 71595 & 68986\tablefootmark{b}      &  1094      & -0.79 \\
  HV1954 & 1.222 & 219.8 & 54265 & 57702\tablefootmark{b}      &  1027      & -0.73 \\
  HV2064 & 1.527 & 184.7 & 65503 & 58973\tablefootmark{c}      &  1182      & -0.61 \\
  HV2195 & 1.621 & 176.3 & 58568 & 58703\tablefootmark{c}      &  1163      & -0.64 \\
  HV2209 & 1.355 & 202.8 & 56286 & 56543\tablefootmark{b}      &  1006      & -0.62 \\
 HV11211 & 1.330 & 205.8 & 54140 & 53884\tablefootmark{c}      &   966      & -0.80 \\
\hline                                      
   HV822 & 1.224 & 219.6 & 67447 & 64428\tablefootmark{b}      &  1022      & -0.70 \\
  HV1328 & 1.200 & 223.0 & 61750 & 61526\tablefootmark{b}      &   976      & -0.63 \\
  HV1333 & 1.212 & 221.2 & 69244 & 69002\tablefootmark{b}      &  1094      & -0.86 \\
  HV1335 & 1.158 & 229.3 & 69493 & 68231\tablefootmark{b}      &  1214      & -0.78 \\
\hline
\end{tabular}
\tablefoot{
\tablefoottext{a}{Period-age relation from \citet{Bono2005}.}
\tablefoottext{b}{Distance based on IRSF/SIRIUS near-infrared photometry \citep{Kato2007}.}
\tablefoottext{c}{Distance based on 2MASS near-infrared photometry \citep{Skru2006}.}}
\end{table*}
%

\section{Conclusions}

\par In this paper we conducted a spectroscopic analysis of Cepheids in the LMC and in the SMC. We provide abundances for a good number of $\alpha$, iron-peak, and neutron-capture elements. Our sample increases by 20\% (respectively 25\%) the number of Cepheids with known metallicities and by 46\% (respectively 50\%) the number of Cepheids with detailed chemical composition in these galaxies.

\par For the first time, we study the chemical composition of several Cepheids located in the same populous cluster NGC~1866, in the Large Magellanic Cloud. We find that the six Cepheids we studied have a very homogeneous chemical composition, which is also consistent with RGB stars already analyzed in this cluster. Our results are also in good agreement with theoretical models accounting for luminosity and radial velocity variations for the two stars (HV~12197, HV~12199) for which such measurements are available. Using various versions of period--age relations with no ($\omega$=0) or average rotation ($\omega$=0.5$\omega_{c}$) we find a similar age for all the Cepheids in NGC~1866, indicating that they all belong to the same stellar population.

\par Using near- or mid-infrared photometry and period-luminosity relations \citep{Inno2016,Scow2016}, we computed the distances for Cepheids in the SMC. Combining our abundances for Cepheids in the SMC with those of \citet{Roma2008}, we study for the first time the metallicity distribution of the young population in the SMC in the depth direction. We find no metallicity gradient in the SMC, but our data include only a small number of stars and do not contain Cepheids in the inner few degrees of the SMC.

\begin{acknowledgements}
\par The authors would like to thank the referee, M. Van der Swaelmen, for his careful reading of the manuscript and for his valuable comments that helped to improve the quality of this paper.\\
\par This work was supported by Sonderforschungsbereich SFB 881 "The Milky Way System" (subproject A5) of the German Research Foundation (DFG). GF has been supported by the Futuro in Ricerca 2013 (grant RBFR13J716). This work has made use of the VALD database, operated at Uppsala University, the Institute of Astronomy RAS in Moscow, and the University of Vienna.     
\end{acknowledgements}


\FloatBarrier

\newpage

\onecolumn
\Online
\begin{appendix}

\section{Abundances of the Cepheids in the LMC cluster NGC~1866}
\label{LMC_abund}

\begin{table*}[htp]
\caption{Chemical composition of HV~12197. For a given element, the abundance (computed as the mean value of the abundances determined for each individual line of this element), rms, and number of lines (N) used is given for each individual spectrum analyzed. The final abundance of a star is computed as the weighted mean (and standard deviation) for the three spectra analyzed, where the weight is the number of lines of a given element measured in each spectrum. We remind the reader that hfs was not taken into account and that the hfs correction might be negligible to severe \citep[up to $-0.2$~dex][]{daSilva2016} for La and (up to $-0.65$~dex, Lemasle et al., in prep) for Mn, while it is expected to be negligible for Y, Zr, Nd, Eu.}
\label{abund_12197}
\centering
\begin{tabular}{r|rcr|rcr|rcr|rc}
\hline    
\hline    
HV~12197  & \multicolumn{3}{c|}{MJD=54806.02457622} & \multicolumn{3}{c|}{MJD=54806.08071473} & \multicolumn{3}{c|}{MJD=54806.13684767}  & \multicolumn{2}{c}{Abundance} \\
\hline    
Element  & [X/H] & $\sigma$ &  N & [X/H] & $\sigma$ &  N & [X/H] & $\sigma$ &  N & [X/H] & $\sigma$ \\
         & (dex) &  (dex)   &    & (dex) &  (dex)   &    & (dex) &   (dex)  &    & (dex) &   (dex) \\
\hline    
 [NaI/H] & -0.33 &          &  1 & -0.37 &          &  1 & -0.46 &          &  1 & {\bf -0.39} & {\bf 0.04} \\{}
 [MgI/H] & -0.36 &          &  1 & -0.43 &          &  1 & -0.29 &          &  1 & {\bf -0.36} & {\bf 0.04} \\{}
 [SiI/H] & -0.16 &     0.18 &  3 & -0.21 &     0.19 &  5 & -0.31 &     0.10 &  5 & {\bf -0.24} & {\bf 0.09} \\{}
[SiII/H] &       &          &    &       &          &    &       &          &    &             &            \\{}
  [SI/H] & -0.06 &          &  1 & -0.24 &          &  1 & -0.06 &          &  1 & {\bf -0.12} & {\bf 0.06} \\{}
 [CaI/H] & -0.19 &     0.06 & 10 & -0.26 &     0.06 & 10 & -0.23 &     0.08 & 11 & {\bf -0.23} & {\bf 0.04} \\{}
[ScII/H] & -0.30 &     0.19 &  6 & -0.34 &     0.12 &  6 & -0.37 &     0.16 &  6 & {\bf -0.34} & {\bf 0.09} \\{}
 [TiI/H] & -0.55 &          &  1 & -0.27 &     0.35 &  2 &       &          &    & {\bf -0.36} & {\bf 0.22} \\{}
[TiII/H] & -0.52 &          &  1 & -0.36 &     0.07 &  2 & -0.31 &     0.16 &  2 & {\bf -0.37} & {\bf 0.07} \\{}
 [CrI/H] & -0.24 &          &  1 &       &          &    &       &          &    & {\bf -0.24} &            \\{}
[CrII/H] & -0.16 &     0.11 &  3 & -0.39 &     0.13 &  3 & -0.35 &     0.07 &  4 & {\bf -0.30} & {\bf 0.06} \\{}
 [MnI/H] &       &          &    & -0.55 &     0.04 &  2 & -0.52 &          &  1 & {\bf -0.54} & {\bf 0.03} \\{}
 [FeI/H] & -0.31 &     0.10 & 27 & -0.34 &     0.11 & 50 & -0.34 &     0.13 & 59 & {\bf -0.33} & {\bf 0.07} \\{}
[FeII/H] & -0.34 &     0.16 &  8 & -0.33 &     0.09 &  8 & -0.33 &     0.12 &  9 & {\bf -0.33} & {\bf 0.07} \\{}
 [NiI/H] & -0.38 &          &  1 & -0.49 &     0.17 &  3 & -0.49 &     0.17 &  5 & {\bf -0.48} & {\bf 0.10} \\{}
 [YII/H] & -0.45 &          &  1 & -0.46 &     0.06 &  2 & -0.41 &     0.15 &  2 & {\bf -0.44} & {\bf 0.05} \\{}
[ZrII/H] &       &          &    & -0.31 &          &  1 & -0.27 &          &  1 & {\bf -0.29} & {\bf 0.02} \\{}
[LaII/H] &       &          &    & -0.17 &          &  1 & -0.09 &     0.16 &  2 & {\bf -0.12} & {\bf 0.10} \\{}
[NdII/H] &       &          &    & -0.09 &     0.08 &  3 & -0.12 &     0.06 &  2 & {\bf -0.10} & {\bf 0.05} \\{}
[EuII/H] &  0.48 &          &  1 &  0.29 &          &  1 &  0.26 &          &  1 & {\bf  0.34} & {\bf 0.07} \\  
\hline                                                                                          
\end{tabular}                                                                                   
\end{table*}                                                                                    

\begin{table*}[hbp]                                                                                  
\caption{Same as Table~\ref{abund_12197} for HV~12198.}                                                                   
\label{abund_12198}                                                                                
\centering                                                                                      
\begin{tabular}{r|rcr|rcr|rcr|rc}                                                               
\hline                                                                                          
\hline                                                                                          
HV~12198  & \multicolumn{3}{c|}{MJD=54806.02457622} & \multicolumn{3}{c|}{MJD=54806.08071473} & \multicolumn{3}{c|}{MJD=54806.13684767}  & \multicolumn{2}{c}{Abundance} \\\hline                                                                                          
Element  & [X/H] & $\sigma$ &  N & [X/H] & $\sigma$ &  N & [X/H] & $\sigma$ &  N & [X/H] & $\sigma$ \\
         & (dex) &  (dex)   &    & (dex) &  (dex)   &    & (dex) &   (dex)  &    & (dex) &   (dex) \\
\hline                                                                                          
 [NaI/H] & -0.37 &     0.05 &  2 & -0.33 &          &  1 & -0.40 &     0.13 &  2 & {\bf -0.37} & {\bf 0.05} \\{}
 [MgI/H] & -0.34 &          &  1 & -0.27 &          &  1 & -0.23 &          &  1 & {\bf -0.28} & {\bf 0.04} \\{}
 [SiI/H] & -0.34 &     0.04 &  2 & -0.31 &     0.28 &  5 & -0.39 &     0.08 &  8 & {\bf -0.36} & {\bf 0.09} \\{}
[SiII/H] &       &          &    & -0.21 &          &  1 & -0.27 &          &  1 & {\bf -0.24} & {\bf 0.03} \\{}
  [SI/H] &       &          &    &       &          &    & -0.24 &          &  1 & {\bf -0.24} &            \\{}
 [CaI/H] & -0.35 &     0.14 &  3 & -0.26 &     0.08 &  4 & -0.26 &     0.05 &  3 & {\bf -0.29} & {\bf 0.05} \\{}
[ScII/H] & -0.50 &     0.06 &  2 & -0.44 &     0.21 &  3 & -0.44 &     0.10 &  3 & {\bf -0.46} & {\bf 0.08} \\{}
 [TiI/H] & -0.43 &          &  1 & -0.48 &     0.08 &  3 & -0.46 &     0.15 &  3 & {\bf -0.46} & {\bf 0.07} \\{}
[TiII/H] & -0.31 &          &  1 & -0.34 &          &  1 & -0.47 &          &  1 & {\bf -0.37} & {\bf 0.05} \\{}
 [CrI/H] & -0.48 &          &  1 &       &          &    &       &          &    & {\bf -0.48} &            \\{}
[CrII/H] & -0.23 &     0.17 &  2 & -0.35 &     0.16 &  3 & -0.34 &     0.09 &  3 & {\bf -0.32} & {\bf 0.07} \\{}
 [MnI/H] & -0.64 &     0.05 &  2 & -0.71 &     0.04 &  2 & -0.57 &          &  1 & {\bf -0.65} & {\bf 0.03} \\{}
 [FeI/H] & -0.39 &     0.10 & 33 & -0.37 &     0.10 & 50 & -0.40 &     0.07 & 43 & {\bf -0.39} & {\bf 0.05} \\{}
[FeII/H] & -0.41 &     0.16 &  4 & -0.38 &     0.11 &  5 & -0.40 &     0.09 &  8 & {\bf -0.40} & {\bf 0.06} \\{}
 [NiI/H] & -0.51 &     0.16 &  4 & -0.59 &     0.16 &  6 & -0.62 &     0.10 &  8 & {\bf -0.59} & {\bf 0.07} \\{}
 [YII/H] & -0.38 &     0.13 &  2 & -0.46 &     0.15 &  3 & -0.57 &     0.15 &  3 & {\bf -0.48} & {\bf 0.08} \\{}
[ZrII/H] &       &          &    &       &          &    & -0.37 &          &  1 & {\bf -0.37} &            \\{}
[LaII/H] & -0.05 &     0.01 &  2 & -0.07 &     0.03 &  2 & -0.16 &     0.01 &  2 & {\bf -0.09} & {\bf 0.02} \\{}
[NdII/H] & -0.16 &     0.09 &  2 & -0.15 &     0.05 &  3 & -0.15 &     0.08 &  3 & {\bf -0.15} & {\bf 0.04} \\{}
[EuII/H] &  0.10 &          &  1 &  0.12 &          &  1 &  0.09 &     0.16 &  2 & {\bf  0.10} & {\bf 0.07} \\
\hline
\end{tabular}
\end{table*}

\newpage

\begin{table*}[htp]
\caption{Same as Table~\ref{abund_12197} for HV~12199.}
\label{abund_12199}
\centering
\begin{tabular}{r|rcr|rcr|rcr|rc}
\hline
\hline
HV~12199  & \multicolumn{3}{c|}{MJD=54806.02457622} & \multicolumn{3}{c|}{MJD=54806.08071473} & \multicolumn{3}{c|}{MJD=54806.13684767}  & \multicolumn{2}{c}{Abundance} \\\hline
Element  & [X/H] & $\sigma$ &  N & [X/H] & $\sigma$ &  N & [X/H] & $\sigma$ &  N & [X/H] & $\sigma$ \\
         & (dex) &  (dex)   &    & (dex) &  (dex)   &    & (dex) &   (dex)  &    & (dex) &   (dex) \\
\hline
 [NaI/H] & -0.33 &          &  1 & -0.27 &     0.03 &  2 & -0.31 &     0.01 &  2 & {\bf -0.30} & {\bf 0.02} \\{}
 [MgI/H] & -0.26 &          &  1 & -0.31 &          &  1 & -0.27 &          &  1 & {\bf -0.28} & {\bf 0.01} \\{}
 [SiI/H] & -0.23 &     0.11 &  2 & -0.31 &     0.05 &  3 & -0.20 &     0.09 &  3 & {\bf -0.25} & {\bf 0.04} \\{}
[SiII/H] &       &          &    &       &          &    &       &          &    &             &            \\{}
  [SI/H] &       &          &    &       &          &    &       &          &    &             &            \\{}
 [CaI/H] & -0.17 &     0.12 & 10 & -0.22 &     0.06 & 11 & -0.17 &     0.11 & 11 & {\bf -0.19} & {\bf 0.05} \\{}
[ScII/H] & -0.32 &     0.21 &  3 & -0.35 &     0.10 &  5 & -0.34 &     0.11 &  6 & {\bf -0.34} & {\bf 0.07} \\{}
 [TiI/H] &       &          &    & -0.47 &          &  1 &       &          &    & {\bf -0.47} &            \\{}
[TiII/H] & -0.44 &     0.05 &  3 & -0.48 &     0.03 &  2 & -0.40 &          &  1 & {\bf -0.45} & {\bf 0.03} \\{}
 [CrI/H] & -0.25 &          &  1 &       &          &    & -0.36 &          &  1 & {\bf -0.30} & {\bf 0.05} \\{}
[CrII/H] & -0.29 &     0.26 &  3 & -0.16 &     0.08 &  2 & -0.29 &     0.18 &  4 & {\bf -0.26} & {\bf 0.11} \\{}
 [MnI/H] &       &          &    &       &          &    & -0.37 &     0.18 &  2 & {\bf -0.37} & {\bf 0.18} \\{}
 [FeI/H] & -0.29 &     0.10 & 29 & -0.30 &     0.05 & 35 & -0.32 &     0.10 & 49 & {\bf -0.31} & {\bf 0.05} \\{}
[FeII/H] & -0.30 &     0.09 &  8 & -0.30 &     0.11 &  9 & -0.31 &     0.15 & 11 & {\bf -0.30} & {\bf 0.07} \\{}
 [NiI/H] & -0.33 &     0.15 &  3 & -0.42 &          &  1 & -0.53 &     0.06 &  2 & {\bf -0.41} & {\bf 0.08} \\{}
 [YII/H] & -0.32 &          &  1 & -0.48 &          &  1 & -0.48 &     0.08 &  2 & {\bf -0.44} & {\bf 0.05} \\{}
[ZrII/H] &       &          &    &       &          &    &       &          &    &             &            \\{}
[LaII/H] &       &          &    &       &          &    &       &          &    &             &            \\{}
[NdII/H] &       &          &    &       &          &    &       &          &    &             &            \\{}
[EuII/H] &       &          &    &       &          &    &       &          &    &             &            \\
\hline                                                                                               
\end{tabular}                                                                                        
\end{table*}                                        
                                                                                                     
\begin{table*}[hbp]                                                                                       
\caption{Same as Table~\ref{abund_12197} for HV~12202.}                                                                        
\label{abund_12202}                                                                                     
\centering                                                                                           
\begin{tabular}{r|rcr|rcr|rcr|rc}                                                                    
\hline                                                                                               
\hline                                                                                               
HV~12202  & \multicolumn{3}{c|}{MJD=54806.02457622} & \multicolumn{3}{c|}{MJD=54806.08071473} & \multicolumn{3}{c|}{MJD=54806.13684767}  & \multicolumn{2}{c}{Abundance} \\\hline                                                                                               
Element  & [X/H] & $\sigma$ &  N & [X/H] & $\sigma$ &  N & [X/H] & $\sigma$ &  N & [X/H] & $\sigma$ \\
         & (dex) &  (dex)   &    & (dex) &  (dex)   &    & (dex) &   (dex)  &    & (dex) &   (dex) \\
\hline
 [NaI/H] & -0.31 &          &  1 & -0.26 &          &  1 & -0.23 &     0.06 &  2 & {\bf -0.26}      & {\bf 0.03}         \\{}
 [MgI/H] & -0.28 &          &  1 & -0.32 &          &  1 & -0.30 &          &  1 & {\bf -0.30}      & {\bf 0.01}         \\{}
 [SiI/H] & -0.31 &     0.18 &  3 & -0.31 &     0.14 &  4 & -0.33 &     0.03 &  3 & {\bf -0.32}      & {\bf 0.07}         \\{}
[SiII/H] &       &          &    & -0.17 &          &  1 &       &          &    & {\bf -0.17}      &                    \\{}
  [SI/H] &       &          &    & -0.09 &          &  1 & -0.29 &          &  1 & {\bf -0.19}      & {\bf 0.10}         \\{}
 [CaI/H] & -0.35 &     0.13 &  9 & -0.21 &     0.10 &  9 & -0.17 &     0.08 &  9 & {\bf -0.24}      & {\bf 0.06}         \\{}
[ScII/H] & -0.45 &     0.23 &  3 & -0.39 &     0.20 &  3 & -0.30 &     0.03 &  2 & {\bf -0.39}      & {\bf 0.10}         \\{}
 [TiI/H] & -0.68 &          &  1 & -0.48 &          &  1 & -0.40 &          &  1 & {\bf -0.52}      & {\bf 0.08}         \\{}
[TiII/H] &       &          &    & -0.40 &          &  1 &       &          &    & {\bf -0.40}      &                    \\{}
 [CrI/H] &       &          &    &       &          &    &       &          &    &                  &                    \\{}
[CrII/H] & -0.32 &     0.14 &  4 & -0.40 &    0.19  &  3 & -0.40 &     0.14 &  3 & {\bf -0.37}      & {\bf 0.08}         \\{}
 [MnI/H] & -0.71 &     0.02 &  1 & -0.58 &    0.03  &  3 &       &          &    & {\bf -0.61}      & {\bf 0.04}         \\{}
 [FeI/H] & -0.38 &     0.08 & 29 & -0.41 &    0.07  & 44 & -0.36 &     0.10 & 41 & {\bf -0.38}      & {\bf 0.05}         \\{}
[FeII/H] & -0.39 &     0.05 &  6 & -0.40 &    0.15  &  5 & -0.38 &     0.14 &  6 & {\bf -0.39}      & {\bf 0.06}         \\{}
 [NiI/H] & -0.54 &     0.03 &  3 & -0.53 &    0.08  &  2 & -0.41 &     0.12 &  4 & {\bf -0.48}      & {\bf 0.06}         \\{}
 [YII/H] & -0.40 &     0.01 &    & -0.54 &          &  1 & -0.58 &          &  1 & {\bf -0.51}      & {\bf 0.05}         \\{}
[ZrII/H] &       &          &    &       &          &    &       &          &    &                  &                    \\{}
[LaII/H] &       &          &    &       &          &    &       &          &    &                  &                    \\{}
[NdII/H] & -0.25 &          &  1 &       &          &    & -0.39 &          &  1 & {\bf -0.32}      & {\bf 0.07}         \\{}
[EuII/H] &       &          &    &  0.08 &          &  1 &       &          &    & {\bf  0.08}      &                    \\
\hline
\end{tabular}
\end{table*}

\newpage

\begin{table*}[htp]
\caption{Same as Table~\ref{abund_12197} for HV~12203.}
\label{abund_12203}
\centering
\begin{tabular}{r|rcr|rcr|rcr|rc}
\hline
\hline
HV~12203  & \multicolumn{3}{c|}{MJD=54806.02457622} & \multicolumn{3}{c|}{MJD=54806.08071473} & \multicolumn{3}{c|}{MJD=54806.13684767}  & \multicolumn{2}{c}{Abundance} \\\hline
Element  & [X/H] & $\sigma$ &  N & [X/H] & $\sigma$ &  N & [X/H] & $\sigma$ &  N & [X/H] & $\sigma$ \\
         & (dex) &  (dex)   &    & (dex) &  (dex)   &    & (dex) &   (dex)  &    & (dex) &   (dex) \\
\hline
 [NaI/H] &       &          &    & -0.30 &          &  1 &       &          &    & {\bf -0.30} &            \\{}
 [MgI/H] & -0.18 &          &  1 & -0.18 &          &  1 & -0.34 &          &  1 & {\bf -0.23} & {\bf 0.05} \\{}
 [SiI/H] & -0.28 &     0.12 &  3 & -0.38 &     0.05 &  3 & -0.41 &     0.10 &  5 & {\bf -0.37} & {\bf 0.05} \\{}
[SiII/H] &       &          &    &       &          &    & -0.23 &          &  1 & {\bf -0.23} &            \\{}
  [SI/H] &       &          &    &       &          &    &       &          &    &             &            \\{}
 [CaI/H] & -0.25 &     0.13 &  4 & -0.18 &     0.10 &  9 & -0.25 &     0.12 &  6 & {\bf -0.22} & {\bf 0.06} \\{}
[ScII/H] & -0.33 &     0.22 &  5 & -0.55 &     0.20 &  3 & -0.57 &     0.02 &  3 & {\bf -0.46} & {\bf 0.11} \\{}
 [TiI/H] &       &          &    & -0.46 &          &  1 & -0.53 &          &  1 & {\bf -0.49} & {\bf 0.04} \\{}
[TiII/H] &       &          &    &       &          &    &       &          &    &             &            \\{}
 [CrI/H] &       &          &    &       &          &    &       &          &    &             &            \\{}
[CrII/H] & -0.25 &     0.33 &  2 & -0.46 &     0.13 &  4 & -0.42 &     0.22 &  3 & {\bf -0.40} & {\bf 0.11} \\{}
 [MnI/H] & -0.71 &          &  1 & -0.59 &     0.11 &  2 & -0.64 &     0.06 &  2 & {\bf -0.63} & {\bf 0.05} \\{}
 [FeI/H] & -0.33 &     0.14 & 32 & -0.36 &     0.08 & 37 & -0.36 &     0.12 & 45 & {\bf -0.35} & {\bf 0.07} \\{}
[FeII/H] & -0.36 &     0.06 &  5 & -0.39 &     0.11 &  2 & -0.35 &     0.16 &  8 & {\bf -0.36} & {\bf 0.09} \\{}
 [NiI/H] & -0.51 &     0.20 &  5 & -0.45 &     0.06 &  3 & -0.56 &     0.20 &  4 & {\bf -0.51} & {\bf 0.10} \\{}
 [YII/H] & -0.34 &     0.19 &  2 & -0.62 &     0.11 &  2 & -0.49 &     0.11 &  2 & {\bf -0.48} & {\bf 0.08} \\{}
[ZrII/H] &       &          &    &       &          &    &       &          &    &             &            \\{}
[LaII/H] &       &          &    & -0.37 &     0.01 &  2 &       &          &    & {\bf -0.37} & {\bf 0.01} \\{}
[NdII/H] &       &          &    & -0.24 &     0.16 &  2 & -0.12 &     0.13 &  2 & {\bf -0.18} & {\bf 0.09} \\{}
[EuII/H] &       &          &    &  0.13 &     0.06 &  2 &       &          &    & {\bf  0.13} & {\bf 0.06} \\
\hline
\end{tabular}
\end{table*}

\begin{table*}[hbp]
\caption{Same as Table~\ref{abund_12197} for HV~12204.}
\label{abund_12204}
\centering
\begin{tabular}{r|rcr|rcr|rcr|rc}
\hline
\hline
HV~12204  & \multicolumn{3}{c|}{MJD=54806.02457622} & \multicolumn{3}{c|}{MJD=54806.08071473} & \multicolumn{3}{c|}{MJD=54806.13684767}  & \multicolumn{2}{c}{Abundance} \\\hline
Element  & [X/H] & $\sigma$ &  N & [X/H] & $\sigma$ &  N & [X/H] & $\sigma$ &  N & [X/H] & $\sigma$ \\
         & (dex) &  (dex)   &    & (dex) &  (dex)   &    & (dex) &   (dex)  &    & (dex) &   (dex) \\
\hline
 [NaI/H] & -0.37 &     0.08 &  2 & -0.32 &     0.05 &  2 & -0.34 &          &  1 & {\bf -0.34} & {\bf 0.03} \\{}
 [MgI/H] & -0.34 &          &  1 & -0.26 &          &  1 & -0.32 &          &  1 & {\bf -0.31} & {\bf 0.02} \\{}
 [SiI/H] & -0.34 &     0.08 &  6 & -0.36 &     0.02 &  6 & -0.30 &     0.12 &  6 & {\bf -0.33} & {\bf 0.05} \\{}
[SiII/H] & -0.21 &          &  1 & -0.18 &          &  1 & -0.22 &          &  1 & {\bf -0.20} & {\bf 0.01} \\{}
  [SI/H] &       &          &    &       &          &    &       &          &    &             &            \\{}
 [CaI/H] & -0.28 &     0.05 &  9 & -0.21 &     0.07 & 10 & -0.26 &     0.05 &  6 & {\bf -0.25} & {\bf 0.03} \\{}
[ScII/H] & -0.47 &     0.19 &  5 & -0.47 &     0.15 &  6 & -0.44 &     0.08 &  4 & {\bf -0.46} & {\bf 0.08} \\{}
 [TiI/H] & -0.46 &     0.13 &  2 & -0.42 &     0.11 &  3 & -0.33 &     0.27 &  4 & {\bf -0.39} & {\bf 0.12} \\{}
[TiII/H] & -0.64 &          &  1 & -0.45 &     0.08 &  3 & -0.43 &     0.06 &  3 & {\bf -0.47} & {\bf 0.05} \\{}
 [CrI/H] &       &          &    & -0.56 &          &  1 & -0.53 &          &  1 & {\bf -0.55} & {\bf 0.02} \\{}
[CrII/H] & -0.47 &     0.06 &  2 & -0.38 &     0.13 &  4 & -0.46 &     0.28 &  2 & {\bf -0.42} & {\bf 0.08} \\{}
 [MnI/H] & -0.64 &          &  1 & -0.61 &          &  1 & -0.60 &          &  1 & {\bf -0.62} & {\bf 0.01} \\{}
 [FeI/H] & -0.38 &     0.07 & 44 & -0.34 &     0.10 & 56 & -0.38 &     0.10 & 57 & {\bf -0.37} & {\bf 0.05} \\{}
[FeII/H] & -0.38 &     0.14 &  9 & -0.35 &     0.09 & 10 & -0.39 &     0.10 & 11 & {\bf -0.37} & {\bf 0.06} \\{}
 [NiI/H] & -0.51 &     0.10 &  5 & -0.60 &     0.10 &  7 & -0.60 &     0.14 &  7 & {\bf -0.58} & {\bf 0.07} \\{}
 [YII/H] & -0.54 &     0.02 &  3 & -0.43 &     0.11 &  2 & -0.62 &     0.09 &  4 & {\bf -0.55} & {\bf 0.05} \\{}
[ZrII/H] &       &          &    & -0.42 &          &  1 & -0.43 &          &  1 & {\bf -0.42} & {\bf 0.01} \\{}
[LaII/H] & -0.17 &          &  1 & -0.25 &     0.18 &  3 & -0.21 &     0.15 &  5 & {\bf -0.22} & {\bf 0.10} \\{}
[NdII/H] & -0.06 &     0.26 &  2 & -0.11 &     0.04 &  3 & -0.15 &     0.02 &  3 & {\bf -0.11} & {\bf 0.06} \\{}
[EuII/H] &       &          &    &  0.18 &          &  1 &       &          &    & {\bf  0.18} &            \\
\hline                                                                                           
\end{tabular}                                                                                    
\end{table*}                                            

\FloatBarrier
\newpage

\section{Abundances of the Cepheids in the SMC field}
\label{SMC_abund}

\begin{table*}[htp]
\caption{Same as Table~\ref{abund_12197} for HV~822.}
\label{abund_822}
\centering
\begin{tabular}{r|rcr|rcr|rcr|rc}
\hline
\hline
HV~822  & \multicolumn{3}{c|}{MJD=54785.04026326} & \multicolumn{3}{c|}{MJD=54785.05240828} & \multicolumn{3}{c|}{MJD=54785.06455378}  & \multicolumn{2}{c}{Abundance} \\
\hline
Element  & [X/H] & $\sigma$ &  N & [X/H] & $\sigma$ &  N & [X/H] & $\sigma$ &  N & [X/H] & $\sigma$ \\
         & (dex) &  (dex)   &    & (dex) &  (dex)   &    & (dex) &   (dex)  &    & (dex) &   (dex) \\
\hline
 [NaI/H] &       &          &    & -0.90 &          &  1 & -0.83 &          &  1 & {\bf -0.86} & {\bf 0.04} \\{}
 [MgI/H] &       &          &    & -0.79 &          &  1 & -0.75 &          &  1 & {\bf -0.77} & {\bf 0.02} \\{}
 [SiI/H] & -0.62 &     0.04 &  2 & -0.67 &     0.13 &  3 & -0.75 &     0.03 &  2 & {\bf -0.68} & {\bf 0.05} \\{}
[SiII/H] &       &          &    &       &          &    &       &          &    &             &            \\{}
  [SI/H] &       &          &    & -0.54 &          &  1 &       &          &    & {\bf -0.54} &            \\{}
 [CaI/H] & -0.52 &     0.15 &  9 & -0.58 &     0.14 & 10 & -0.59 &     0.13 &  9 & {\bf -0.56} & {\bf 0.08} \\{}
[ScII/H] & -0.96 &     0.18 &  6 & -0.96 &     0.15 &  6 & -1.00 &     0.18 &  5 & {\bf -0.97} & {\bf 0.09} \\{}
 [TiI/H] &       &          &    &       &          &    &       &          &    &             &            \\{}
[TiII/H] & -0.80 &     0.21 &  4 & -0.78 &     0.24 &  3 & -0.80 &     0.21 &  3 & {\bf -0.79} & {\bf 0.11} \\{}
 [CrI/H] &       &          &    & -0.66 &          &  1 &       &          &    & {\bf -0.66} &            \\{}
[CrII/H] & -0.76 &     0.16 &  4 & -0.77 &     0.13 &  4 & -0.77 &     0.21 &  3 & {\bf -0.77} & {\bf 0.09} \\{}
 [MnI/H] &       &          &    &       &          &    &       &          &    &             &            \\{}
 [FeI/H] & -0.66 &     0.08 & 35 & -0.71 &     0.06 & 31 & -0.63 &     0.13 & 38 & {\bf -0.66} & {\bf 0.06} \\{}
[FeII/H] & -0.72 &     0.12 & 12 & -0.77 &     0.11 & 11 & -0.74 &     0.09 & 11 & {\bf -0.74} & {\bf 0.06} \\{}
 [NiI/H] & -0.82 &          &  1 & -0.87 &          &  1 &       &          &    & {\bf -0.84} & {\bf 0.03} \\{}
 [YII/H] & -1.18 &          &  1 & -1.00 &     0.27 &  2 &       &          &    & {\bf -1.06} & {\bf 0.17} \\{}
[ZrII/H] &       &          &    &       &          &    & -0.73 &          &  1 & {\bf -0.73} &            \\{}
[LaII/H] &       &          &    &       &          &    &       &          &    &             &            \\{}
[NdII/H] &       &          &    &       &          &    &       &          &    &             &            \\{}
[EuII/H] &       &          &    &       &          &    &       &          &    &             &            \\
\hline                                                                                           
\end{tabular}                                                                                    
\end{table*}     
                                                                                                 
\begin{table*}[hbp]
\caption{Same as Table~\ref{abund_12197} for HV~1328.}
\label{abund_1328}
\centering
\begin{tabular}{r|rcr|rcr|rcr|rc}
\hline
\hline
HV~1328  & \multicolumn{3}{c|}{MJD=54785.00779033} & \multicolumn{3}{c|}{MJD=54785.01762254} & \multicolumn{3}{c|}{MJD=54785.02744873}  & \multicolumn{2}{c}{Abundance} \\
\hline
Element  & [X/H] & $\sigma$ &  N & [X/H] & $\sigma$ &  N & [X/H] & $\sigma$ &  N & [X/H] & $\sigma$ \\
         & (dex) &  (dex)   &    & (dex) &  (dex)   &    & (dex) &   (dex)  &    & (dex) &   (dex) \\
\hline
 [NaI/H] & -0.79 &          &  1 & -0.80 &          &  1 & -0.79 &          &  1 & {\bf -0.79} & {\bf 0.01} \\{}
 [MgI/H] & -0.70 &          &  1 & -0.69 &          &  1 & -0.71 &          &  1 & {\bf -0.70} & {\bf 0.01} \\{}
 [SiI/H] & -0.59 &     0.09 &  4 & -0.68 &     0.14 &  2 & -0.60 &     0.17 &  4 & {\bf -0.61} & {\bf 0.07} \\{}
[SiII/H] & -0.32 &          &  1 &       &          &    &       &          &    & {\bf -0.32} &            \\{}
  [SI/H] & -0.55 &          &  1 & -0.55 &          &  1 &       &          &    & {\bf -0.55} &            \\{}
 [CaI/H] & -0.58 &     0.07 & 10 & -0.61 &     0.06 & 10 & -0.64 &     0.05 & 10 & {\bf -0.61} & {\bf 0.04} \\{}
[ScII/H] & -0.66 &     0.17 &  5 & -0.68 &     0.14 &  6 & -0.71 &     0.16 &  4 & {\bf -0.68} & {\bf 0.08} \\{}
 [TiI/H] & -0.76 &          &  1 & -0.79 &          &  1 & -0.83 &          &  1 & {\bf -0.79} & {\bf 0.02} \\{}
[TiII/H] & -0.74 &     0.05 &  4 & -0.67 &     0.08 &  4 & -0.67 &     0.13 &  5 & {\bf -0.69} & {\bf 0.06} \\{}
 [CrI/H] &       &          &    &       &          &    & -0.69 &     0.08 &  2 & {\bf -0.69} & {\bf 0.08} \\{}
[CrII/H] & -0.54 &     0.11 &  2 & -0.62 &     0.15 &  4 & -0.63 &     0.15 &  4 & {\bf -0.61} & {\bf 0.08} \\{}
 [MnI/H] & -0.85 &          &  1 & -0.92 &          &  1 &       &          &    & {\bf -0.89} & {\bf 0.04} \\{}
 [FeI/H] & -0.65 &     0.05 & 43 & -0.66 &     0.08 & 47 & -0.66 &     0.09 & 54 & {\bf -0.66} & {\bf 0.05} \\{}
[FeII/H] & -0.62 &     0.07 & 11 & -0.59 &     0.09 & 10 & -0.60 &     0.10 & 12 & {\bf -0.60} & {\bf 0.05} \\{}
 [NiI/H] & -1.00 &     0.06 &  2 & -1.02 &     0.11 &  2 & -1.01 &     0.05 &  2 & {\bf -1.01} & {\bf 0.03} \\{}
 [YII/H] & -0.89 &     0.16 &  2 & -0.87 &     0.14 &  2 & -0.94 &     0.11 &  2 & {\bf -0.90} & {\bf 0.06} \\{}
[ZrII/H] & -0.66 &          &  1 & -0.66 &          &  1 & -0.70 &          &  1 & {\bf -0.67} & {\bf 0.01} \\{}
[LaII/H] & -0.21 &          &  1 &       &          &    & -0.38 &     0.18 &  2 & {\bf -0.32} & {\bf 0.12} \\{}
[NdII/H] & -0.50 &          &  1 & -0.47 &          &  1 & -0.49 &     0.04 &  2 & {\bf -0.49} & {\bf 0.02} \\{}
[EuII/H] & -0.04 &          &  1 &       &          &    & -0.08 &          &  1 & {\bf -0.06} & {\bf 0.02}  \\
\hline                                                                                           
\end{tabular}                                                                                    
\end{table*}     

\newpage

\begin{table*}[htp]
\caption{Same as Table~\ref{abund_12197} for HV~1333.}
\label{abund_1333}
\centering
\begin{tabular}{r|rcr|rcr|rcr|rc}
\hline
\hline
HV~1333  & \multicolumn{3}{c|}{MJD=54785.07912217} & \multicolumn{3}{c|}{MJD=54785.09358357} & \multicolumn{3}{c|}{MJD=54785.10804394}  & \multicolumn{2}{c}{Abundance} \\
\hline
Element  & [X/H] & $\sigma$ &  N & [X/H] & $\sigma$ &  N & [X/H] & $\sigma$ &  N & [X/H] & $\sigma$ \\
         & (dex) &  (dex)   &    & (dex) &  (dex)   &    & (dex) &   (dex)  &    & (dex) &   (dex) \\
\hline
 [NaI/H] & -0.95 &          &  1 & -0.97 &          &  1 &       &          &    & {\bf -0.96} & {\bf 0.01} \\{}
 [MgI/H] & -0.89 &          &  1 & -0.67 &          &  1 & -0.82 &          &  1 & {\bf -0.79} & {\bf 0.06} \\{}
 [SiI/H] & -0.88 &     0.10 &  4 & -0.68 &     0.13 &  3 &       &          &    & {\bf -0.79} & {\bf 0.08} \\{}
[SiII/H] &       &          &    &       &          &    &       &          &    &             &            \\{}
  [SI/H] &       &          &    &       &          &    &       &          &    &             &            \\{}
 [CaI/H] & -0.81 &     0.09 &  8 & -0.69 &     0.10 &  8 & -0.82 &     0.12 &  8 & {\bf -0.77} & {\bf 0.06} \\{}
[ScII/H] & -0.96 &     0.10 &  2 & -0.92 &     0.10 &  2 & -0.81 &          &  1 & {\bf -0.91} & {\bf 0.05} \\{}
 [TiI/H] & -1.00 &     0.08 &  2 & -0.91 &          &  1 & -0.97 &     0.06 &  2 & {\bf -0.97} & {\bf 0.04} \\{}
[TiII/H] & -0.94 &     0.13 &  3 & -0.83 &     0.20 &  3 & -0.92 &     0.15 &  5 & {\bf -0.90} & {\bf 0.09} \\{}
 [CrI/H] &       &          &    &       &          &    &       &          &    &             &            \\{}
[CrII/H] & -0.89 &     0.13 &  3 & -0.92 &     0.16 &  4 & -0.89 &     0.08 &  4 & {\bf -0.90} & {\bf 0.07} \\{}
 [MnI/H] & -1.19 &     0.02 &  2 & -1.15 &          &  1 & -1.20 &          &  1 & {\bf -1.18} & {\bf 0.02} \\{}
 [FeI/H] & -0.90 &     0.08 & 40 & -0.80 &     0.05 & 34 & -0.88 &     0.08 & 41 & {\bf -0.86} & {\bf 0.04} \\{}
[FeII/H] & -0.88 &     0.10 & 10 & -0.80 &     0.10 &  8 & -0.89 &     0.12 &  7 & {\bf -0.86} & {\bf 0.06} \\{}
 [NiI/H] & -1.09 &     0.15 &  4 & -1.06 &     0.14 &  5 & -1.09 &     0.15 &  4 & {\bf -1.08} & {\bf 0.08} \\{}
 [YII/H] & -1.10 &     0.07 &  5 & -1.01 &     0.08 &  4 & -1.10 &     0.09 &  3 & {\bf -1.07} & {\bf 0.04} \\{}
[ZrII/H] & -0.76 &          &  1 & -0.57 &          &  1 & -0.71 &          &  1 & {\bf -0.68} & {\bf 0.06} \\{}
[LaII/H] & -0.78 &     0.24 &  3 & -0.52 &     0.17 &  3 & -0.54 &     0.08 &  3 & {\bf -0.61} & {\bf 0.10} \\{}
[NdII/H] & -0.50 &     0.13 &  3 & -0.50 &     0.16 &  4 & -0.49 &     0.10 &  3 & {\bf -0.50} & {\bf 0.07} \\{}
[EuII/H] & -0.36 &          &  1 & -0.33 &          &  1 &       &          &    & {\bf -0.34} & {\bf 0.01} \\
\hline                                                                                           
\end{tabular}                                                                                    
\end{table*}     

\begin{table*}[hbp]
\caption{Same as Table~\ref{abund_12197} for HV~1335.}
\label{abund_1335}
\centering
\begin{tabular}{r|rcr|rcr|rcr|rc}
\hline
\hline
HV~1335  & \multicolumn{3}{c|}{MJD=54785.12765152} & \multicolumn{3}{c|}{MJD=54785.14326987} & \multicolumn{3}{c|}{MJD=54785.15889887}  & \multicolumn{2}{c}{Abundance} \\
\hline
Element  & [X/H] & $\sigma$ &  N & [X/H] & $\sigma$ &  N & [X/H] & $\sigma$ &  N & [X/H] & $\sigma$ \\
         & (dex) &  (dex)   &    & (dex) &  (dex)   &    & (dex) &   (dex)  &    & (dex) &   (dex) \\
\hline
 [NaI/H] & -0.99 &          &  1 & -1.02 &          &  1 & -0.87 &          &  1 & {\bf -0.96} & {\bf 0.04} \\{}
 [MgI/H] & -0.90 &          &  1 & -0.78 &          &  1 & -0.81 &          &  1 & {\bf -0.83} & {\bf 0.04} \\{}
 [SiI/H] & -0.92 &          &  1 & -0.73 &          &  2 &       &          &    & {\bf -0.79} & {\bf 0.06} \\{}
[SiII/H] & -0.88 &          &  1 & -1.02 &          &  1 & -0.78 &          &  1 & {\bf -0.89} & {\bf 0.07} \\{}
  [SI/H] & -0.71 &          &  1 &       &          &    &       &          &    & {\bf -0.71} &            \\{}
 [CaI/H] & -0.77 &     0.06 &  9 & -0.71 &     0.10 & 10 & -0.75 &     0.06 &  9 & {\bf -0.74} & {\bf 0.04} \\{}
[ScII/H] & -0.93 &     0.16 &  5 & -0.80 &     0.17 &  7 & -0.92 &     0.20 &  7 & {\bf -0.88} & {\bf 0.10} \\{}
 [TiI/H] & -0.91 &     0.21 &  2 & -0.81 &          &  1 & -1.05 &          &  1 & {\bf -0.92} & {\bf 0.10} \\{}
[TiII/H] & -0.90 &     0.13 &  4 & -0.87 &     0.11 &  5 & -0.90 &     0.09 &  6 & {\bf -0.89} & {\bf 0.06} \\{}
 [CrI/H] & -0.78 &          &  1 & -0.65 &          &  1 &       &          &    & {\bf -0.72} & {\bf 0.07} \\{}
[CrII/H] & -0.85 &     0.22 &  3 & -0.83 &     0.10 &  4 & -0.87 &     0.16 &  3 & {\bf -0.85} & {\bf 0.08} \\{}
 [MnI/H] & -1.06 &     0.03 &  2 & -1.08 &     0.02 &  2 & -1.17 &          &  1 & {\bf -1.09} & {\bf 0.03} \\{}
 [FeI/H] & -0.80 &     0.08 & 48 & -0.72 &     0.06 & 47 & -0.82 &     0.06 & 46 & {\bf -0.78} & {\bf 0.04} \\{}
[FeII/H] & -0.80 &     0.07 & 13 & -0.73 &     0.09 & 14 & -0.80 &     0.09 & 14 & {\bf -0.78} & {\bf 0.05} \\{}
 [NiI/H] & -1.01 &     0.03 &  3 & -0.97 &     0.10 &  4 & -1.01 &     0.09 &  2 & {\bf -0.99} & {\bf 0.05} \\{}
 [YII/H] & -1.00 &     0.10 &  4 & -0.93 &     0.09 &  3 & -1.02 &     0.18 &  4 & {\bf -0.99} & {\bf 0.07} \\{}
[ZrII/H] & -0.64 &          &  1 & -0.58 &          &  1 & -0.72 &          &  1 & {\bf -0.65} & {\bf 0.04} \\{}
[LaII/H] & -0.58 &     0.13 &  4 & -0.57 &     0.11 &  4 & -0.67 &     0.12 &  4 & {\bf -0.61} & {\bf 0.07} \\{}
[NdII/H] & -0.51 &     0.05 &  3 & -0.49 &     0.04 &  3 & -0.49 &     0.06 &  3 & {\bf -0.50} & {\bf 0.03} \\{}
[EuII/H] & -0.23 &     0.08 &  2 & -0.15 &          &  1 &       &          &    & {\bf -0.20} & {\bf 0.05}  \\
\hline                                                                                           
\end{tabular}                                                                                    
\end{table*}                                                                                                     

\FloatBarrier
\section{List of lines measured}
\label{lines}

\begin{table*}[hbp]
\label{lines_meas1}
\centering
\begin{tabular}{cccc|cccc}
\hline
\hline
Wavelength & Element & chi$_{ex}$ & $log~gf$ & Wavelength & Element& chi$_{ex}$ & $log~gf$ \\
  (\AA)    &         &  (     )   &    ()    &  (\AA)     &        &  (     )   &    ()    \\
\hline
4874.010 & Ti2 & 3.09 & -0.809 & 5627.497 & Fe2 & 3.39 & -4.10 \\
4892.859 & Fe1 & 4.22 & -1.292 & 5633.946 & Fe1 & 4.99 & -0.23 \\
4893.820 & Fe2 & 2.83 & -4.273 & 5638.262 & Fe1 & 4.22 & -0.77 \\
4917.230 & Fe1 & 4.19 & -1.089 & 5641.000 & Sc2 & 1.50 & -1.13 \\
4950.106 & Fe1 & 3.42 & -1.672 & 5641.434 & Fe1 & 4.26 & -1.08 \\
4959.115 & Nd2 & 0.06 & -0.806 & 5645.613 & Si1 & 4.93 & -2.04 \\
5005.157 & Ti2 & 1.57 & -2.737 & 5657.896 & Sc2 & 1.51 & -0.60 \\
5017.570 & Ni1 & 3.54 & -0.024 & 5658.816 & Fe1 & 3.40 & -0.81 \\
5044.211 & Fe1 & 2.85 & -2.045 & 5665.555 & Si1 & 4.92 & -1.94 \\
5084.089 & Ni1 & 3.68 & -0.088 & 5667.149 & Sc2 & 1.50 & -1.31 \\
5092.788 & Nd2 & 0.38 & -0.618 & 5669.042 & Sc2 & 1.50 & -1.20 \\
5112.270 & Zr2 & 1.66 & -0.856 & 5679.023 & Fe1 & 4.65 & -0.82 \\
5114.560 & La2 & 0.23 & -1.033 & 5682.633 & Na1 & 2.10 & -0.71 \\
5119.112 &  Y2 & 0.99 & -1.369 & 5684.202 & Sc2 & 1.51 & -1.07 \\
5129.152 & Ti2 & 1.89 & -1.249 & 5686.530 & Fe1 & 4.55 & -0.66 \\
5130.586 & Nd2 & 1.30 &  0.450 & 5688.205 & Na1 & 2.10 & -0.40 \\
5133.688 & Fe1 & 4.18 &  0.148 & 5690.430 & Si1 & 4.93 & -1.77 \\
5153.402 & Na1 & 2.10 & -1.740 & 5693.620 & Fe1 & 4.96 & -2.59 \\
5155.125 & Ni1 & 3.90 & -0.560 & 5701.544 & Fe1 & 2.56 & -2.16 \\
5155.762 & Ni1 & 3.90 &  0.070 & 5705.464 & Fe1 & 4.30 & -1.35 \\
5159.058 & Fe1 & 4.28 & -0.828 & 5708.400 & Si1 & 4.95 & -1.37 \\
5176.559 & Ni1 & 3.90 & -0.300 & 5711.088 & Mg1 & 4.34 & -1.83 \\
5196.059 & Fe1 & 4.26 & -0.496 & 5717.833 & Fe1 & 4.28 & -1.03 \\
5210.385 & Ti1 & 0.05 & -0.835 & 5731.762 & Fe1 & 4.26 & -1.20 \\
5216.274 & Fe1 & 1.61 & -2.081 & 5732.860 & Fe1 & 4.10 & -2.90 \\
5217.389 & Fe1 & 3.21 & -1.121 & 5737.059 &  V1 & 1.06 & -0.74 \\
5305.853 & Cr2 & 3.83 & -2.363 & 5740.858 & Nd2 & 1.16 & -0.53 \\
5307.361 & Fe1 & 1.61 & -2.911 & 5741.860 & Fe1 & 4.26 & -1.67 \\
5329.138 & Cr1 & 2.91 & -0.061 & 5752.032 & Fe1 & 4.55 & -1.18 \\
5329.990 & Fe1 & 4.08 & -1.198 & 5753.120 & Fe1 & 4.26 & -0.69 \\
5334.869 & Cr2 & 4.07 & -1.617 & 5853.668 & Ba2 & 0.60 & -0.91 \\
5336.771 & Ti2 & 1.58 & -1.638 & 5859.586 & Fe1 & 4.55 & -0.42 \\
5345.796 & Cr1 & 1.00 & -0.950 & 5862.357 & Fe1 & 4.55 & -0.13 \\
5349.465 & Ca1 & 2.71 & -0.311 & 5866.451 & Ti1 & 1.07 & -0.78 \\
5353.370 & Fe1 & 4.10 & -0.840 & 5899.300 & Ti1 & 1.05 & -1.17 \\
5369.961 & Fe1 & 4.37 &  0.537 & 5905.671 & Fe1 & 4.65 & -0.69 \\
5373.709 & Fe1 & 4.47 & -0.767 & 5909.973 & Fe1 & 3.21 & -2.59 \\
5379.574 & Fe1 & 3.69 & -1.519 & 5916.247 & Fe1 & 2.45 & -2.91 \\
5381.015 & Ti2 & 1.57 & -1.977 & 5927.789 & Fe1 & 4.65 & -0.99 \\
5383.369 & Fe1 & 4.31 &  0.641 & 5930.179 & Fe1 & 4.65 & -0.23 \\
5398.279 & Fe1 & 4.44 & -0.634 & 5934.655 & Fe1 & 3.93 & -1.07 \\
5402.774 &  Y2 & 1.84 & -0.634 & 5948.541 & Si1 & 5.08 & -1.13 \\
5410.910 & Fe1 & 4.47 &  0.407 & 5956.694 & Fe1 & 0.86 & -4.55 \\
5414.073 & Fe2 & 3.22 & -3.582 & 5976.777 & Fe1 & 3.94 & -1.24 \\
5420.922 & Cr2 & 3.76 & -2.466 & 5983.680 & Fe1 & 4.55 & -1.47 \\
5425.254 & Fe1 & 1.01 & -2.121 & 5984.815 & Fe1 & 4.73 & -0.20 \\
5445.042 & Fe1 & 4.39 & -0.029 & 5987.065 & Fe1 & 4.80 & -0.43 \\
5454.090 & Ti2 & 1.57 & -3.547 & 5991.376 & Fe2 & 3.15 & -3.65 \\
5462.960 & Fe1 & 4.47 & -0.047 & 6003.011 & Fe1 & 3.88 & -1.12 \\
5463.276 & Fe1 & 4.43 &  0.073 & 6007.960 & Fe1 & 4.65 & -0.60 \\
5501.465 & Fe1 & 0.96 & -3.056 & 6008.556 & Fe1 & 3.88 & -0.99 \\
5502.067 & Cr2 & 4.17 & -2.097 & 6013.513 & Mn1 & 3.07 & -0.35 \\
5509.895 &  Y2 & 0.99 & -0.959 & 6016.673 & Mn1 & 3.07 & -0.18 \\
5526.790 & Sc2 & 1.77 &  0.027 & 6020.169 & Fe1 & 4.61 & -0.27 \\
5528.405 & Mg1 & 4.35 & -0.625 & 6021.819 & Mn1 & 3.08 & -0.05 \\
5572.842 & Fe1 & 3.40 & -0.270 & 6024.058 & Fe1 & 4.55 & -0.12 \\
5576.089 & Fe1 & 3.43 & -0.903 & 6027.051 & Fe1 & 4.08 & -1.09 \\
5581.965 & Ca1 & 2.52 & -0.552 & 6055.990 & Fe1 & 4.73 & -0.46 \\
5590.114 & Ca1 & 2.52 & -0.572 & 6065.482 & Fe1 & 2.61 & -1.47 \\
5591.370 & Fe2 & 3.27 & -4.597 & 6078.491 & Fe1 & 4.80 & -0.32 \\
5601.277 & Ca1 & 2.53 & -0.523 & 6079.008 & Fe1 & 4.65 & -1.02 \\
\hline                                      
\end{tabular}                               
\end{table*}                                
                              
\begin{table*}[hbp]
\label{lines_meas2}
\centering
\begin{tabular}{cccc|cccc}
\hline
\hline
Wavelength & Element& chi$_{ex}$ & $log~gf$ & Wavelength & Element& chi$_{ex}$ & $log~gf$ \\
  (\AA)    &        &  (     )   &    ()    &  (\AA)     &        &  (     )   &    ()    \\
\hline                              
6082.710 & Fe1 & 2.22 & -3.57 & 6335.330 & Fe1 & 2.20 & -2.18 \\
6084.111 & Fe2 & 3.20 & -3.88 & 6336.823 & Fe1 & 3.69 & -0.86 \\
6085.270 & Fe1 & 2.76 & -2.86 & 6344.148 & Fe1 & 2.43 & -2.90 \\
6091.919 & Si1 & 5.87 & -1.47 & 6347.109 & Si2 & 8.12 &  0.17 \\
6096.664 & Fe1 & 3.98 & -1.83 & 6355.028 & Fe1 & 2.85 & -2.32 \\
6102.180 & Fe1 & 4.84 & -0.10 & 6358.697 & Fe1 & 0.86 & -4.47 \\
6102.723 & Ca1 & 1.88 & -0.79 & 6362.338 & Zn1 & 5.80 &  0.14 \\
6108.107 & Ni1 & 1.68 & -2.44 & 6369.462 & Fe2 & 2.89 & -4.11 \\
6113.322 & Fe2 & 3.22 & -4.23 & 6380.743 & Fe1 & 4.19 & -1.37 \\
6122.217 & Ca1 & 1.89 & -0.32 & 6390.477 & La2 & 0.32 & -1.41 \\
6127.906 & Fe1 & 4.14 & -1.40 & 6393.600 & Fe1 & 2.43 & -1.50 \\
6141.713 & Ba2 & 0.70 & -0.03 & 6407.251 & Fe2 & 3.89 & -3.85 \\
6149.258 & Fe2 & 3.89 & -2.84 & 6408.018 & Fe1 & 3.69 & -1.02 \\
6151.617 & Fe1 & 2.18 & -3.31 & 6411.648 & Fe1 & 3.65 & -0.66 \\
6155.134 & Si1 & 5.62 & -0.75 & 6414.980 & Si1 & 5.87 & -1.04 \\
6157.728 & Fe1 & 4.08 & -1.16 & 6416.919 & Fe2 & 3.89 & -2.88 \\
6160.747 & Na1 & 2.10 & -1.25 & 6419.980 & Fe1 & 4.73 & -0.17 \\
6161.297 & Ca1 & 2.52 & -1.27 & 6421.350 & Fe1 & 2.28 & -2.02 \\
6162.173 & Ca1 & 1.90 & -0.09 & 6430.845 & Fe1 & 2.18 & -1.98 \\
6165.360 & Fe1 & 4.14 & -1.47 & 6432.680 & Fe2 & 2.89 & -3.57 \\
6166.439 & Ca1 & 2.52 & -1.14 & 6433.457 & Si1 & 5.96 & -2.06 \\
6169.042 & Ca1 & 2.52 & -0.80 & 6437.640 & Eu2 & 1.32 & -0.32 \\
6169.563 & Ca1 & 2.53 & -0.48 & 6439.075 & Ca1 & 2.52 &  0.39 \\
6170.506 & Fe1 & 4.79 & -0.44 & 6449.808 & Ca1 & 2.52 & -0.50 \\
6173.334 & Fe1 & 2.22 & -2.88 & 6455.598 & Ca1 & 2.52 & -1.29 \\
6175.360 & Ni1 & 4.09 & -0.39 & 6462.567 & Ca1 & 2.52 &  0.26 \\
6176.807 & Ni1 & 4.09 & -0.26 & 6471.662 & Ca1 & 2.53 & -0.69 \\
6180.203 & Fe1 & 2.73 & -2.62 & 6481.870 & Fe1 & 2.28 & -2.99 \\
6187.989 & Fe1 & 3.94 & -1.62 & 6482.796 & Ni1 & 1.93 & -2.63 \\
6191.558 & Fe1 & 2.43 & -1.42 & 6491.561 & Ti2 & 2.06 & -1.94 \\
6200.313 & Fe1 & 2.61 & -2.40 & 6493.781 & Ca1 & 2.52 & -0.11 \\
6213.430 & Fe1 & 2.22 & -2.48 & 6494.980 & Fe1 & 2.40 & -1.26 \\
6215.150 & Fe1 & 4.19 & -1.14 & 6496.897 & Ba2 & 0.60 & -0.41 \\
6216.354 &  V1 & 0.28 & -1.29 & 6498.938 & Fe1 & 0.96 & -4.69 \\
6219.281 & Fe1 & 2.20 & -2.43 & 6499.650 & Ca1 & 2.52 & -0.82 \\
6229.226 & Fe1 & 2.85 & -2.80 & 6516.080 & Fe2 & 2.89 & -3.31 \\
6230.722 & Fe1 & 2.56 & -1.28 & 6518.366 & Fe1 & 2.83 & -2.37 \\
6232.640 & Fe1 & 3.65 & -1.22 & 6559.588 & Ti2 & 2.05 & -2.17 \\
6237.319 & Si1 & 5.61 & -0.98 & 6569.214 & Fe1 & 4.73 & -0.38 \\
6238.392 & Fe2 & 3.89 & -2.60 & 6572.779 & Ca1 & 0.00 & -4.24 \\
6239.370 & Fe2 & 2.81 & -4.76 & 6586.308 & Ni1 & 1.95 & -2.75 \\
6239.953 & Fe2 & 3.89 & -3.57 & 6592.913 & Fe1 & 2.73 & -1.47 \\ 
6240.646 & Fe1 & 2.22 & -3.20 & 6593.870 & Fe1 & 2.43 & -2.39 \\
6243.815 & Si1 & 5.62 & -1.24 & 6604.601 & Sc2 & 1.36 & -1.31 \\
6244.466 & Si1 & 5.62 & -1.09 & 6606.949 & Ti2 & 2.06 & -2.80 \\
6245.637 & Sc2 & 1.51 & -1.02 & 6609.110 & Fe1 & 2.56 & -2.68 \\
6246.318 & Fe1 & 3.60 & -0.80 & 6613.733 &  Y2 & 1.75 & -0.85 \\
6247.557 & Fe2 & 3.89 & -2.43 & 6645.064 & Eu2 & 1.38 &  0.12 \\
6252.555 & Fe1 & 2.40 & -1.72 & 6677.990 & Fe1 & 2.69 & -1.37 \\
6254.258 & Fe1 & 2.28 & -2.43 & 6680.133 & Ti2 & 3.09 & -1.79 \\
6258.100 & Ti1 & 1.44 & -0.30 & 6680.140 & Cr1 & 4.16 & -0.39 \\
6261.100 & Ti1 & 1.43 & -0.42 & 6703.566 & Fe1 & 2.76 & -3.06 \\
6262.290 & La2 & 0.40 & -1.22 & 6715.410 & Fe1 & 4.61 & -1.36 \\
6265.132 & Fe1 & 2.18 & -2.54 & 6717.681 & Ca1 & 2.71 & -0.52 \\
6271.278 & Fe1 & 3.33 & -2.70 & 6721.848 & Si1 & 5.86 & -1.52 \\
6279.753 & Sc2 & 1.50 & -1.25 & 6726.666 & Fe1 & 4.61 & -1.13 \\
6290.965 & Fe1 & 4.73 & -0.77 & 6740.080 & Nd2 & 0.06 & -1.53 \\
6297.793 & Fe1 & 2.22 & -2.70 & 6748.837 &  S1 & 7.87 & -0.64 \\
6301.500 & Fe1 & 3.65 & -0.72 & 6750.152 & Fe1 & 2.42 & -2.60 \\
6318.018 & Fe1 & 2.45 & -1.80 & 6757.171 &  S1 & 7.87 & -0.24 \\
6318.717 & Mg1 & 5.11 & -2.10 & 6767.768 & Ni1 & 1.83 & -2.17 \\
6320.410 & La2 & 0.17 & -1.33 & 6772.313 & Ni1 & 3.66 & -0.80 \\
6320.851 & Sc2 & 1.50 & -1.82 & 6774.270 & La2 & 0.13 & -1.71 \\
6322.685 & Fe1 & 2.59 & -2.45 & 6795.414 &  Y2 & 1.74 & -1.03 \\
6327.593 & Ni1 & 1.68 & -3.15 &          &     &      &       \\
\hline                                                                                           
\end{tabular}                                                                                    
\end{table*}                              

\FloatBarrier

\section{Curves of growth}   
\label{CoG_newmarcs}

\par The curve of growth is a graph showing how the equivalent width (EW) of an absorption line varies with the abundance (A) of the atoms/ions producing the line. When there are few atoms, the line profile is dominated by Doppler broadening and EW~$\propto$~A. When the line reaches saturation, EW~$\propto$~$\sqrt{ln~A}$. When saturation increases, the line profile becomes dominated by the wings due to collisional broadening and EW~$\propto$~$\sqrt{A}$.

\par For weak lines it can be shown that $\log \left(\frac{W_{\lambda}}{\lambda} \right)$ varies linearly with $\log(\alpha_{*}gf)$. The $\Gamma_{*}$ quantity, defined as $\log \left(\frac{W_{\lambda}}{\lambda} \right) = \log(\alpha_{*}gf) + \log \Gamma_{*}$ can be computed for each individual line as a function of the atmosphere model, the element considered and its ionization stage, the line excitation potential etc.  As the value of the stellar abundance $\alpha_{*}$ is the unknown one wants to determine, the abundance of the same element in the Sun $\alpha_{\odot}$ is used instead. The experimental curves of growth will then be shifted from the theoretical curves of growth by a factor $log(\alpha_{*}) -log(\alpha_{\odot})$, which is by definition the abundance ratio [X/H] of the considered element X.

\par The experimental curve of growth is traced using many lines from the same element (e.g., Fe I) with a wide range of excitation potentials and oscillator strengths that lead to a wide range of equivalent widths. The curves of growth plotted below show the theoretical curves of growths computed for a line at $\lambda$=5000 \AA{} with $\chi_{ex}$ = 3 and using the atmospheric parameters derived for the star in the corresponding phase.

\newpage
\begin{sidewaysfigure}
  \centering
  \label{CoG197}\caption{Curves of growth (Fe~I, Fe~II) for the HV~12197 spectra. From left to right, MJD=54806.02457622, 54806.08071473, MJD=54806.13684767}
  \subfloat[][]{\label{CoG197:1}\includegraphics[angle=-90,width=0.33\textwidth]{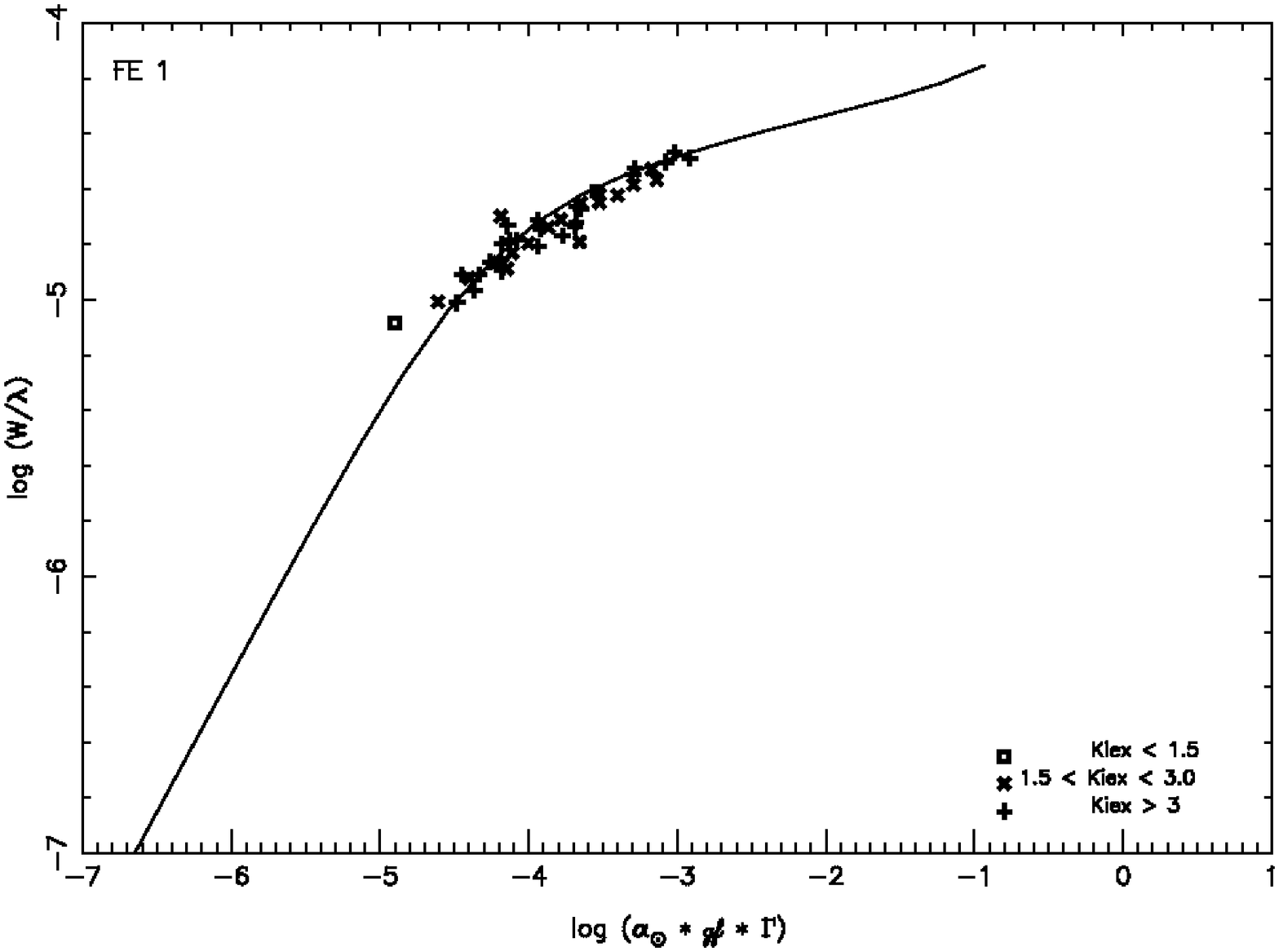}}
  \subfloat[][]{\label{CoG197:2}\includegraphics[angle=-90,width=0.33\textwidth]{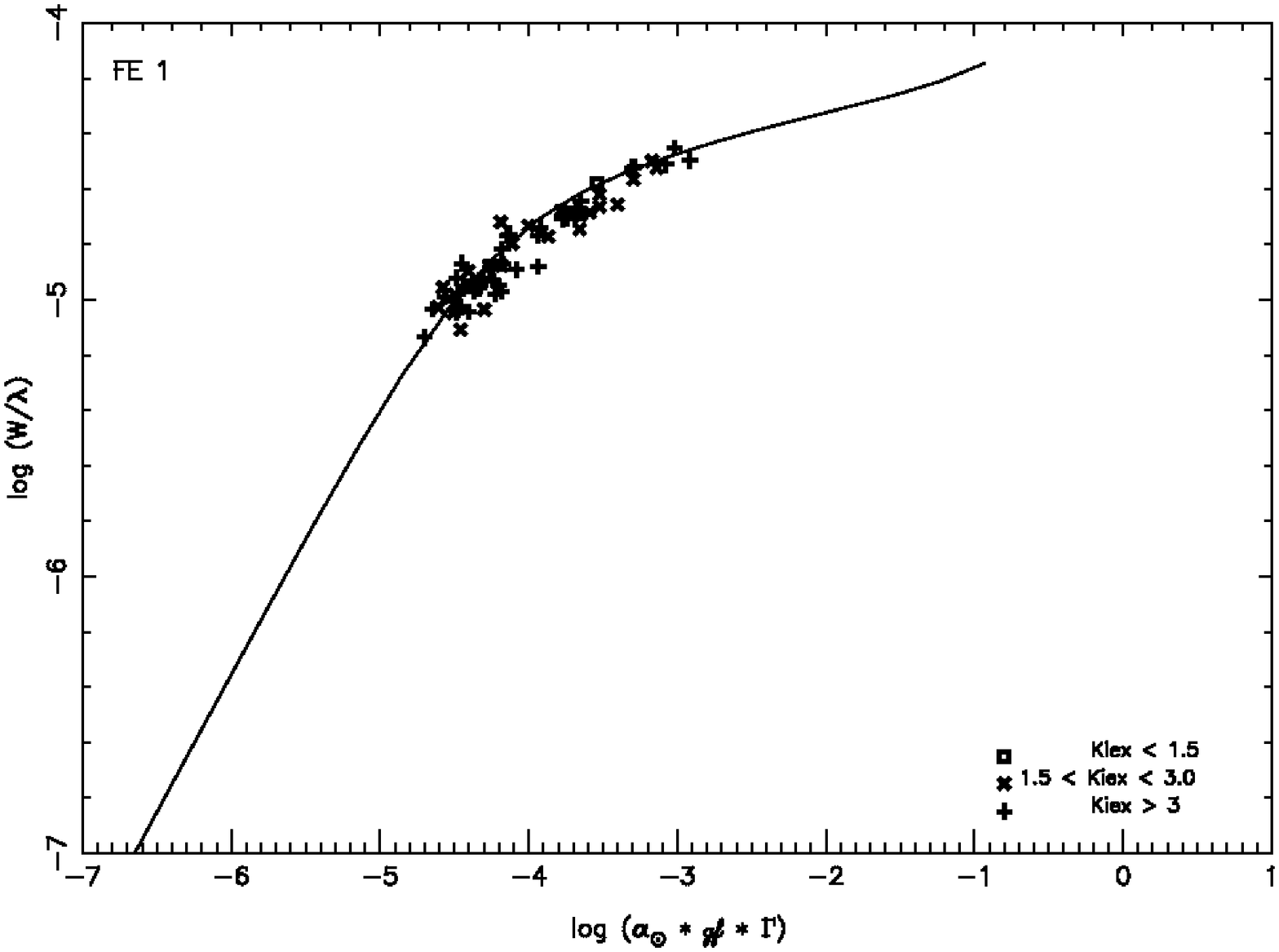}}
  \subfloat[][]{\label{CoG197:3}\includegraphics[angle=-90,width=0.33\textwidth]{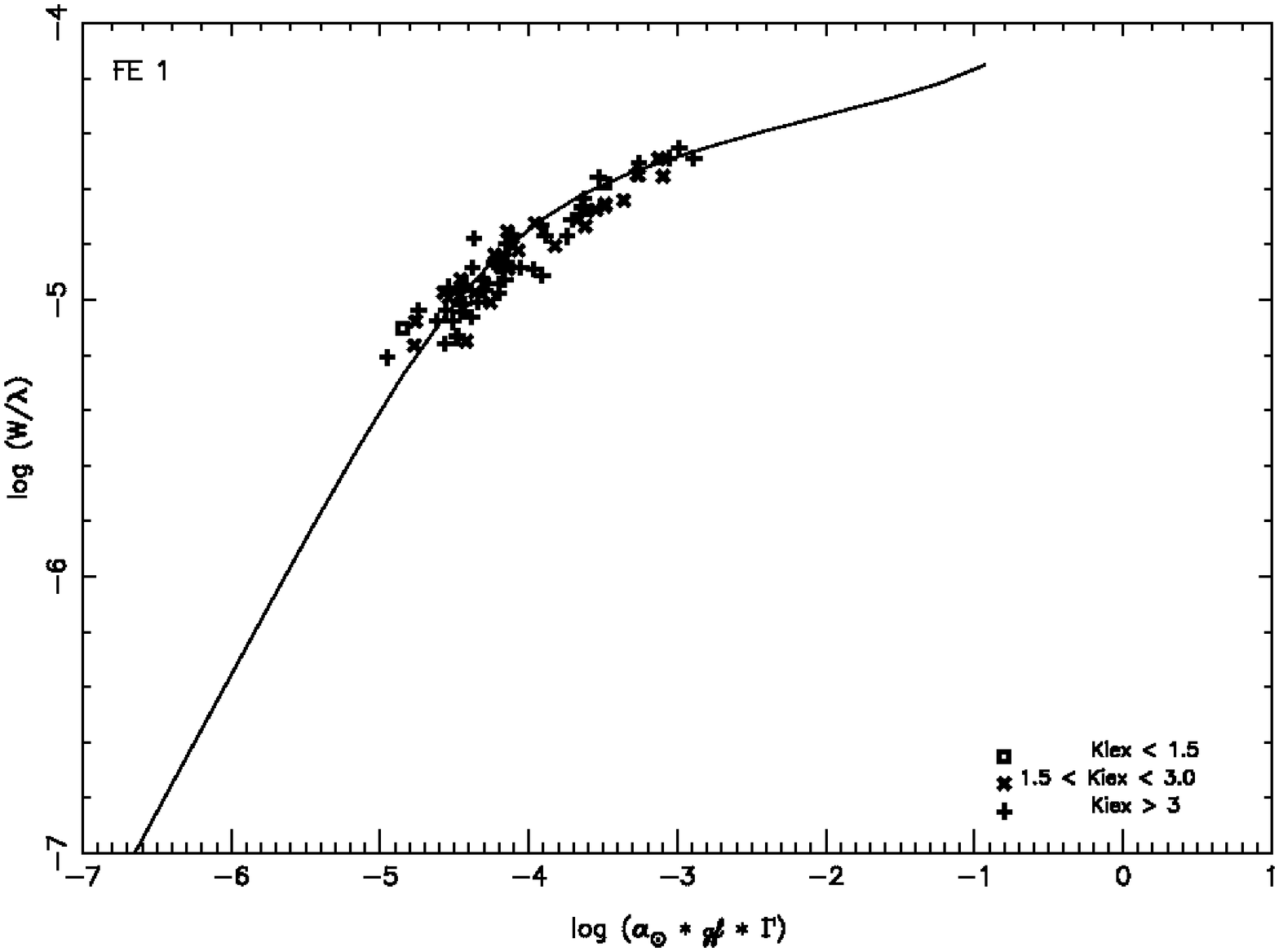}}
  \\
  \subfloat[][]{\label{CoG197:4}\includegraphics[angle=-90,width=0.33\textwidth]{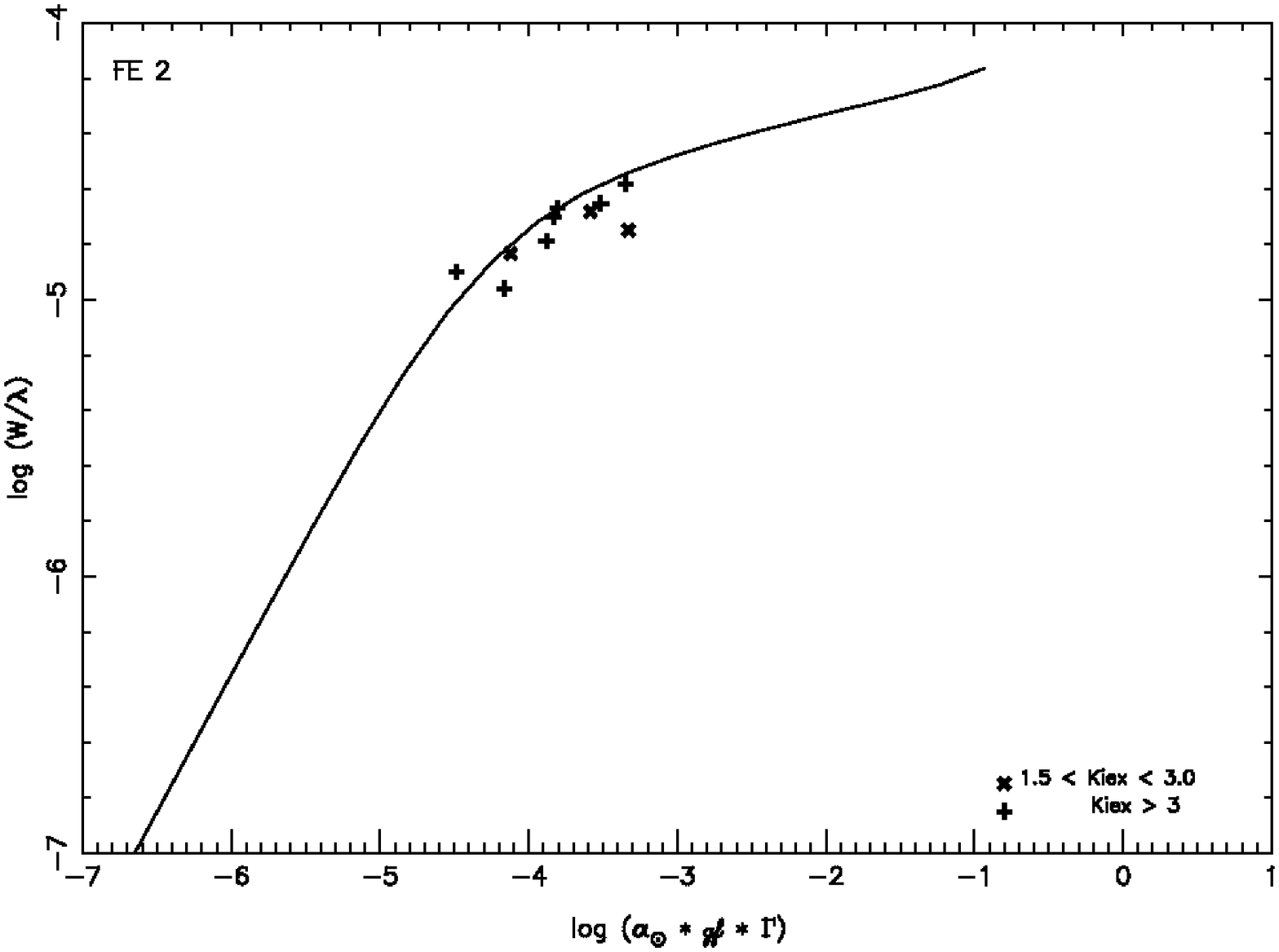}}
  \subfloat[][]{\label{CoG197:5}\includegraphics[angle=-90,width=0.33\textwidth]{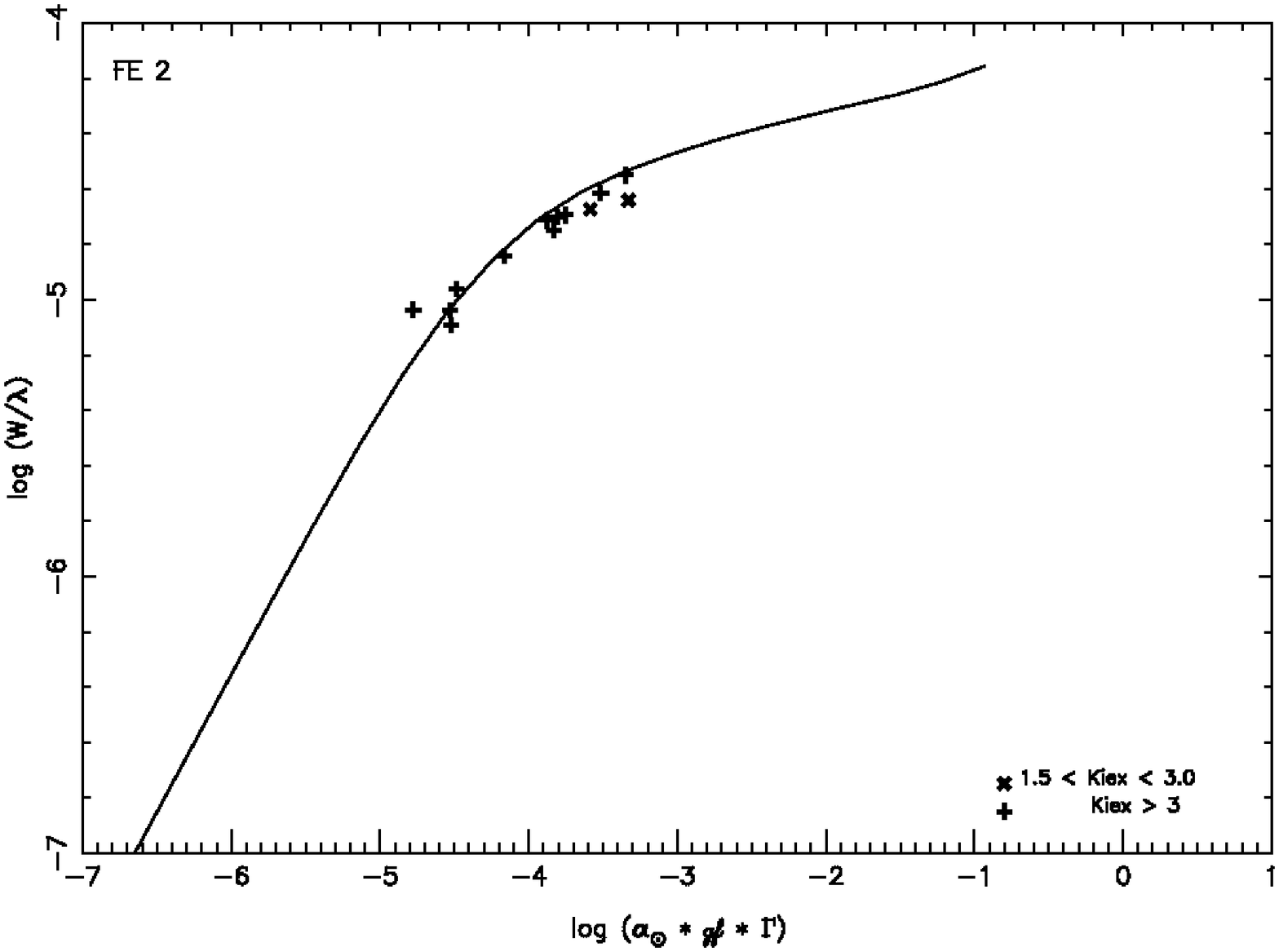}}
  \subfloat[][]{\label{CoG197:6}\includegraphics[angle=-90,width=0.33\textwidth]{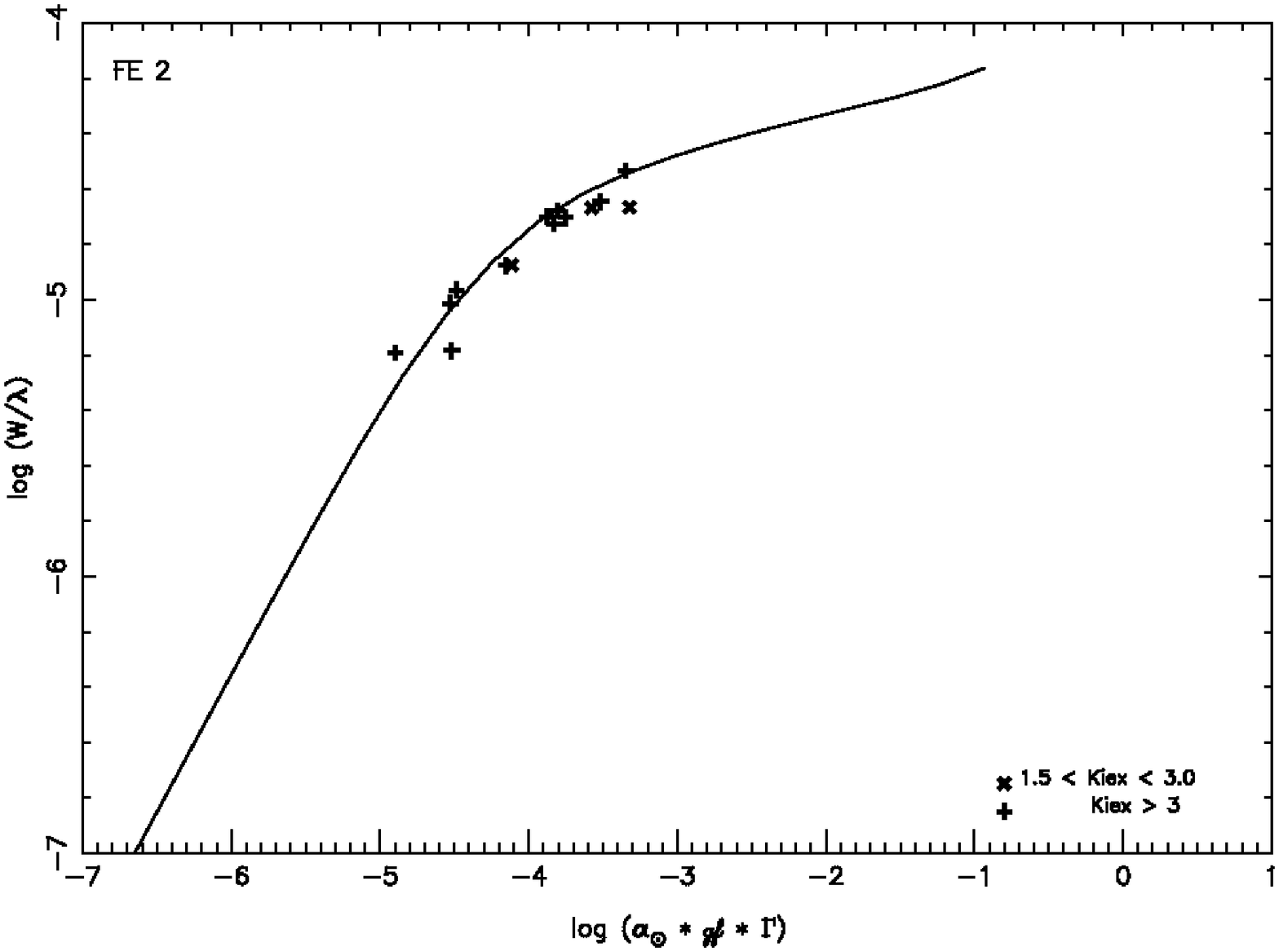}}
\end{sidewaysfigure}


\begin{sidewaysfigure}
  \centering
  \label{CoG198}\caption{Curves of growth (Fe~I, Fe~II) for the HV~12198 spectra. From left to right, MJD=54806.02457622, 54806.08071473, MJD=54806.13684767}
  \subfloat[][]{\label{CoG198:1}\includegraphics[angle=-90,width=0.33\textwidth]{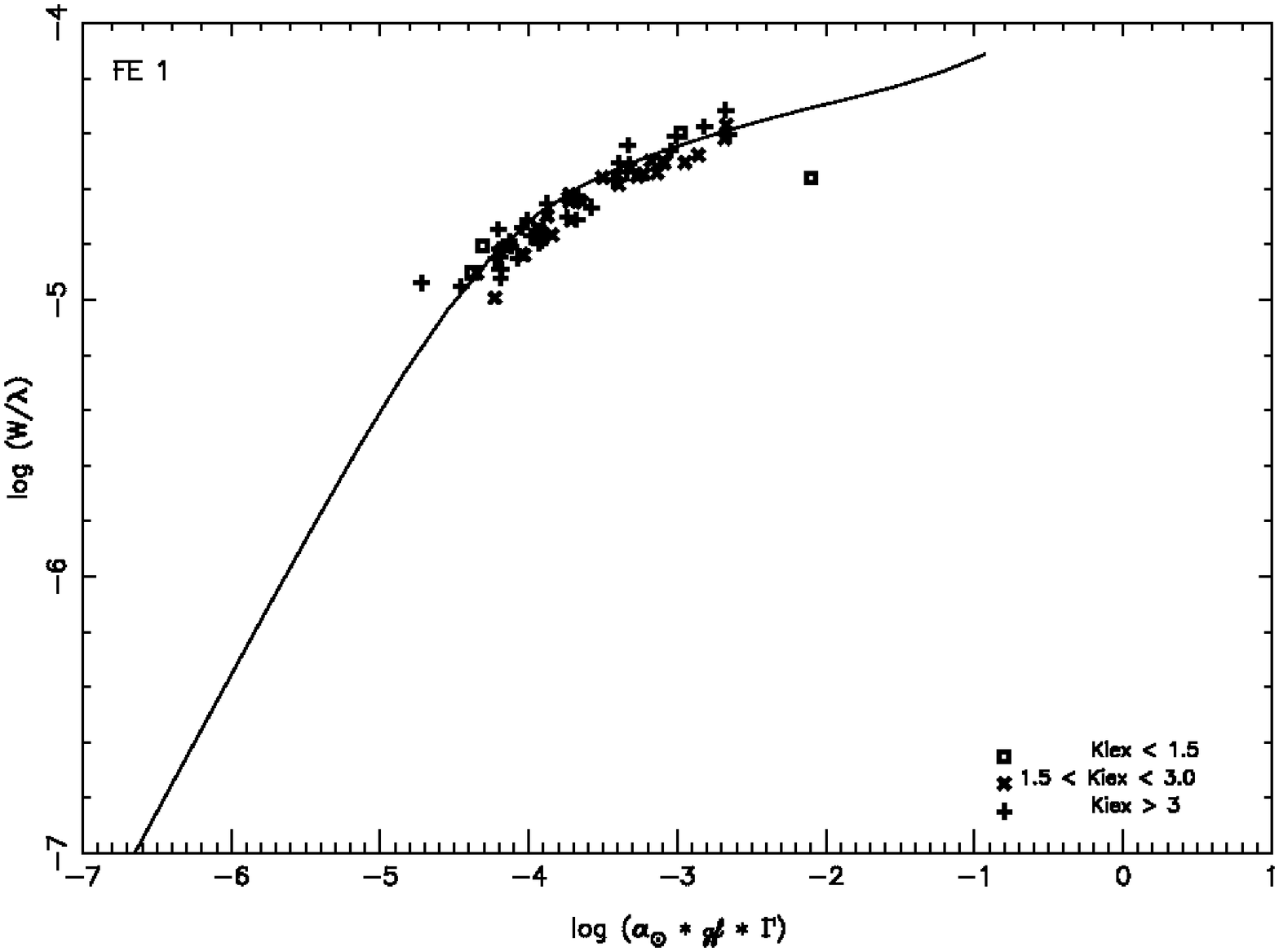}}
  \subfloat[][]{\label{CoG198:2}\includegraphics[angle=-90,width=0.33\textwidth]{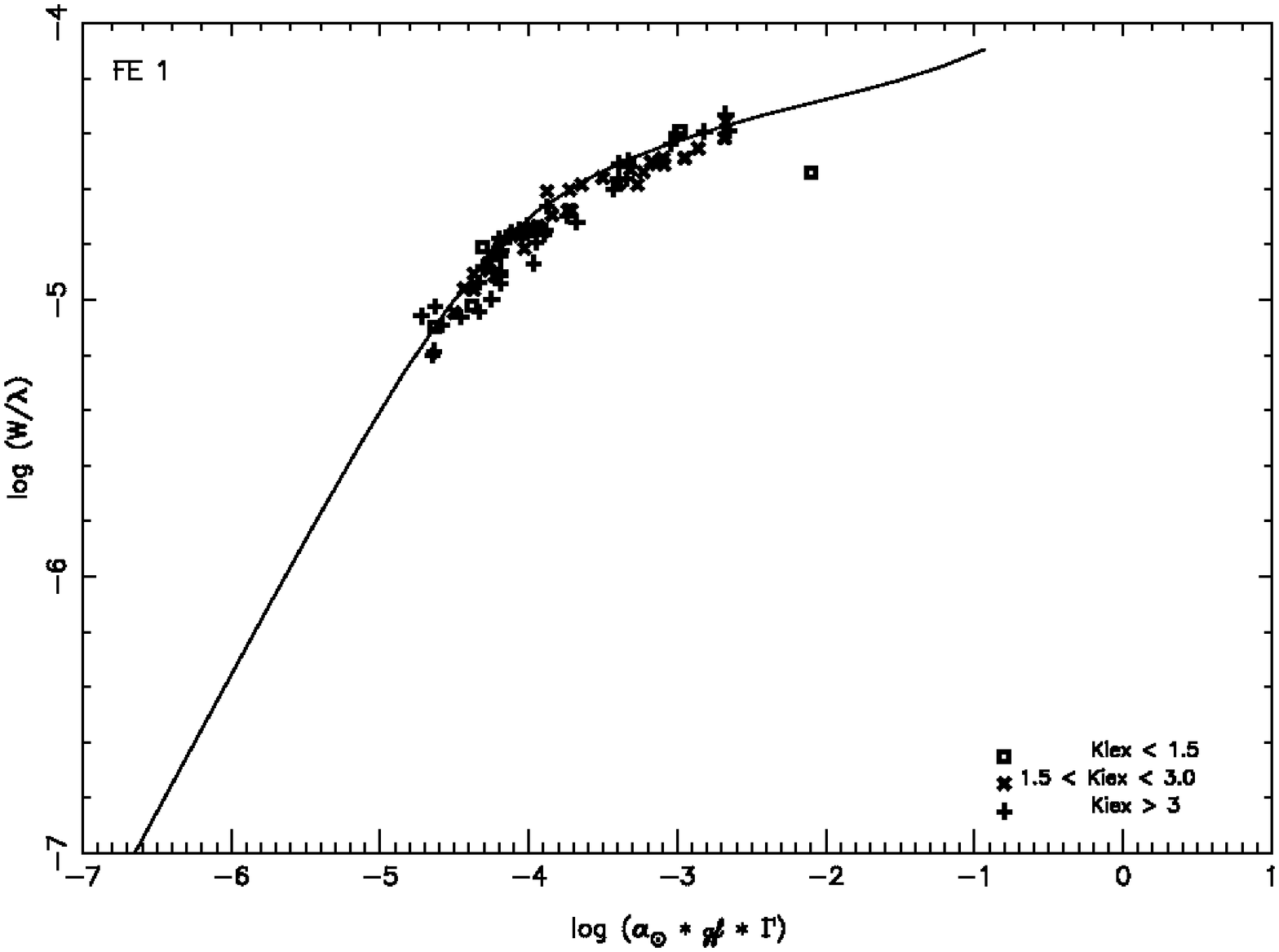}}
  \subfloat[][]{\label{CoG198:3}\includegraphics[angle=-90,width=0.33\textwidth]{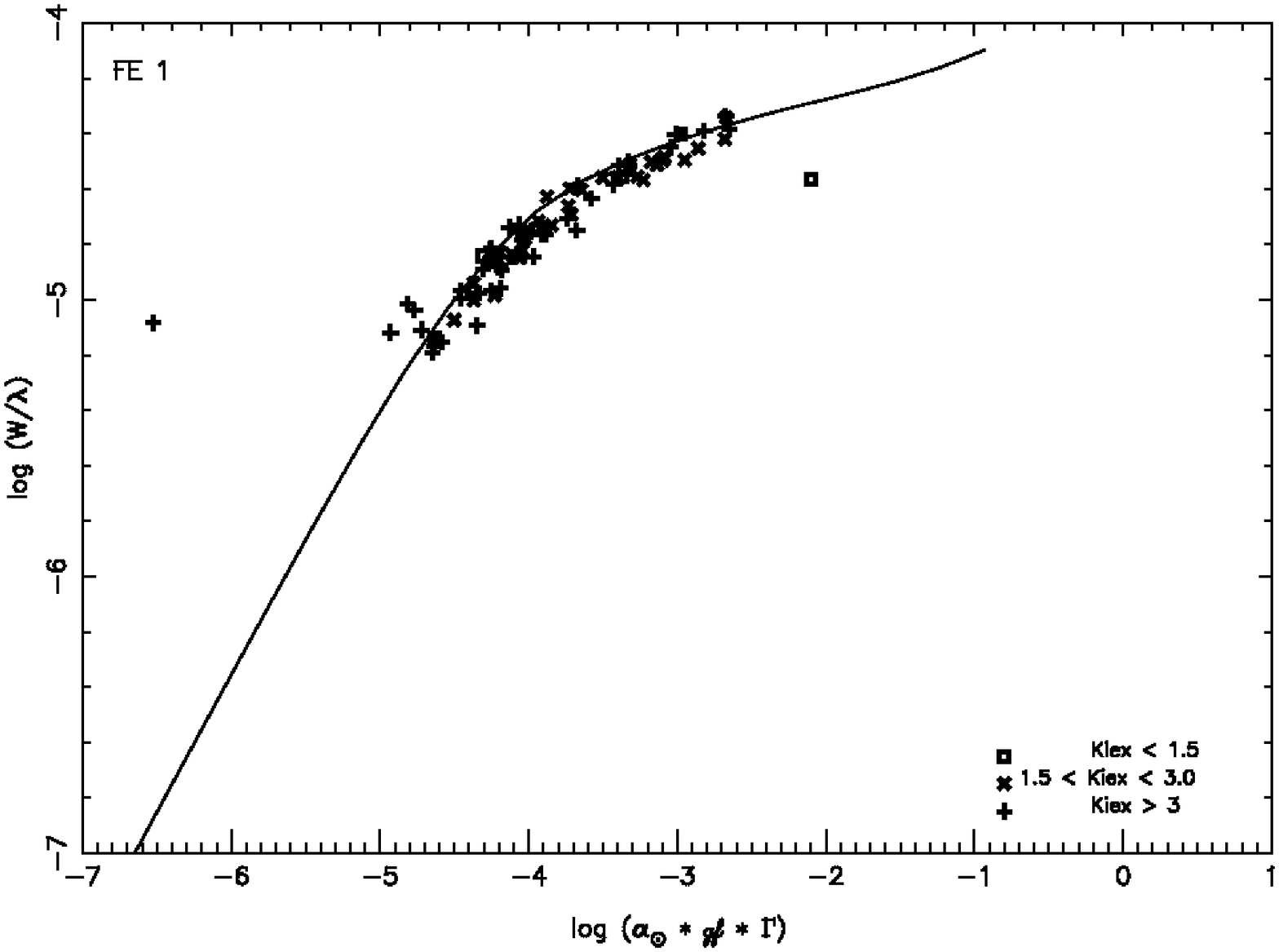}}
  \\
  \subfloat[][]{\label{CoG198:4}\includegraphics[angle=-90,width=0.33\textwidth]{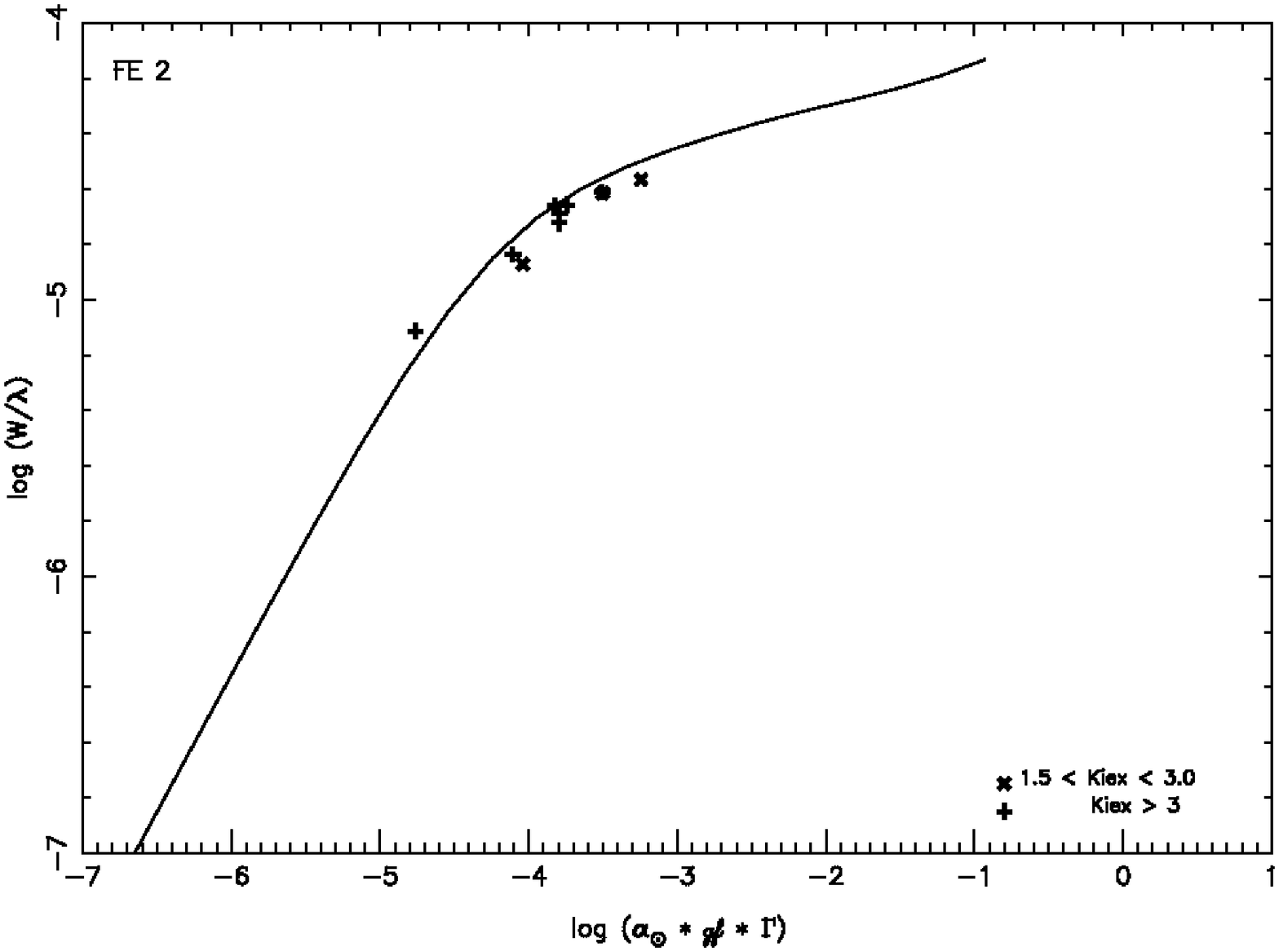}}
  \subfloat[][]{\label{CoG198:5}\includegraphics[angle=-90,width=0.33\textwidth]{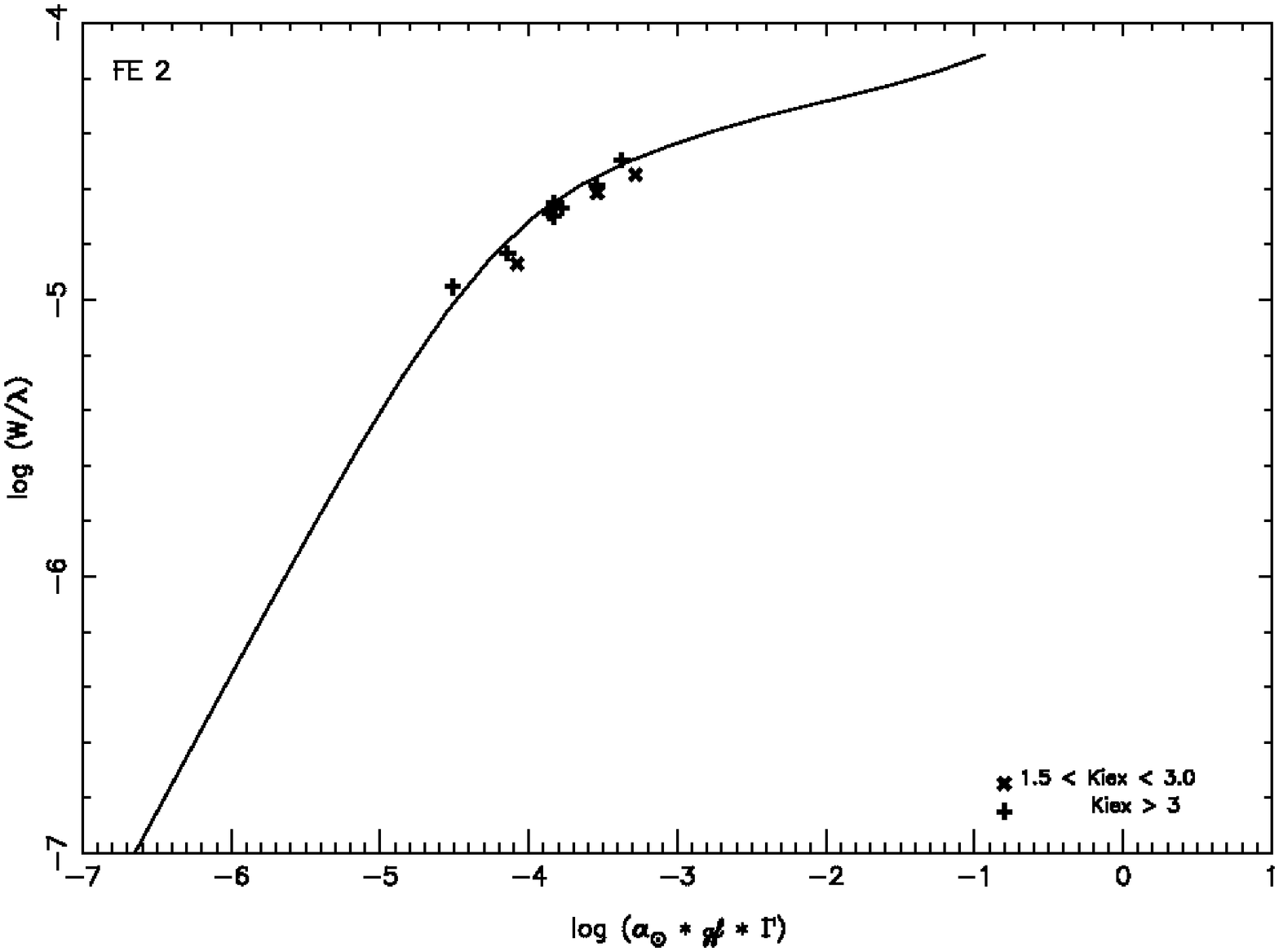}}
  \subfloat[][]{\label{CoG198:6}\includegraphics[angle=-90,width=0.33\textwidth]{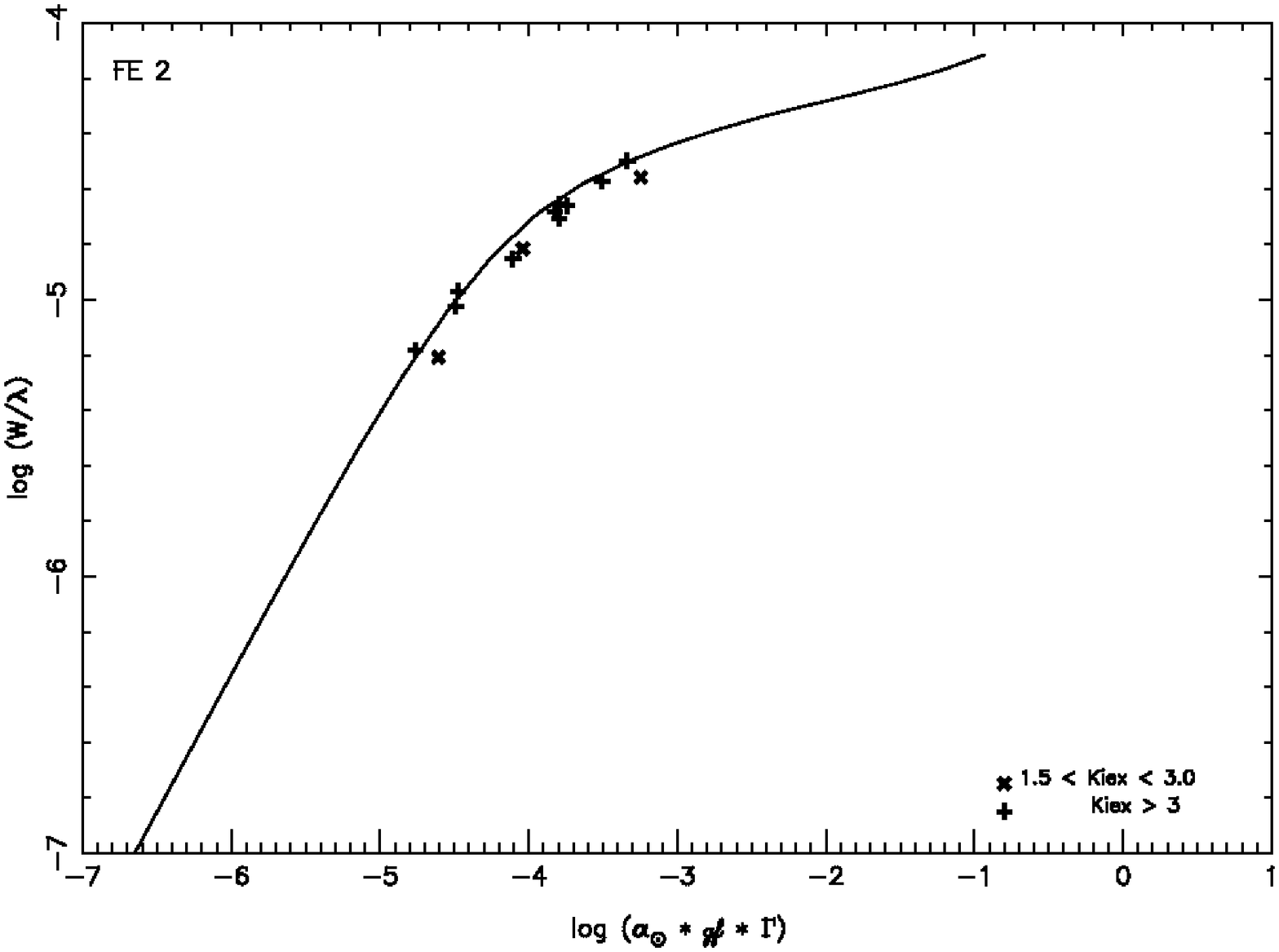}}
\end{sidewaysfigure}


\begin{sidewaysfigure}
  \centering
  \label{CoG199}\caption{Curves of growth (Fe~I, Fe~II) for the HV~12199 spectra. From left to right, MJD=54806.02457622, 54806.08071473, MJD=54806.13684767}
  \subfloat[][]{\label{CoG199:1}\includegraphics[angle=-90,width=0.33\textwidth]{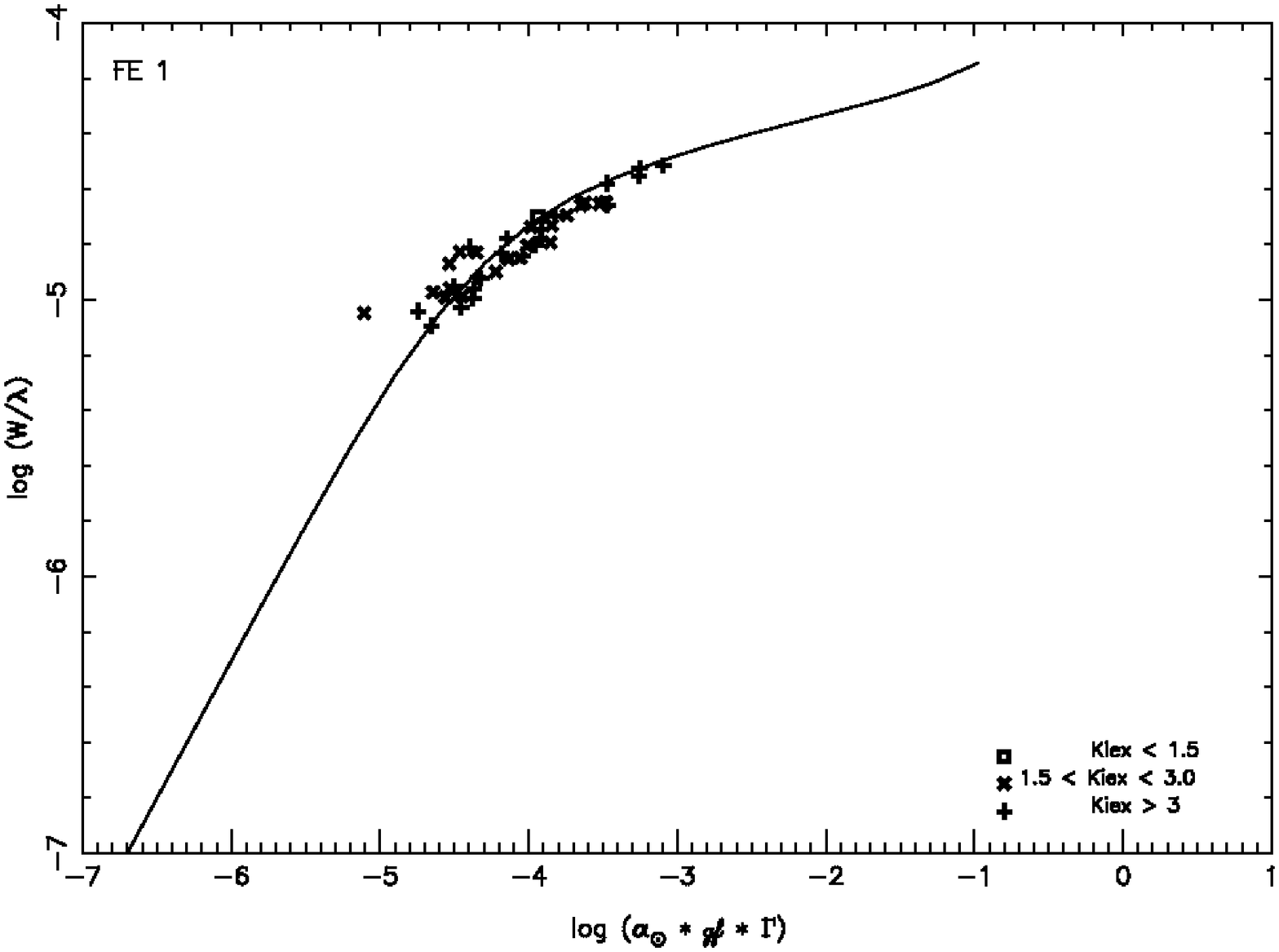}}
  \subfloat[][]{\label{CoG199:2}\includegraphics[angle=-90,width=0.33\textwidth]{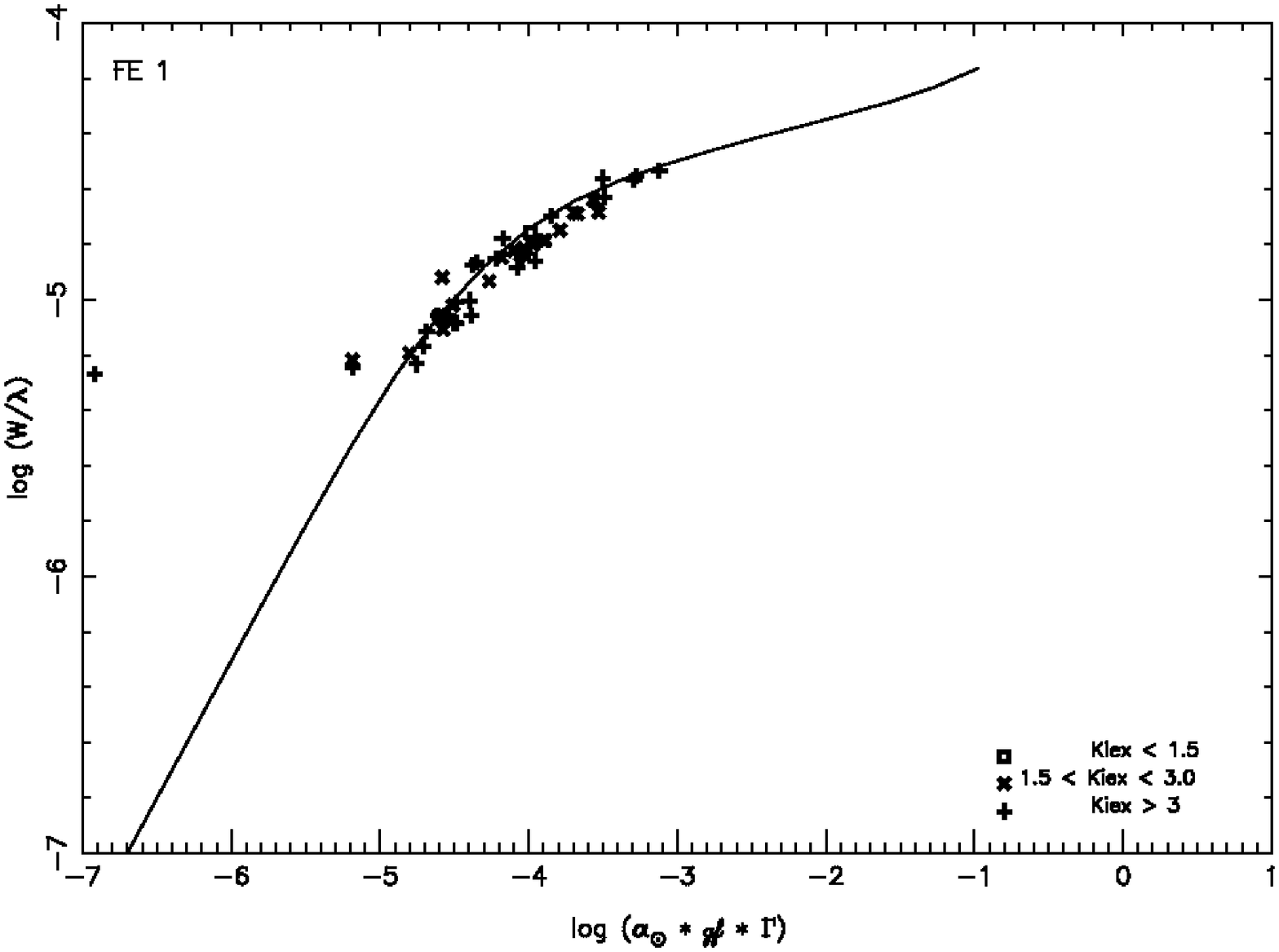}}
  \subfloat[][]{\label{CoG199:3}\includegraphics[angle=-90,width=0.33\textwidth]{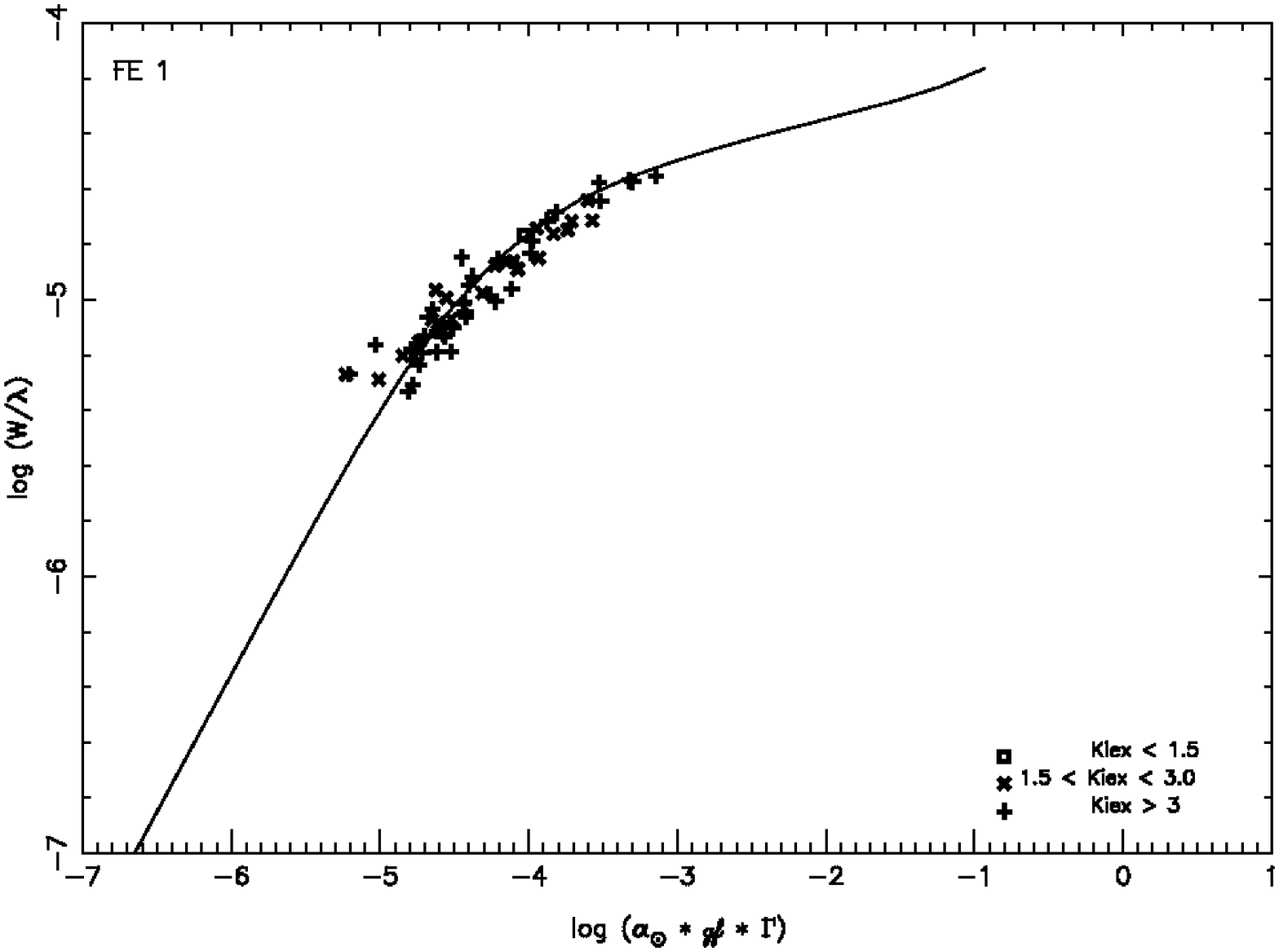}}
  \\
  \subfloat[][]{\label{CoG199:4}\includegraphics[angle=-90,width=0.33\textwidth]{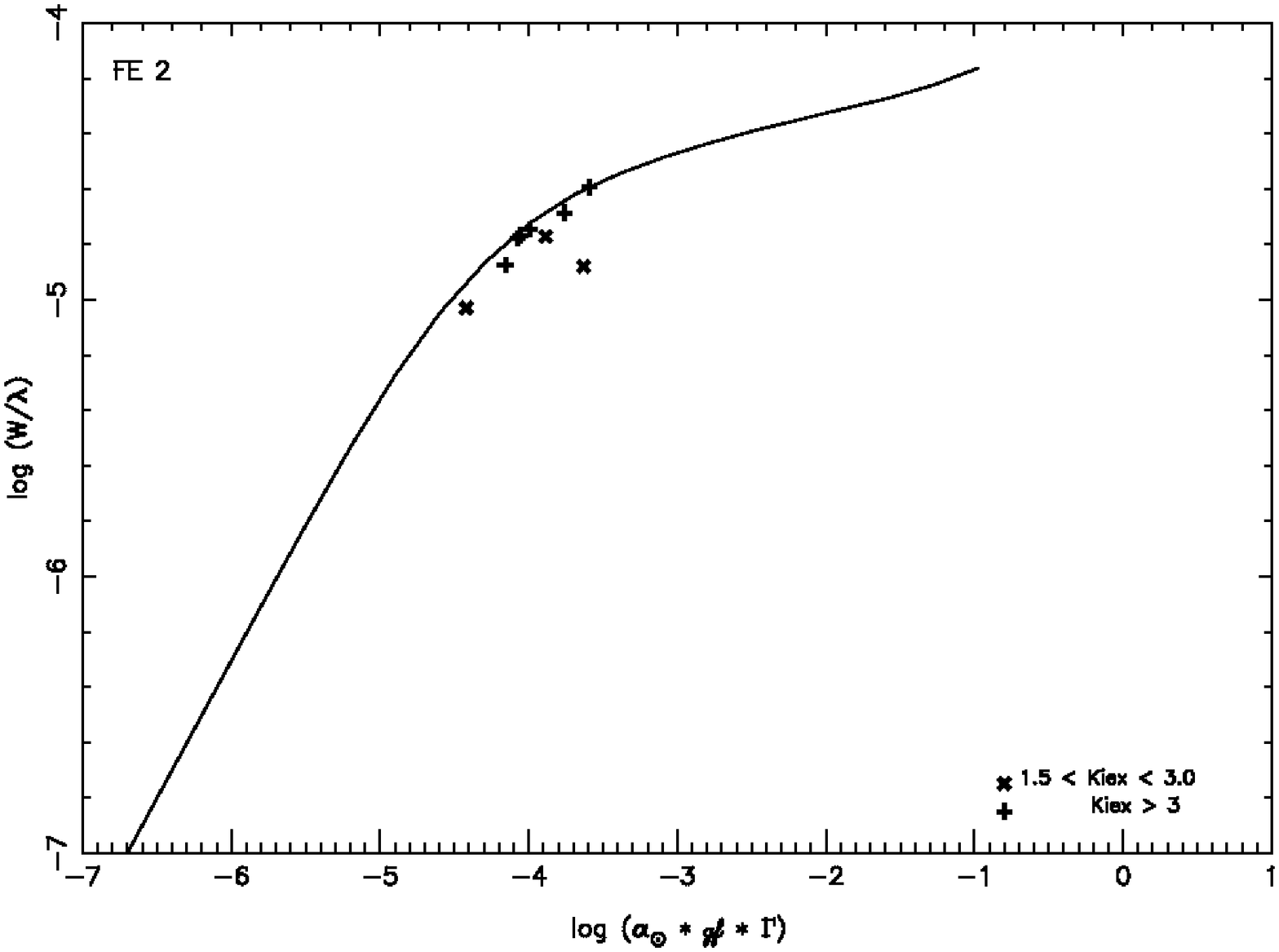}}
  \subfloat[][]{\label{CoG199:5}\includegraphics[angle=-90,width=0.33\textwidth]{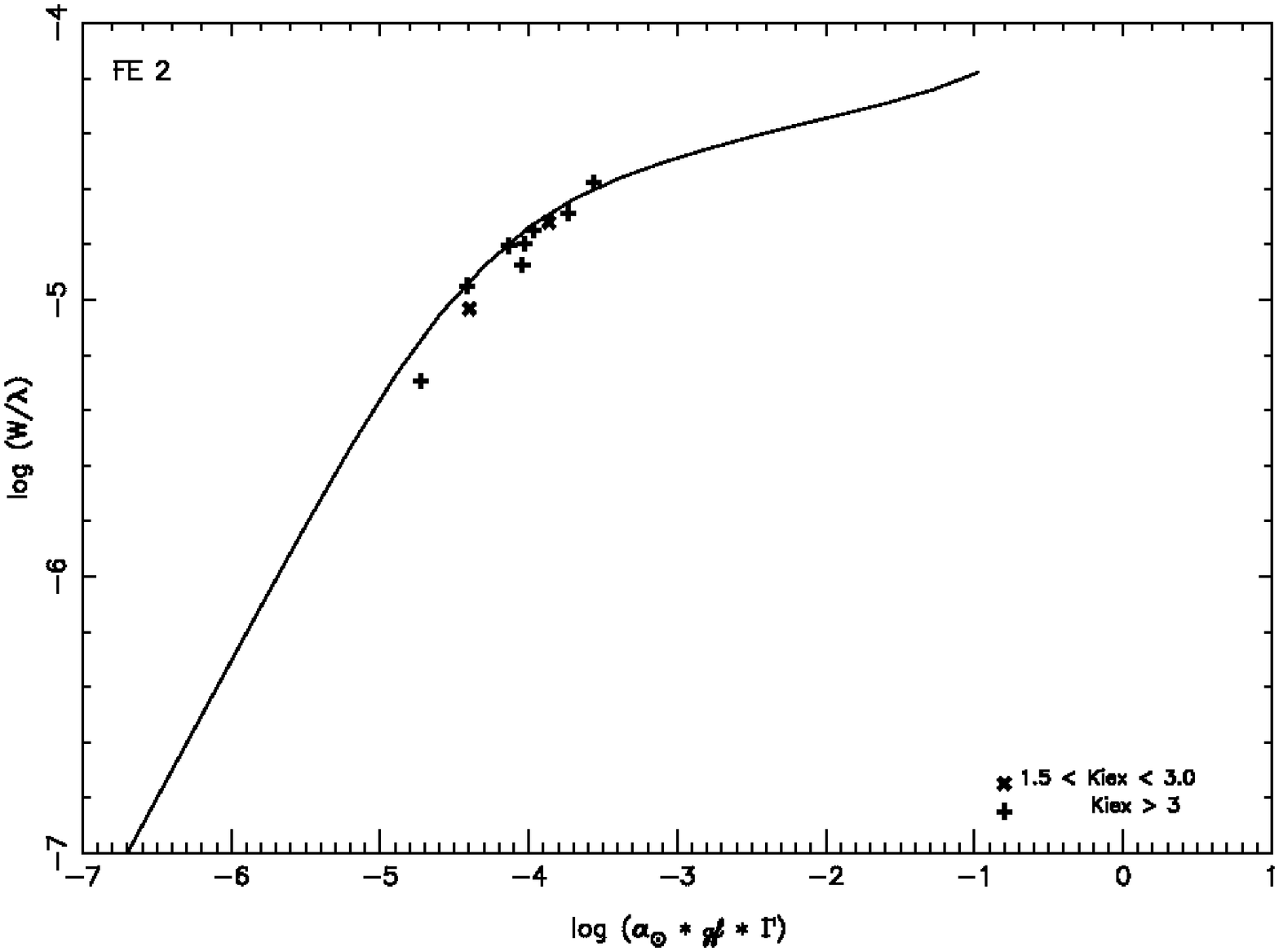}}
  \subfloat[][]{\label{CoG199:6}\includegraphics[angle=-90,width=0.33\textwidth]{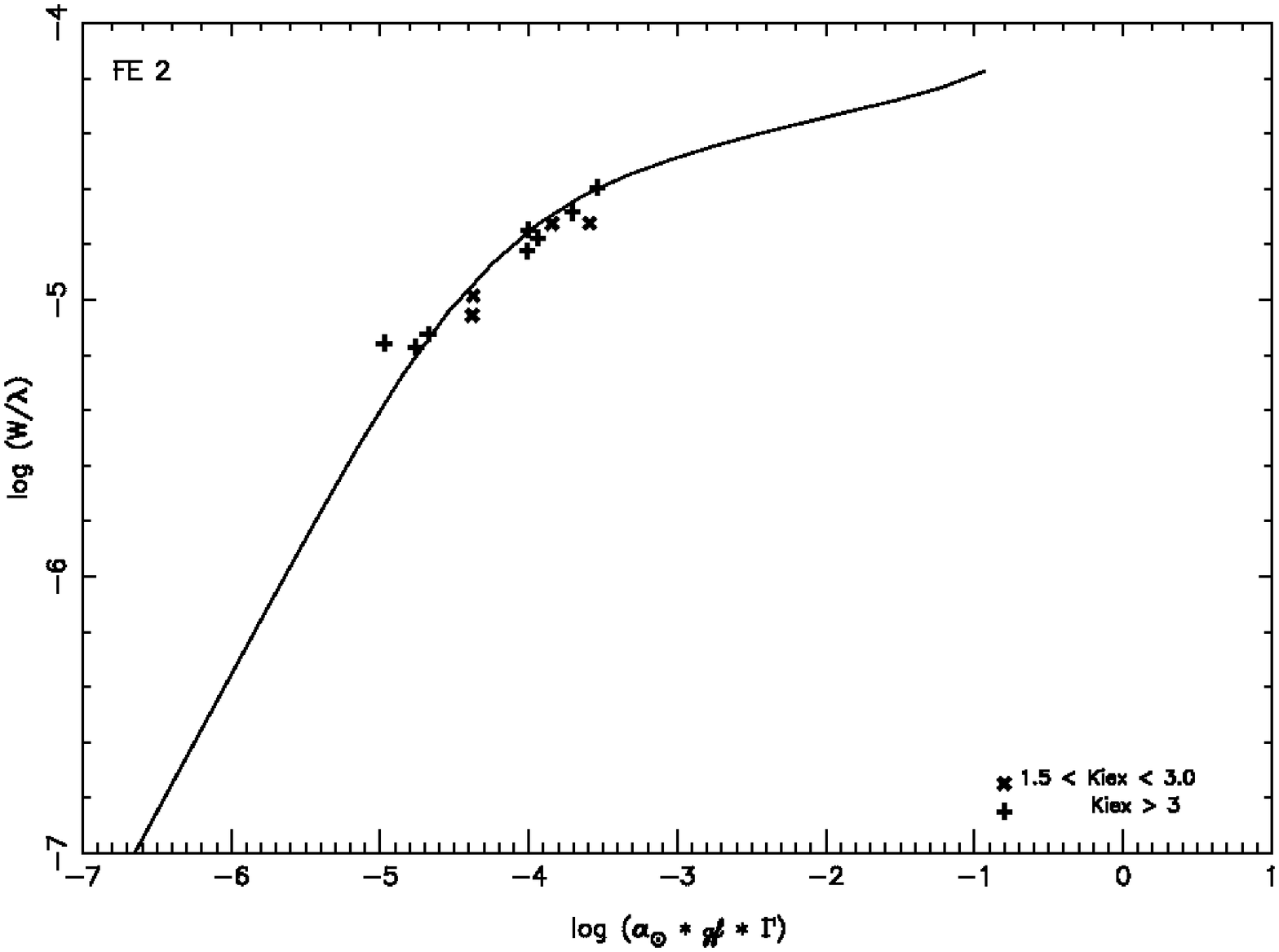}}
\end{sidewaysfigure}


\begin{sidewaysfigure}
  \centering
  \label{CoG202}\caption{Curves of growth (Fe~I, Fe~II) for the HV~12202 spectra. From left to right, MJD=54806.02457622, 54806.08071473, MJD=54806.13684767}
  \subfloat[][]{\label{CoG202:1}\includegraphics[angle=-90,width=0.33\textwidth]{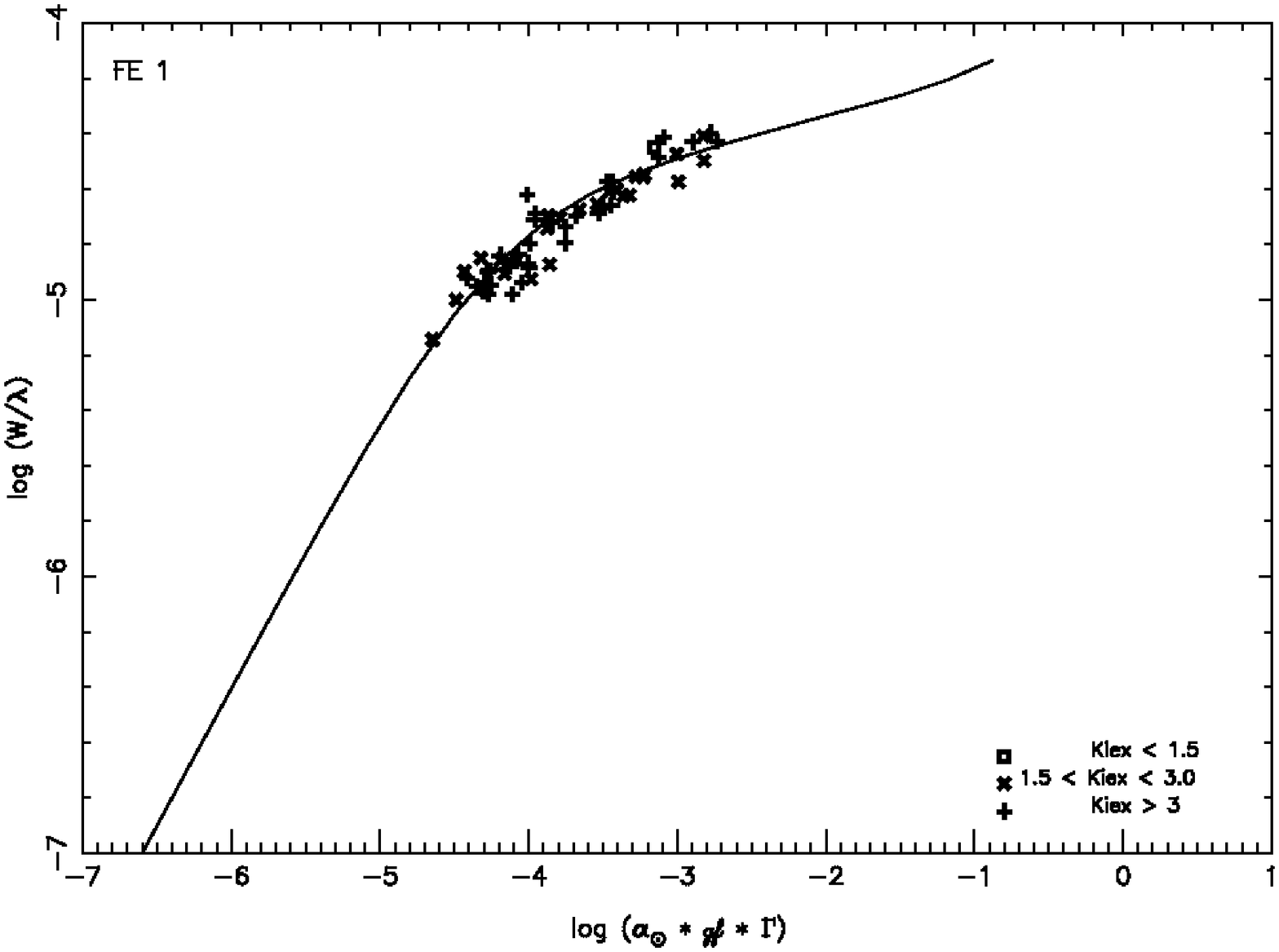}}
  \subfloat[][]{\label{CoG202:2}\includegraphics[angle=-90,width=0.33\textwidth]{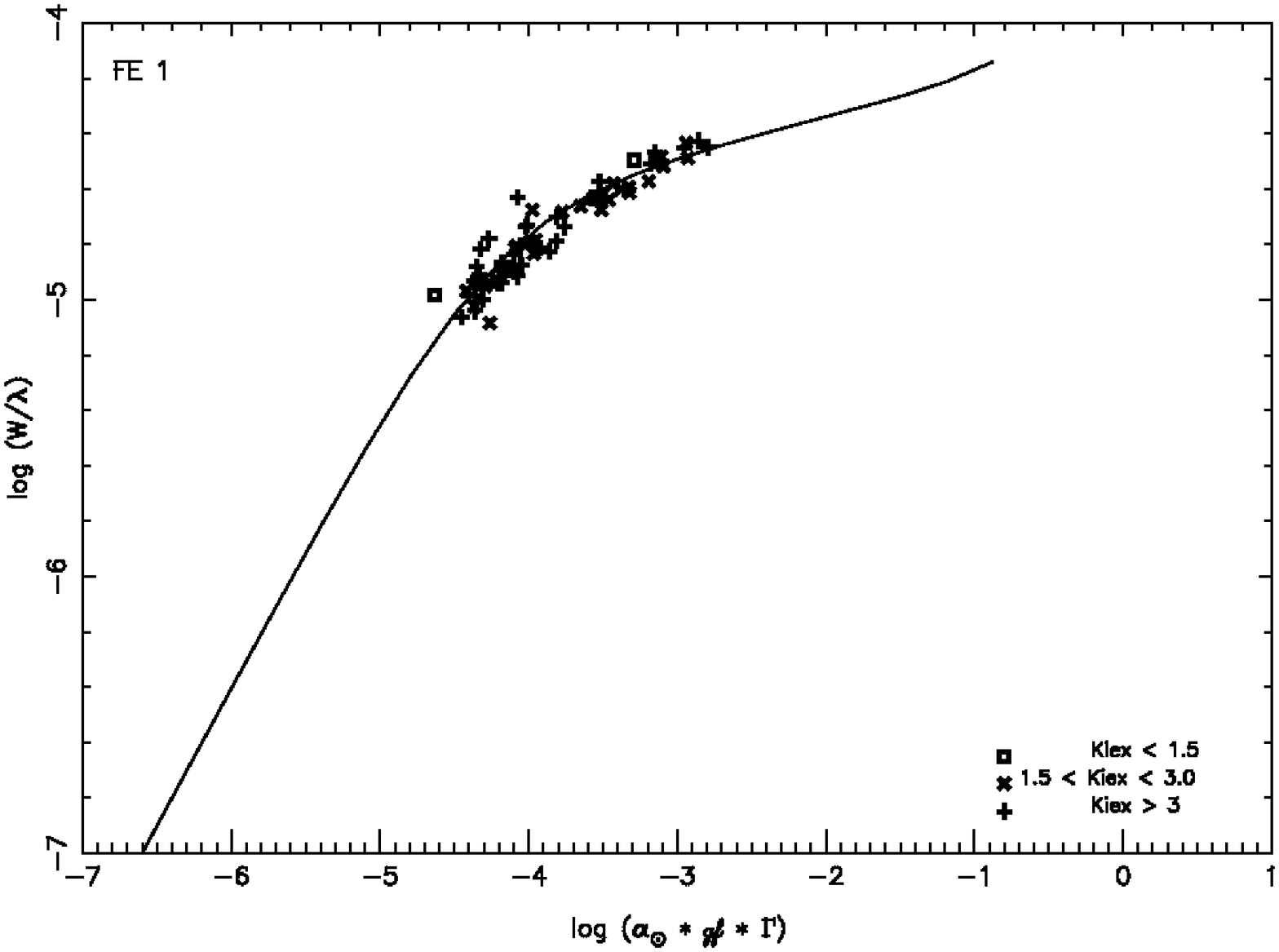}}
  \subfloat[][]{\label{CoG202:3}\includegraphics[angle=-90,width=0.33\textwidth]{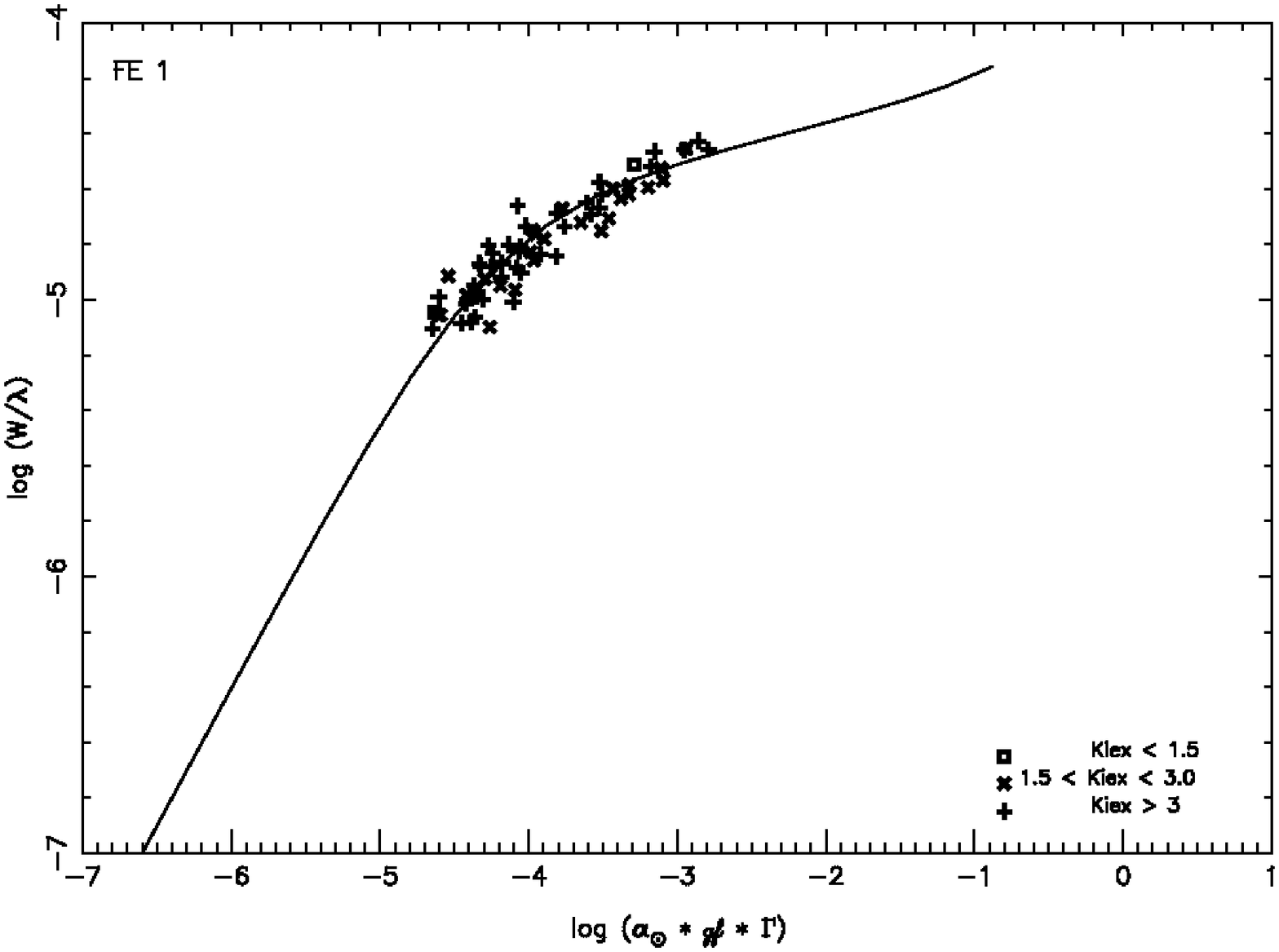}}
  \\
  \subfloat[][]{\label{CoG202:4}\includegraphics[angle=-90,width=0.33\textwidth]{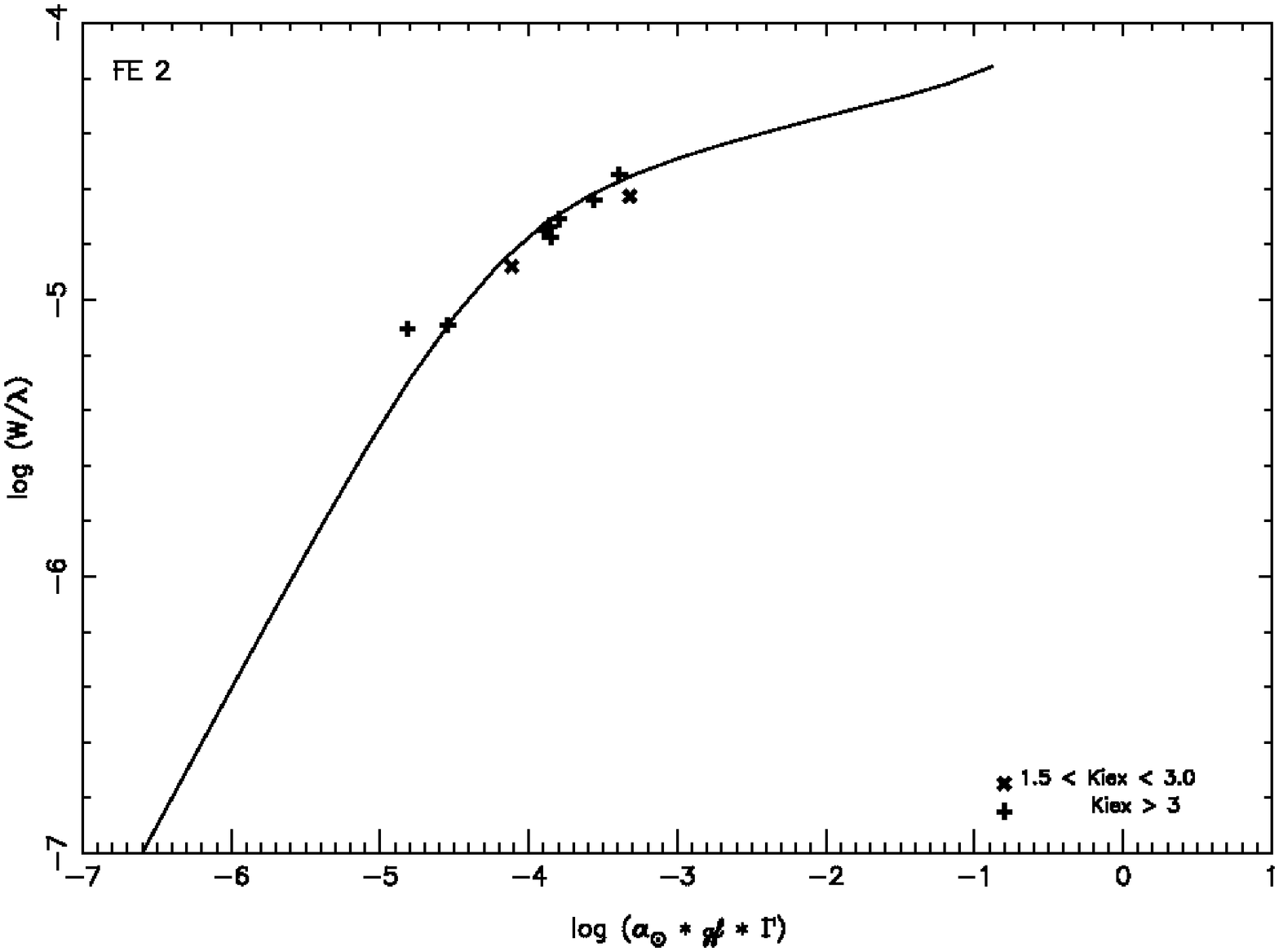}}
  \subfloat[][]{\label{CoG202:5}\includegraphics[angle=-90,width=0.33\textwidth]{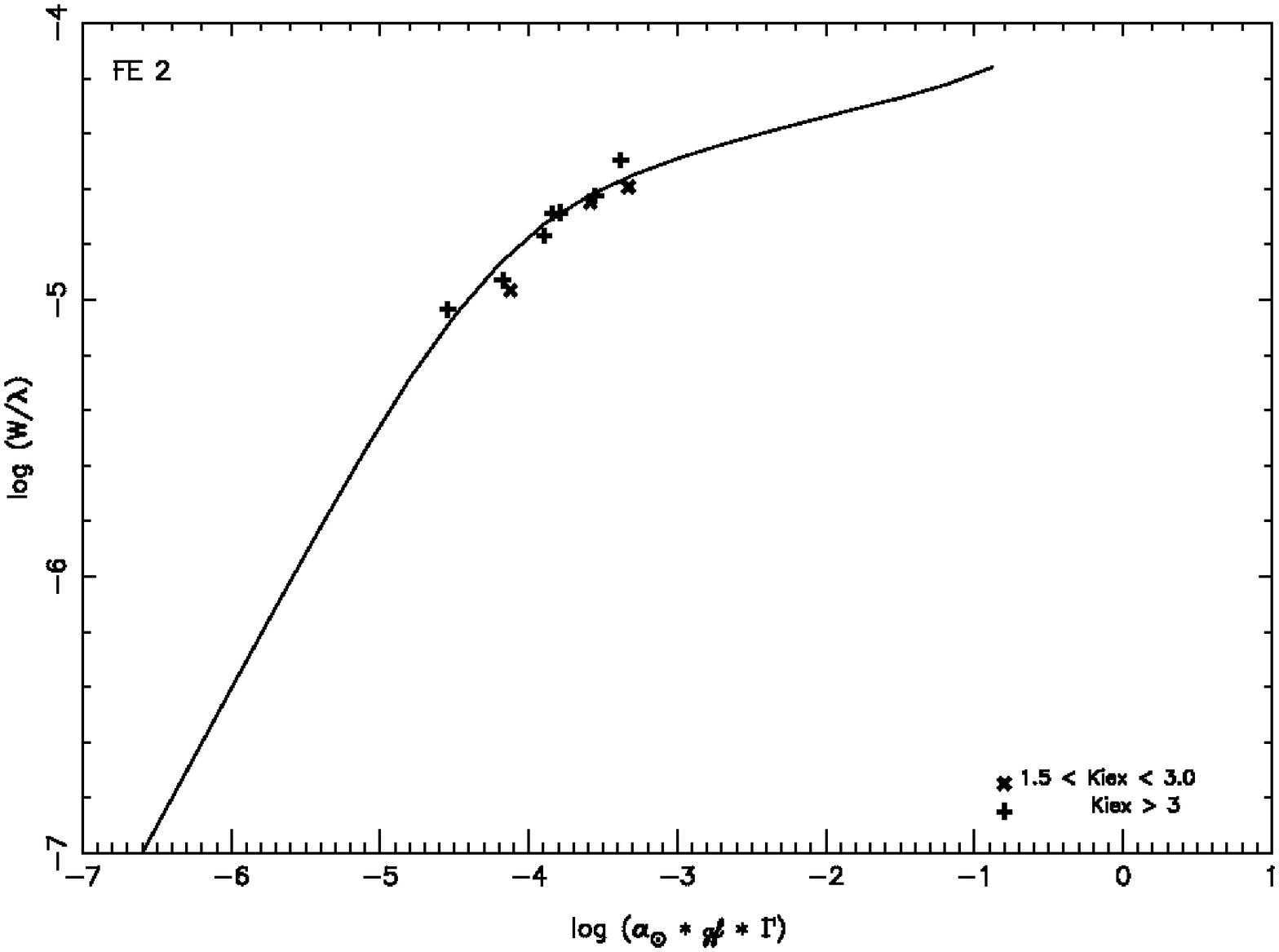}}
  \subfloat[][]{\label{CoG202:6}\includegraphics[angle=-90,width=0.33\textwidth]{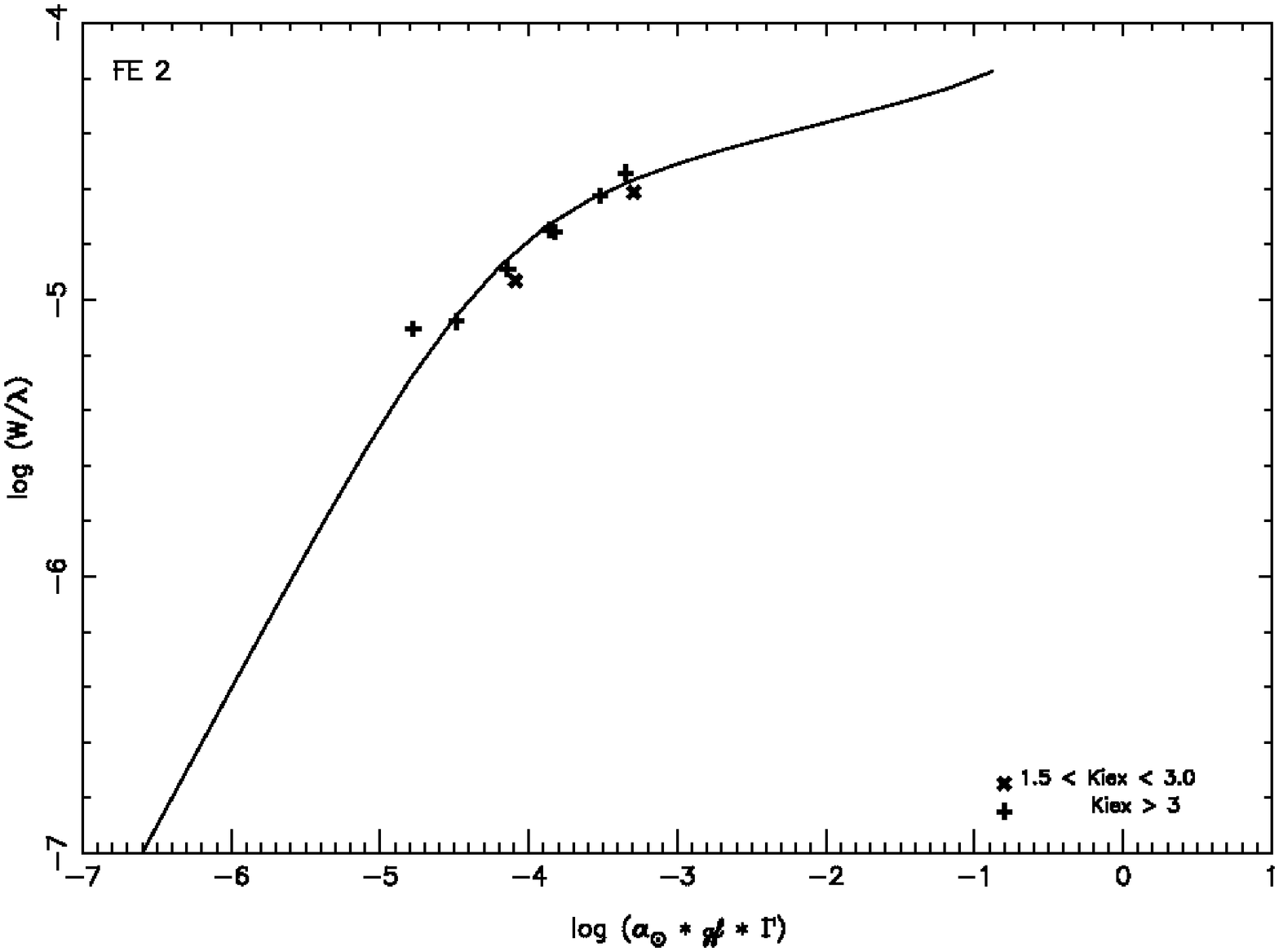}}
\end{sidewaysfigure}


\begin{sidewaysfigure}
  \centering
  \label{CoG203}\caption{Curves of growth (Fe~I, Fe~II) for the HV~12203 spectra. From left to right, MJD=54806.02457622, 54806.08071473, MJD=54806.13684767}
  \subfloat[][]{\label{CoG203:1}\includegraphics[angle=-90,width=0.33\textwidth]{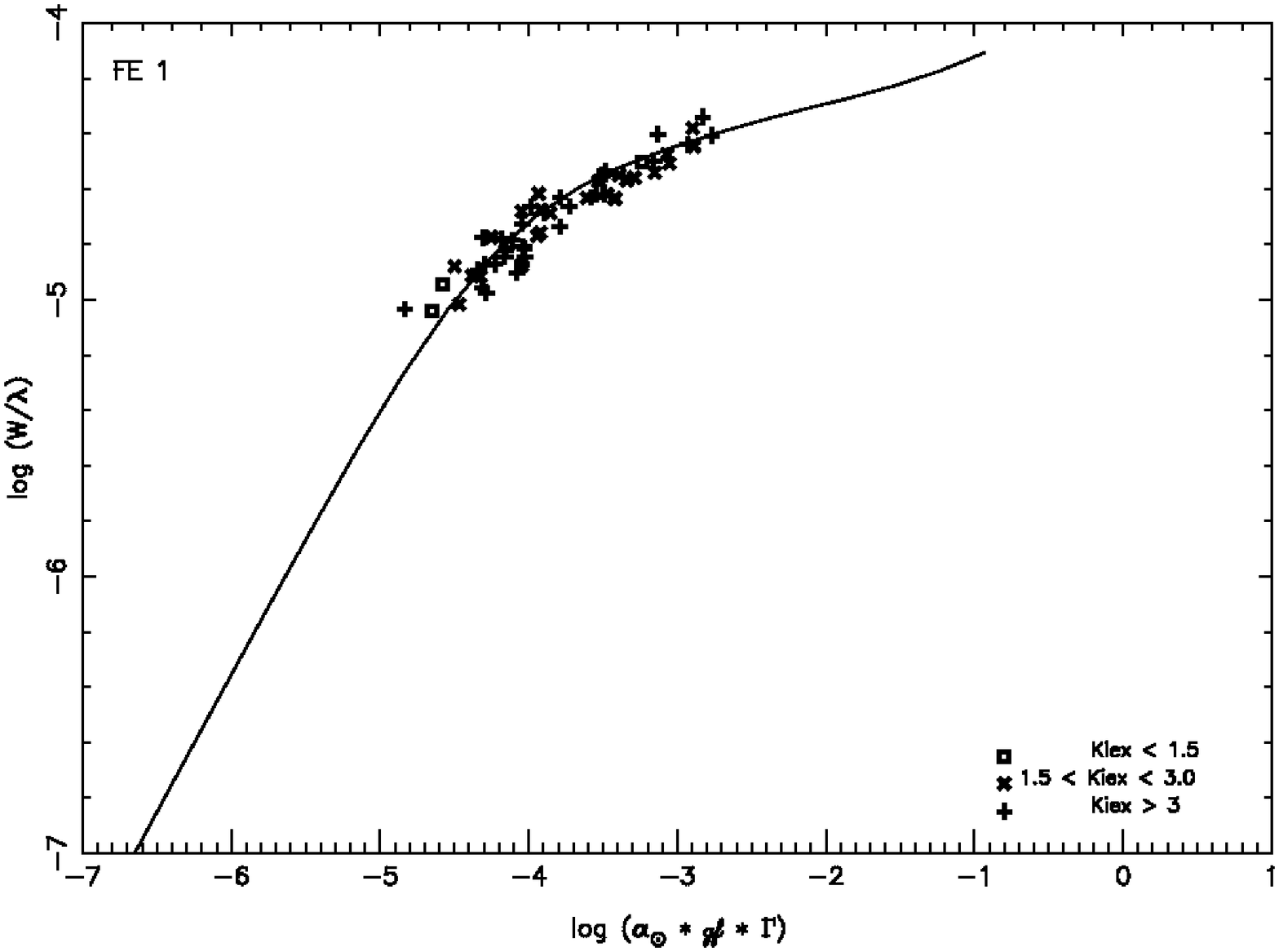}}
  \subfloat[][]{\label{CoG203:2}\includegraphics[angle=-90,width=0.33\textwidth]{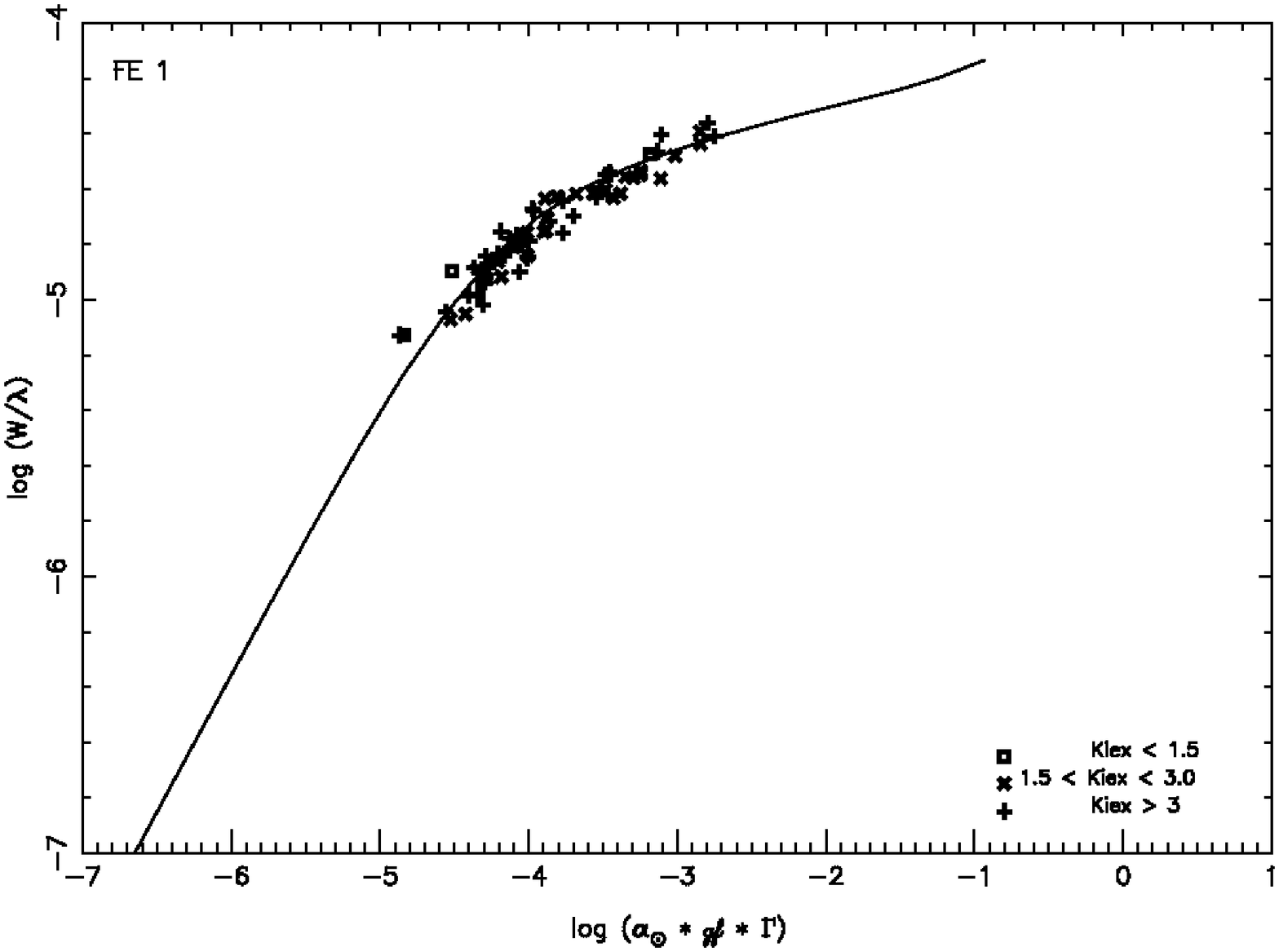}}
  \subfloat[][]{\label{CoG203:3}\includegraphics[angle=-90,width=0.33\textwidth]{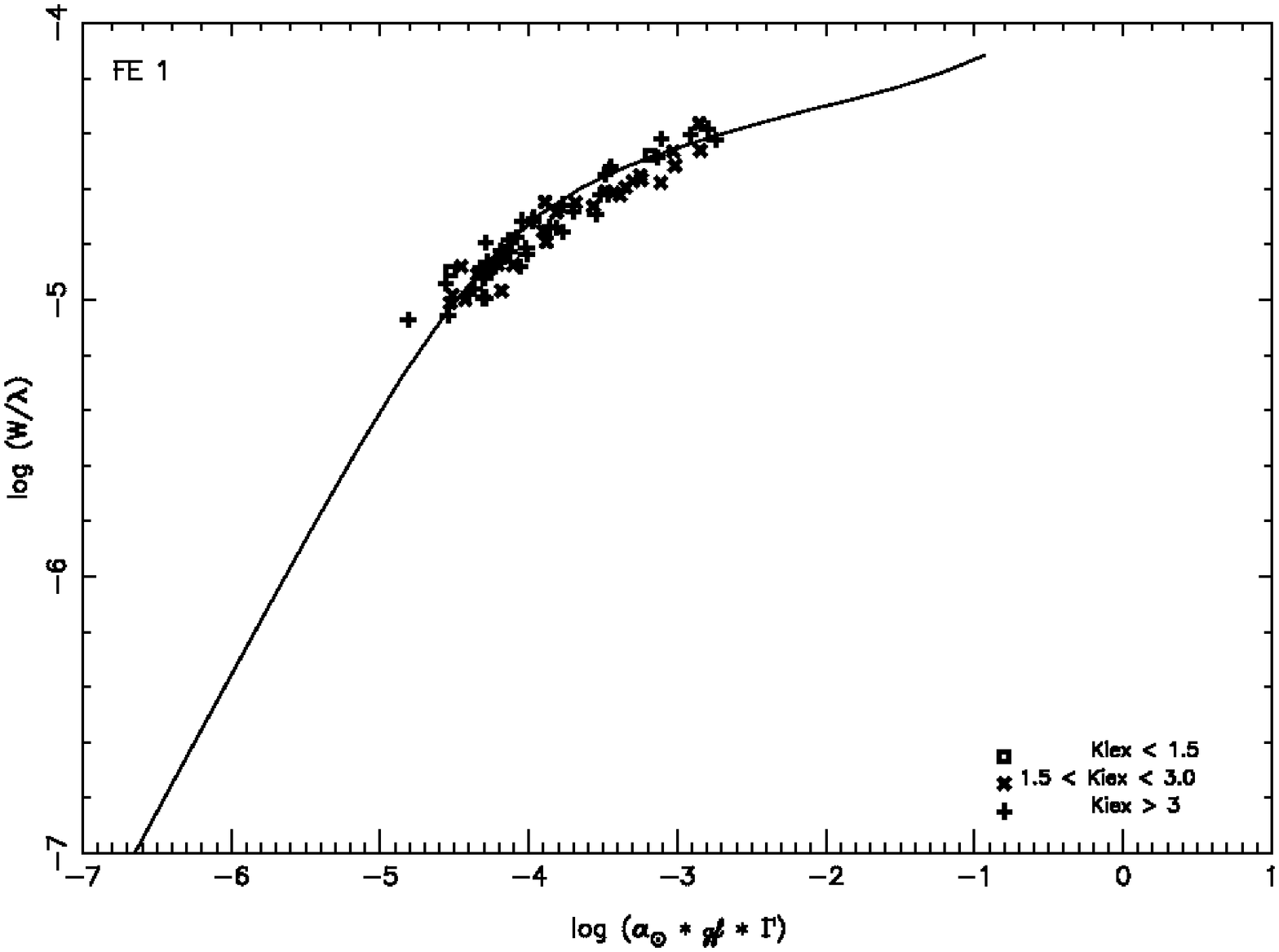}}
  \\
  \subfloat[][]{\label{CoG203:4}\includegraphics[angle=-90,width=0.33\textwidth]{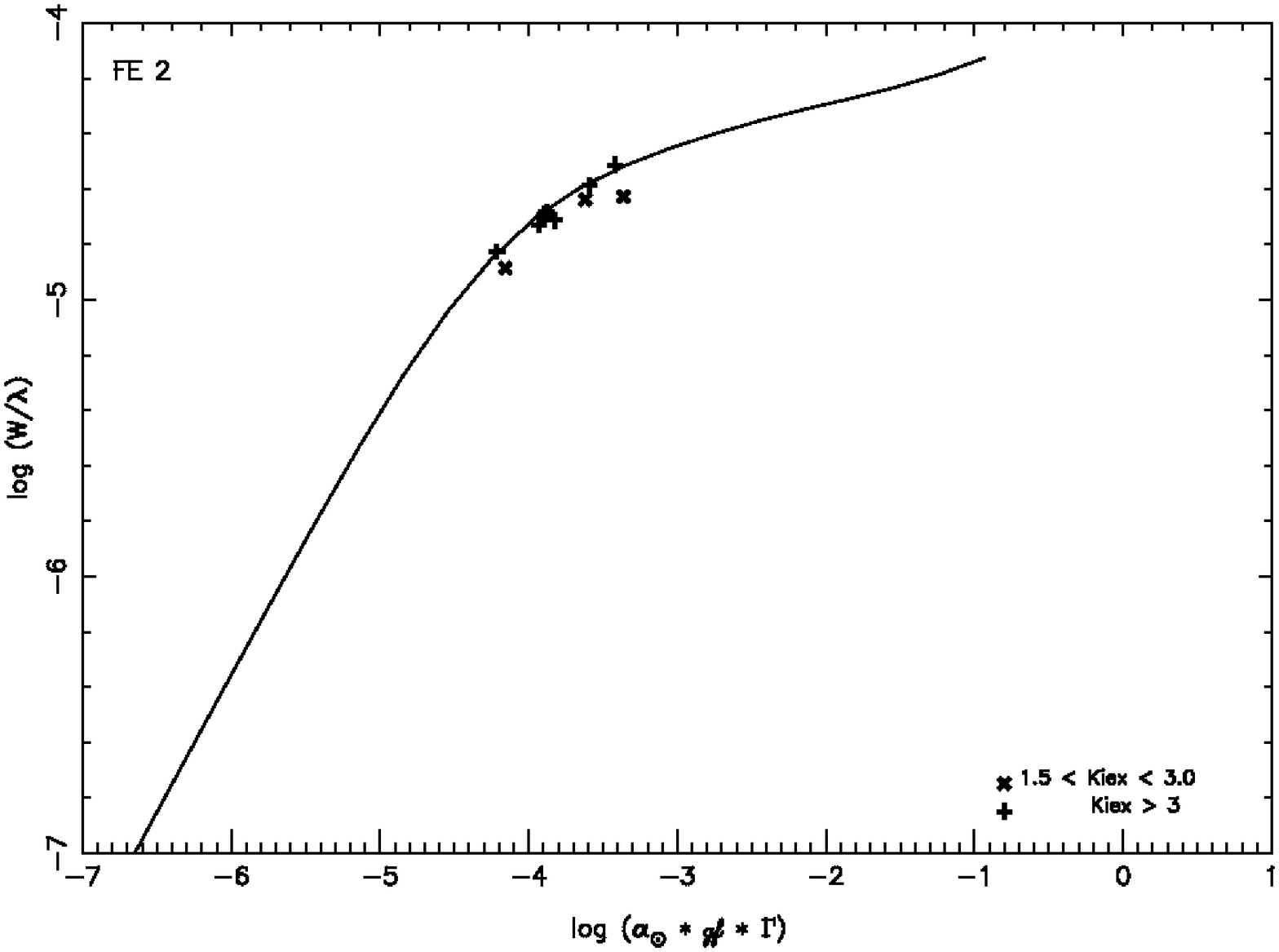}}
  \subfloat[][]{\label{CoG203:5}\includegraphics[angle=-90,width=0.33\textwidth]{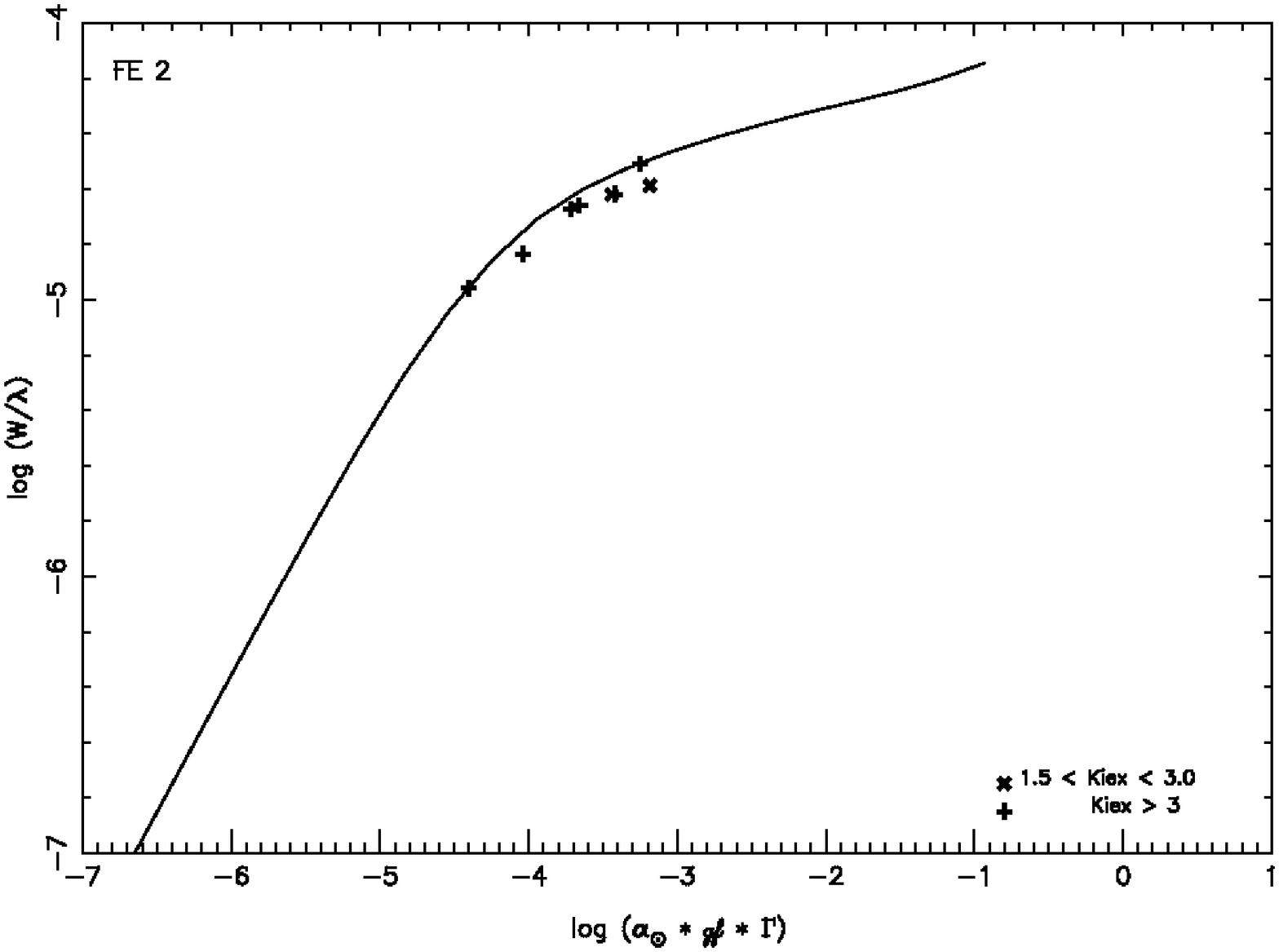}}
  \subfloat[][]{\label{CoG203:6}\includegraphics[angle=-90,width=0.33\textwidth]{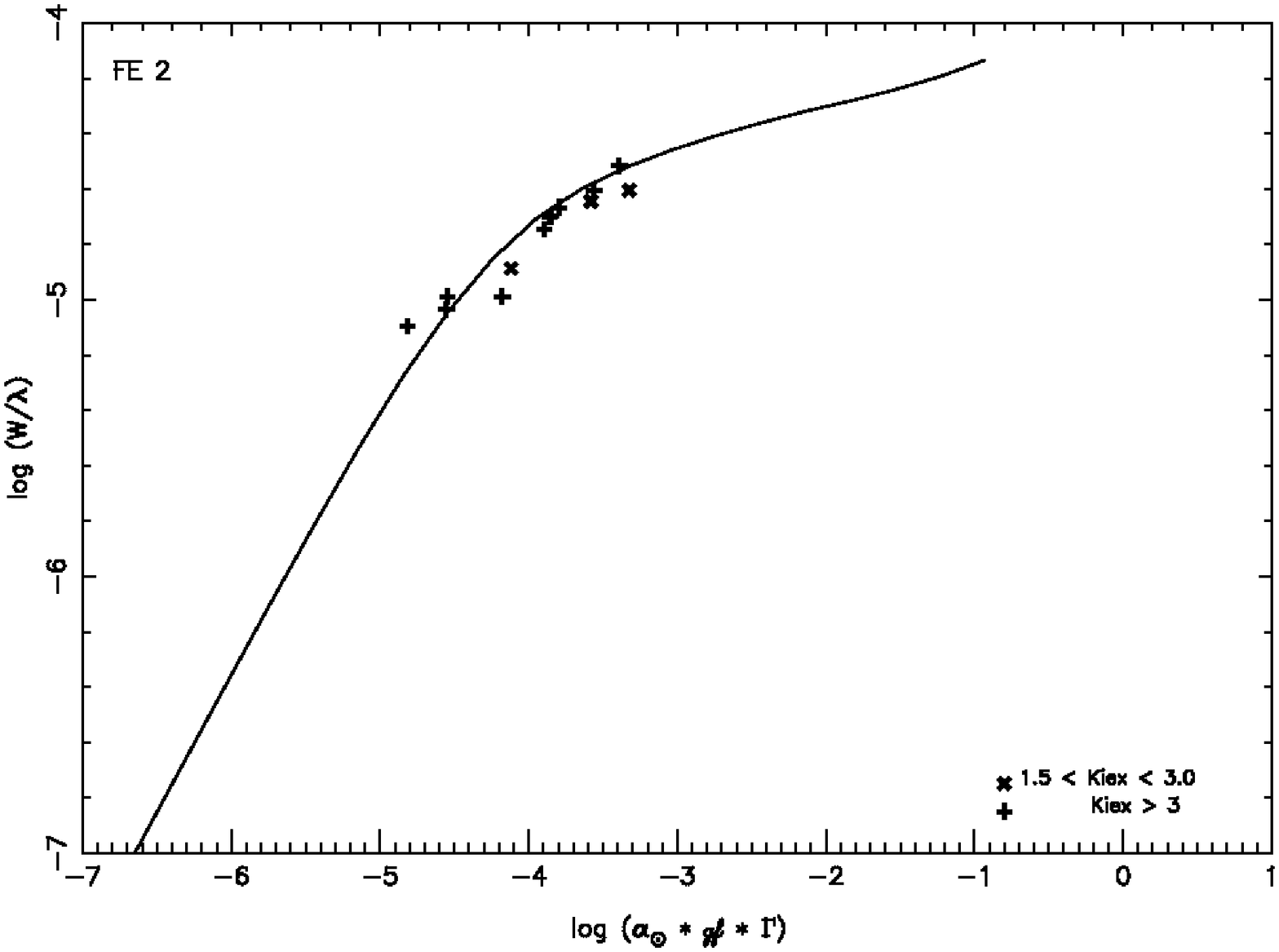}}
\end{sidewaysfigure}


\begin{sidewaysfigure}
  \centering
  \label{CoG204}\caption{Curves of growth (Fe~I, Fe~II) for the HV~12204 spectra. From left to right, MJD=54806.02457622, 54806.08071473, MJD=54806.13684767}
  \subfloat[][]{\label{CoG204:1}\includegraphics[angle=-90,width=0.33\textwidth]{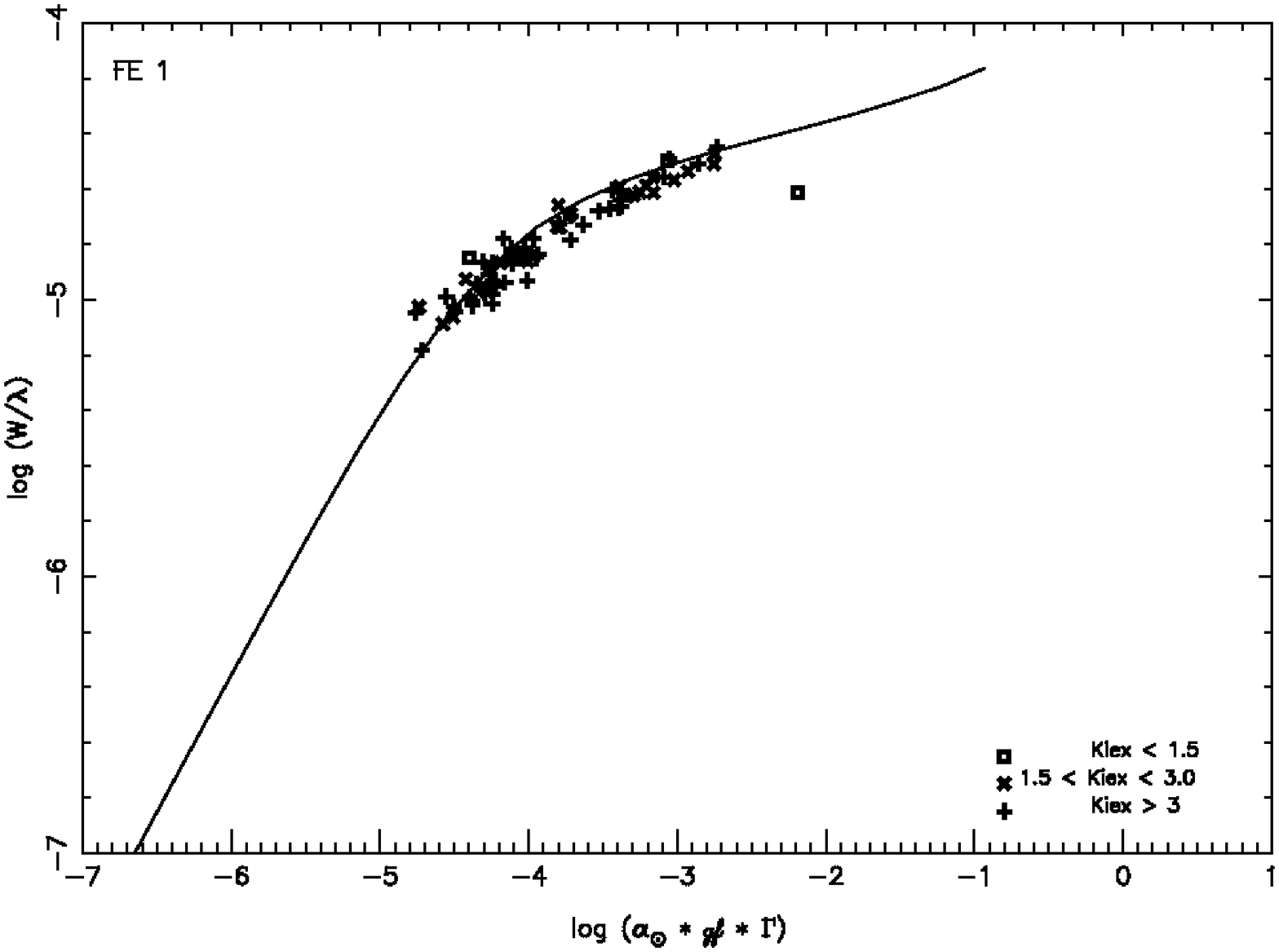}}
  \subfloat[][]{\label{CoG204:2}\includegraphics[angle=-90,width=0.33\textwidth]{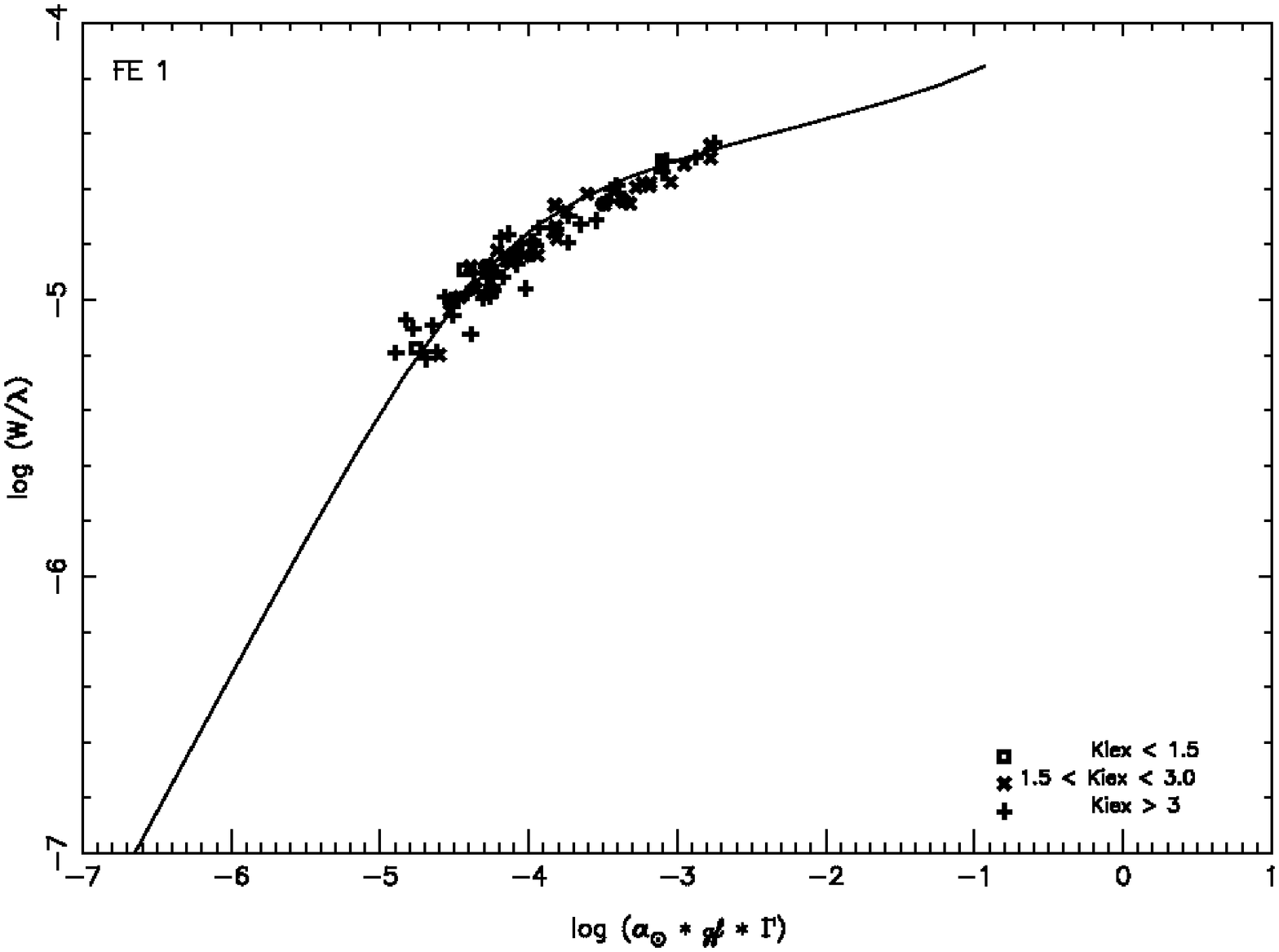}}
  \subfloat[][]{\label{CoG204:3}\includegraphics[angle=-90,width=0.33\textwidth]{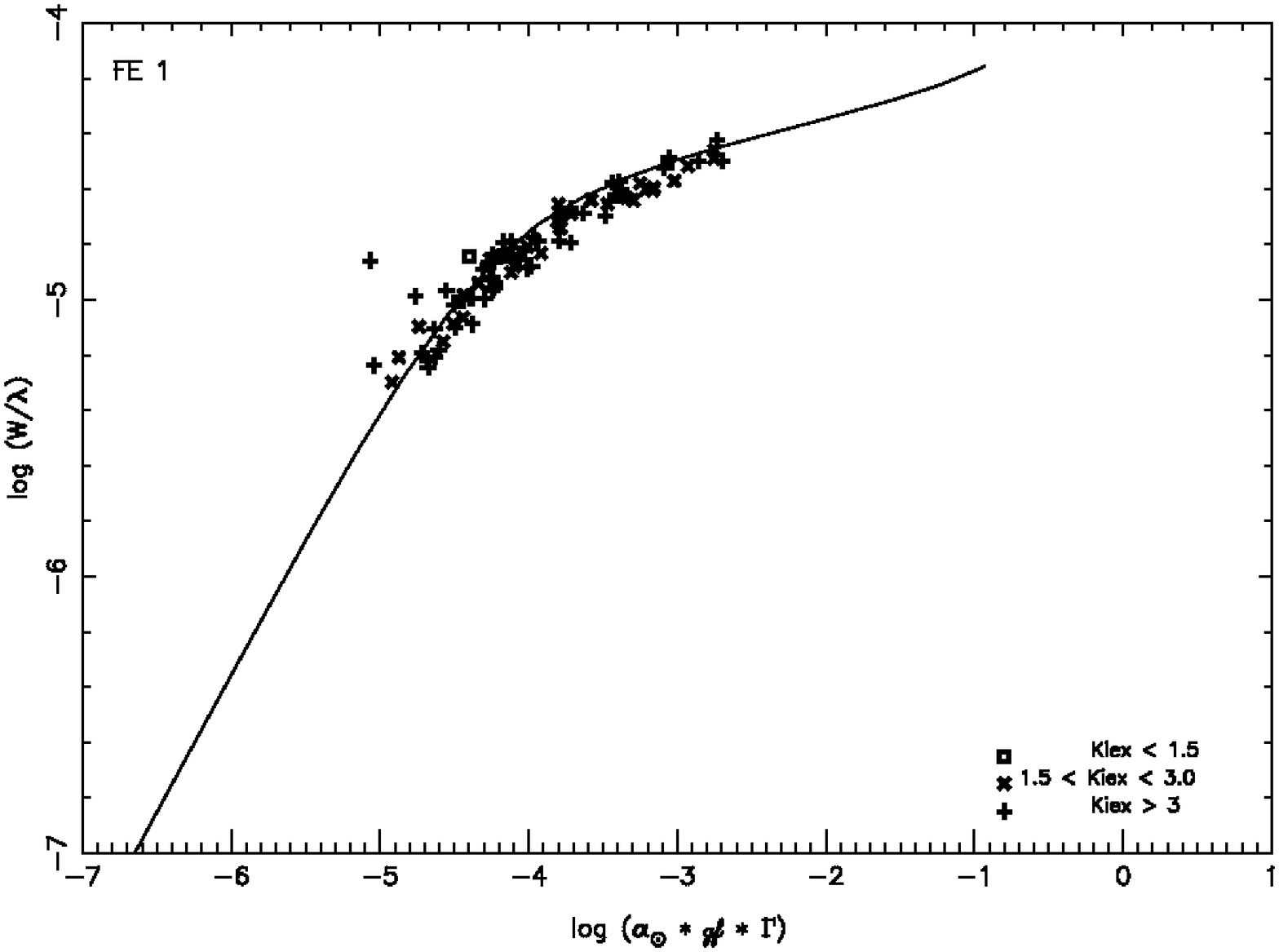}}
  \\
  \subfloat[][]{\label{CoG204:4}\includegraphics[angle=-90,width=0.33\textwidth]{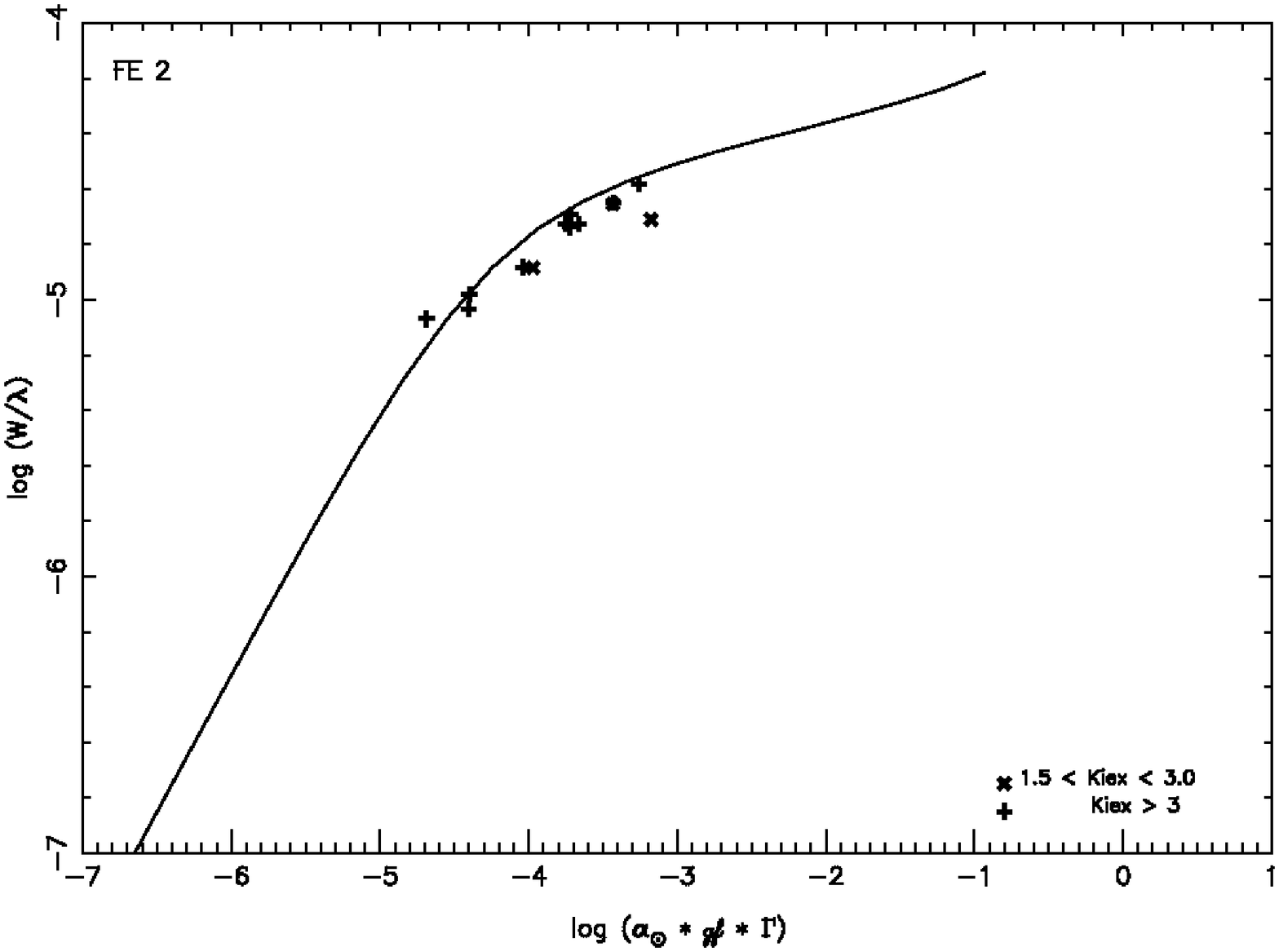}}
  \subfloat[][]{\label{CoG204:5}\includegraphics[angle=-90,width=0.33\textwidth]{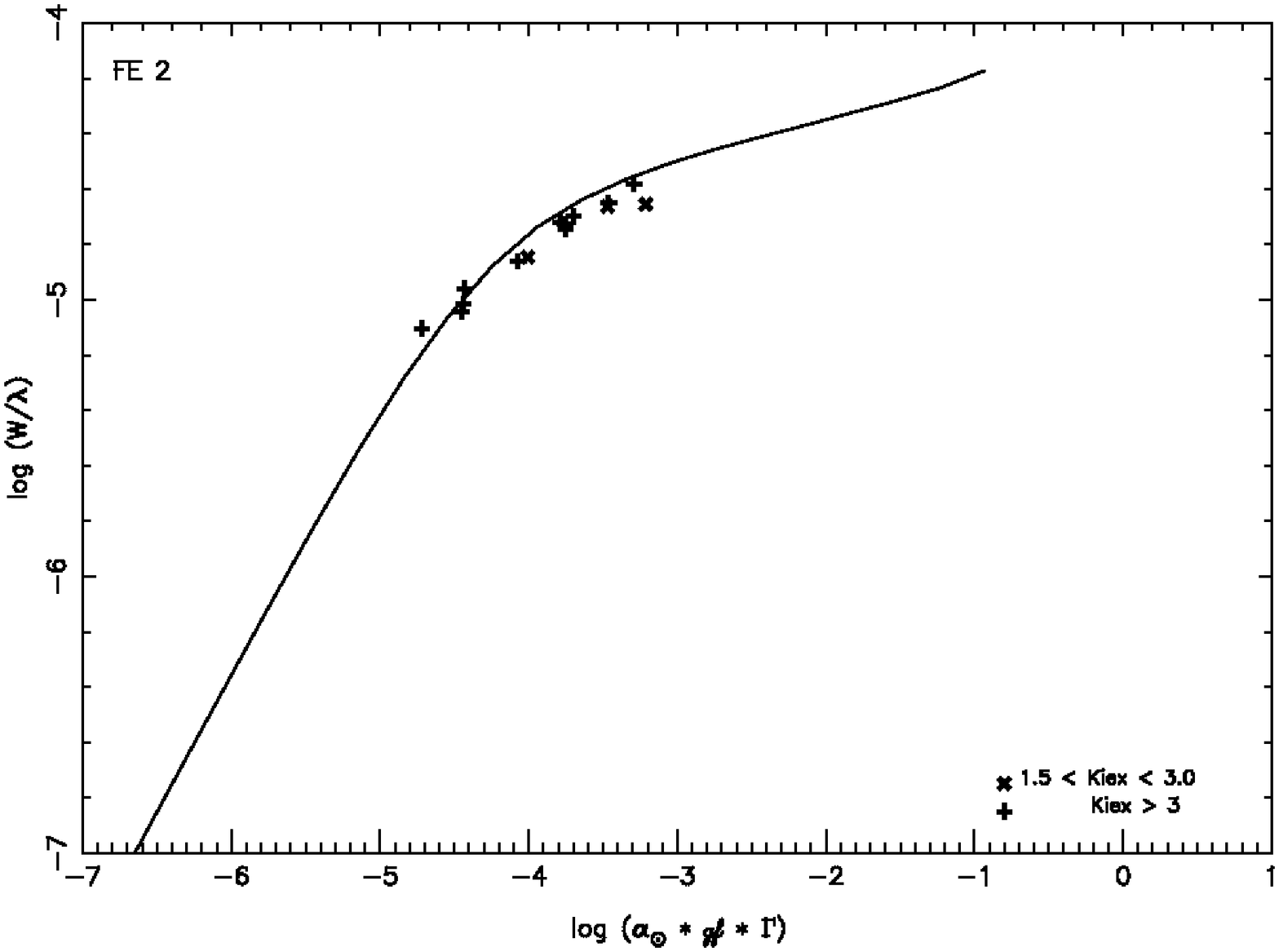}}
  \subfloat[][]{\label{CoG204:6}\includegraphics[angle=-90,width=0.33\textwidth]{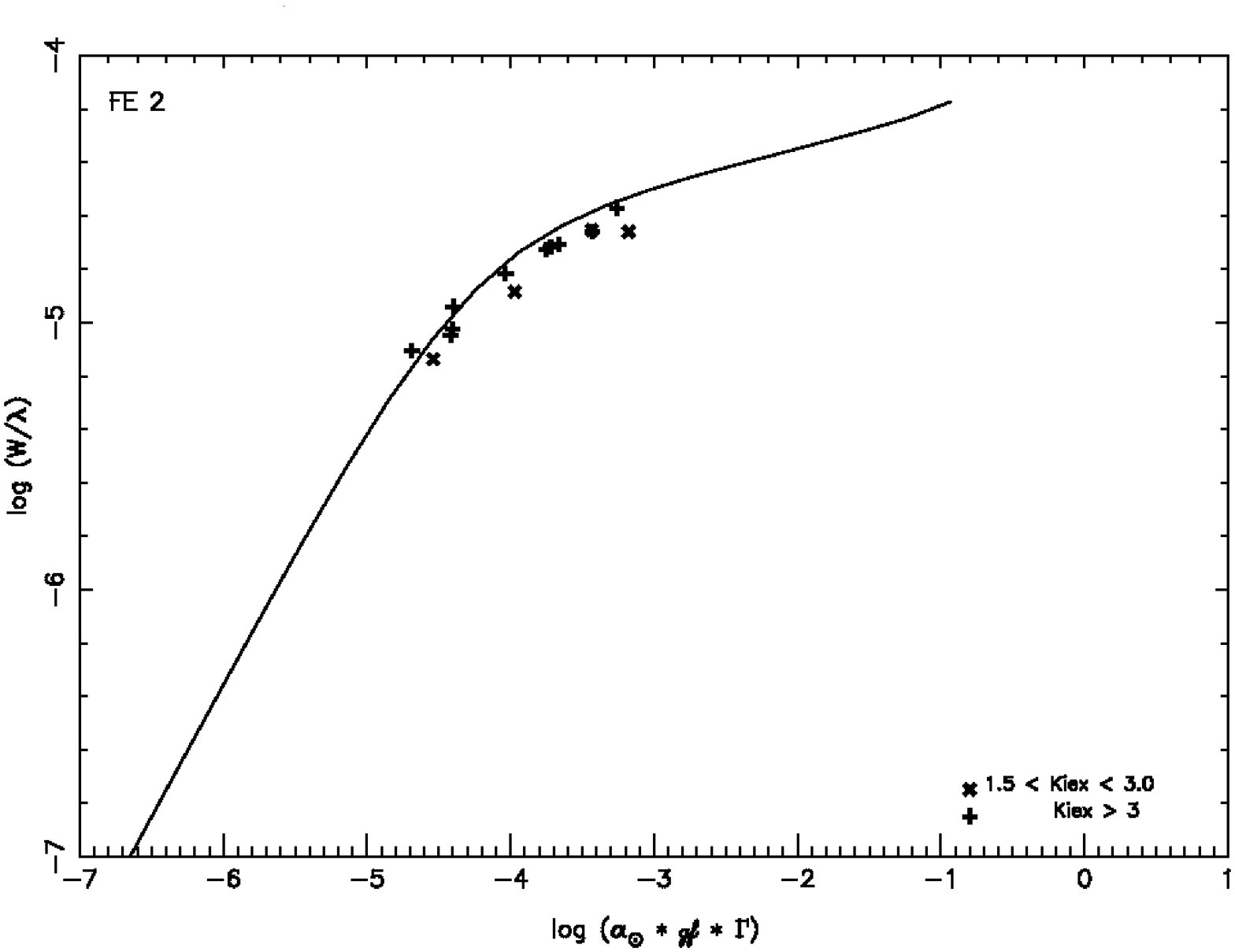}}
\end{sidewaysfigure}


\begin{sidewaysfigure}
  \centering
  \label{CoG822}\caption{Curves of growth (Fe~I, Fe~II) for the HV~822 spectra. From left to right, MJD=54785.04026326, 54785.05240828, 54785.06455378
}
  \subfloat[][]{\label{CoG822:1}\includegraphics[angle=-90,width=0.33\textwidth]{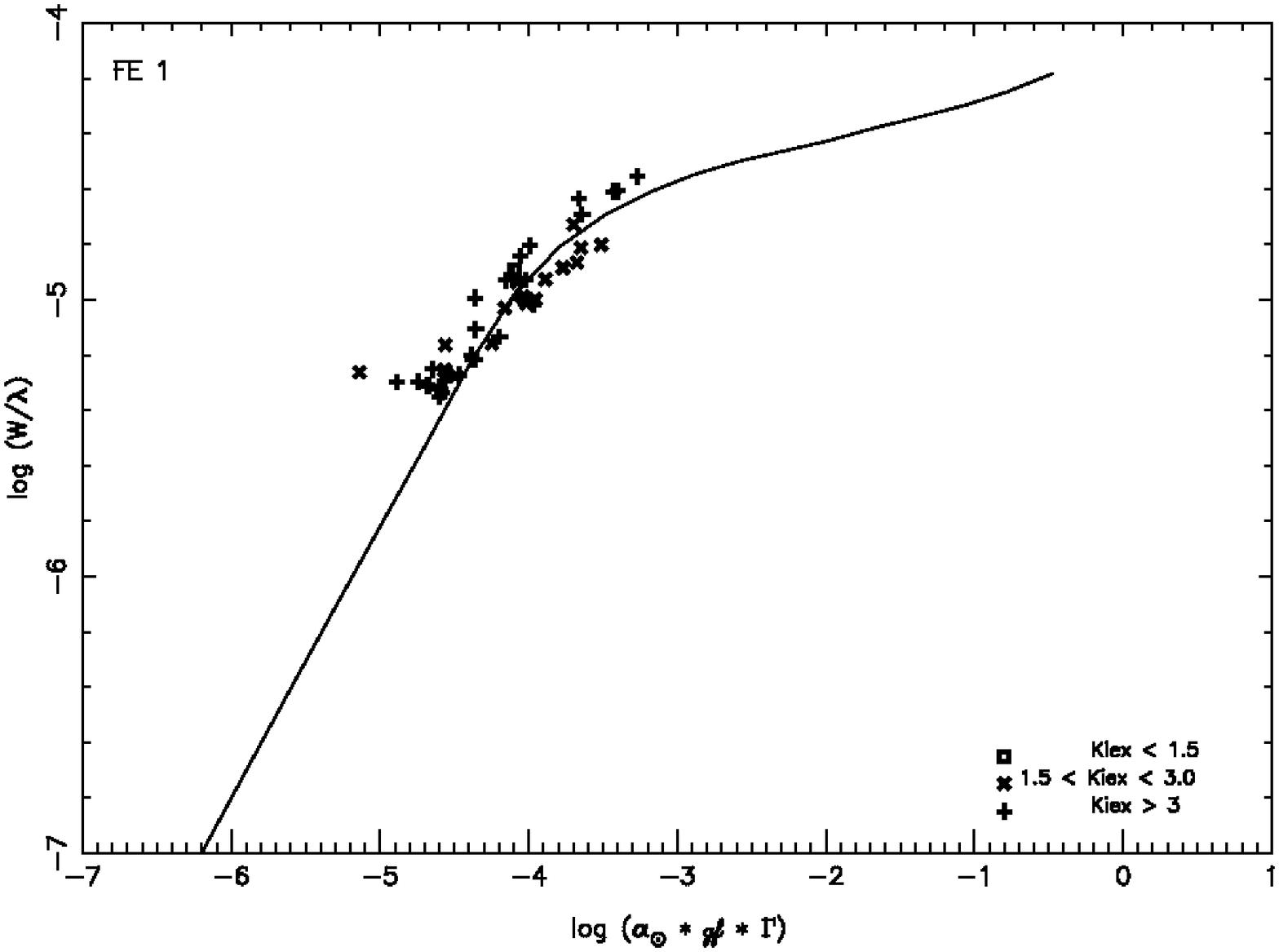}}
  \subfloat[][]{\label{CoG822:2}\includegraphics[angle=-90,width=0.33\textwidth]{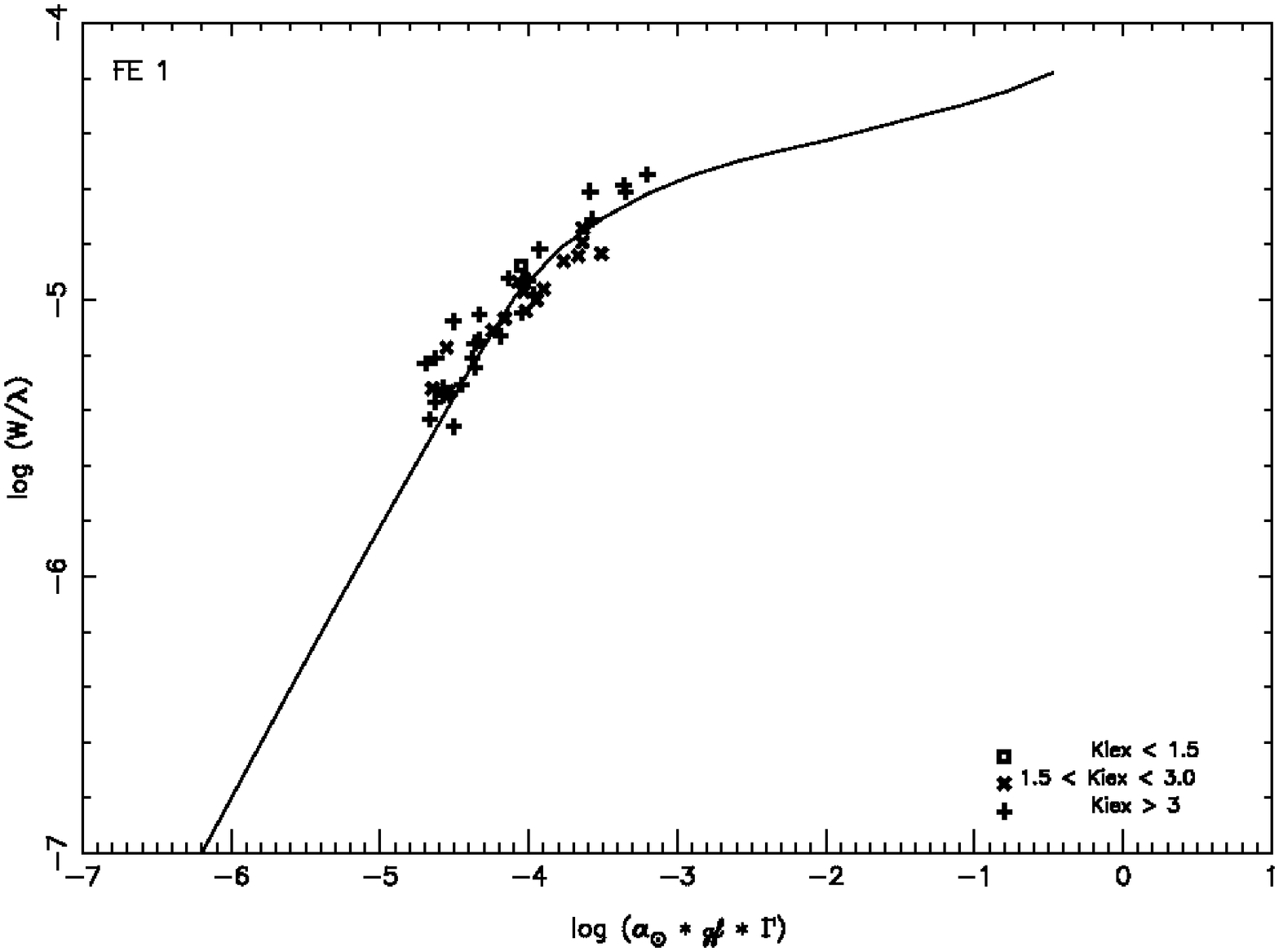}}
  \subfloat[][]{\label{CoG822:3}\includegraphics[angle=-90,width=0.33\textwidth]{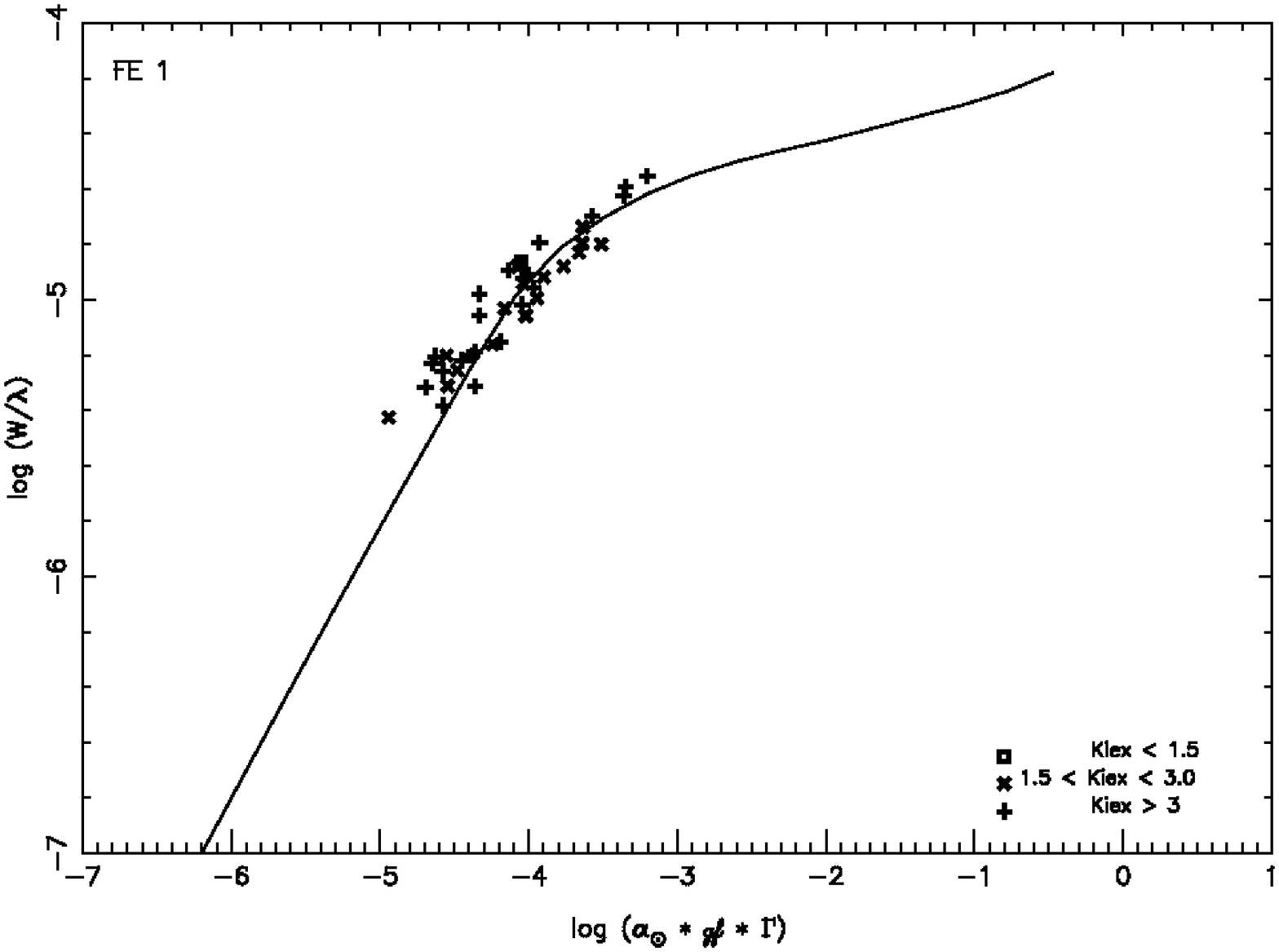}}
  \\
  \subfloat[][]{\label{CoG822:4}\includegraphics[angle=-90,width=0.33\textwidth]{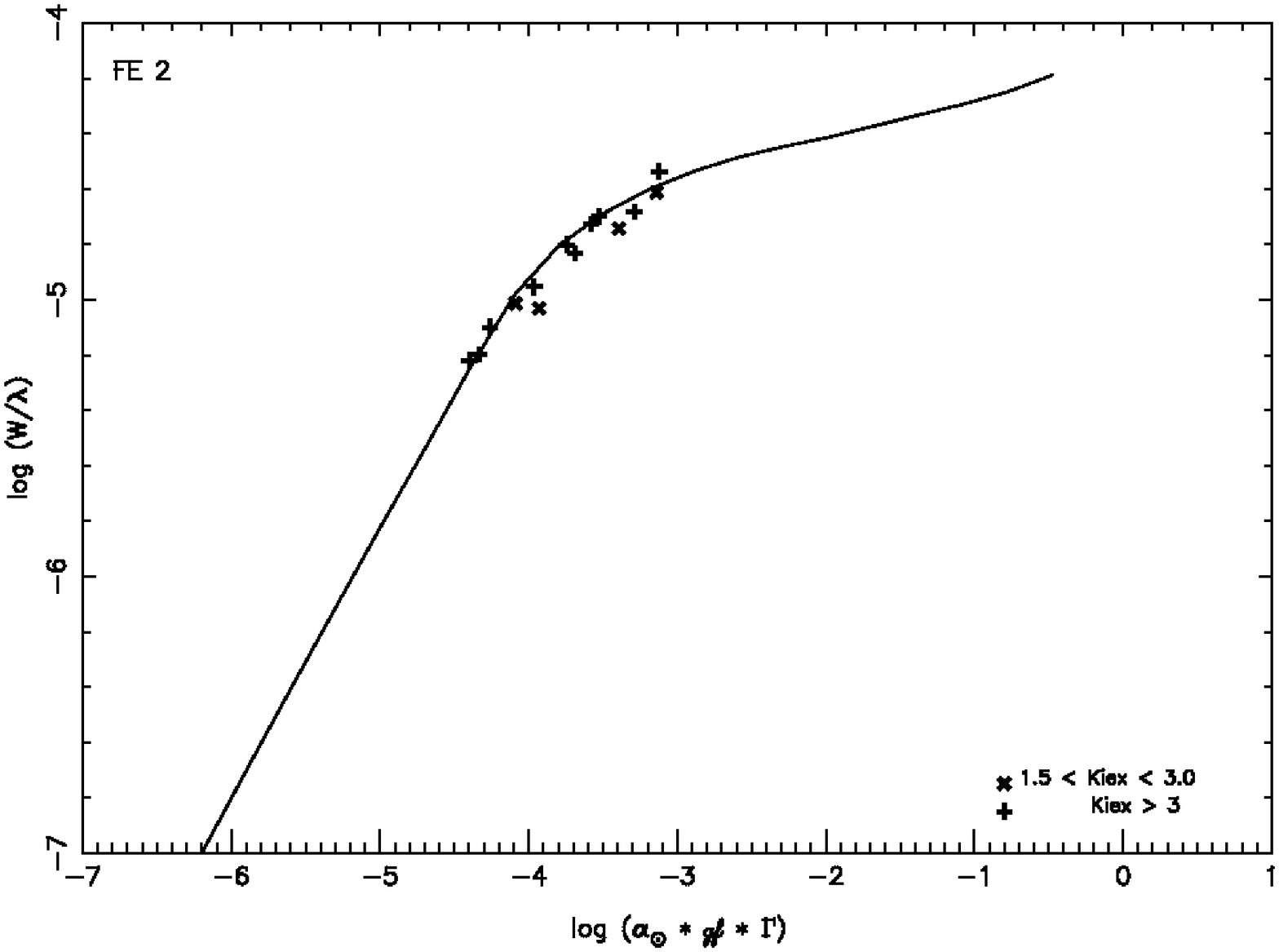}}
  \subfloat[][]{\label{CoG822:5}\includegraphics[angle=-90,width=0.33\textwidth]{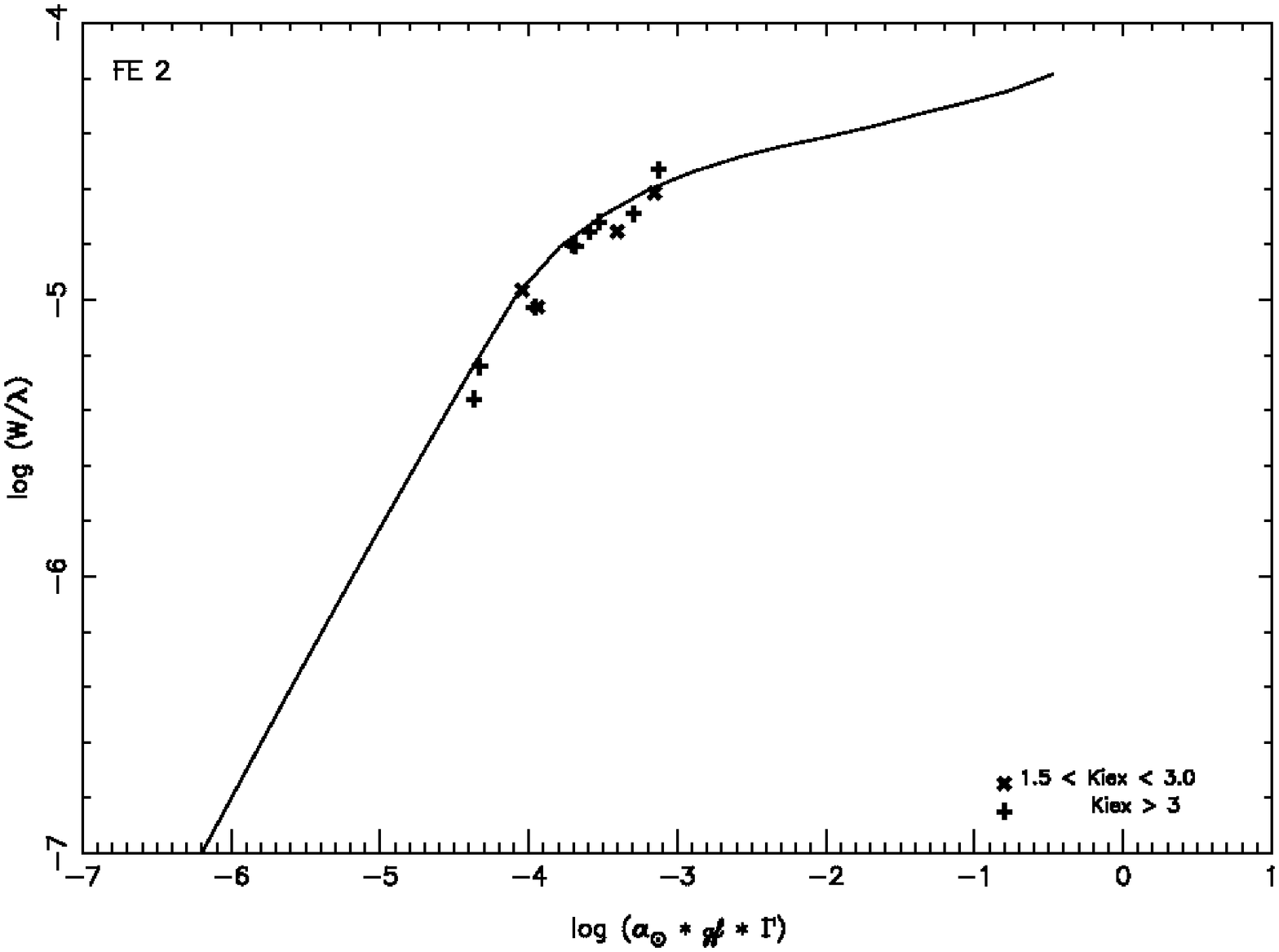}}
  \subfloat[][]{\label{CoG822:6}\includegraphics[angle=-90,width=0.33\textwidth]{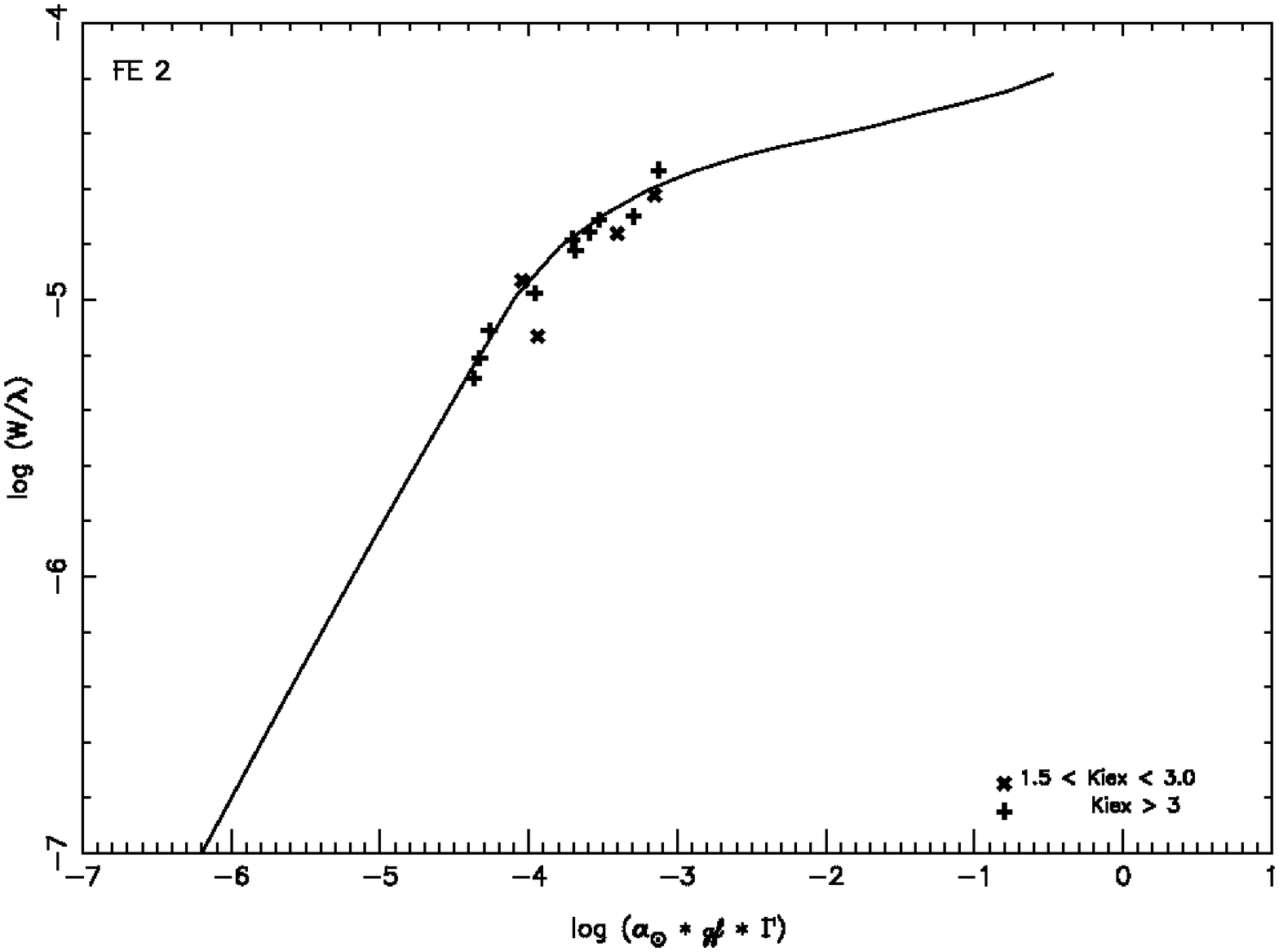}}
\end{sidewaysfigure}


\begin{sidewaysfigure}
  \centering
  \label{CoG1328}\caption{Curves of growth (Fe~I, Fe~II) for the HV~1328 spectra. From left to right, MJD=54785.00779033, 54785.01762254, 54785.02744873
}
  \subfloat[][]{\label{CoG1328:1}\includegraphics[angle=-90,width=0.33\textwidth]{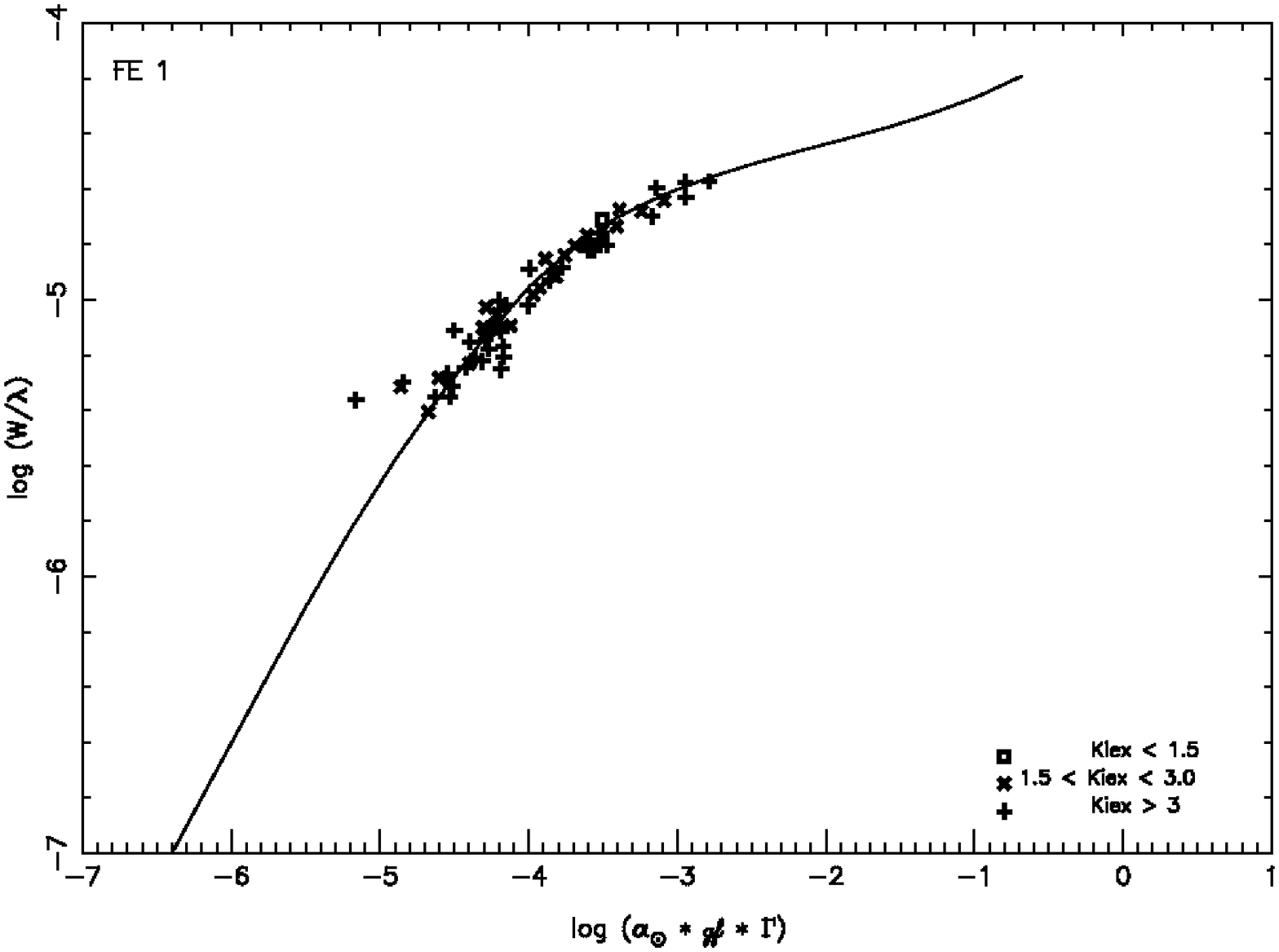}}
  \subfloat[][]{\label{CoG1328:2}\includegraphics[angle=-90,width=0.33\textwidth]{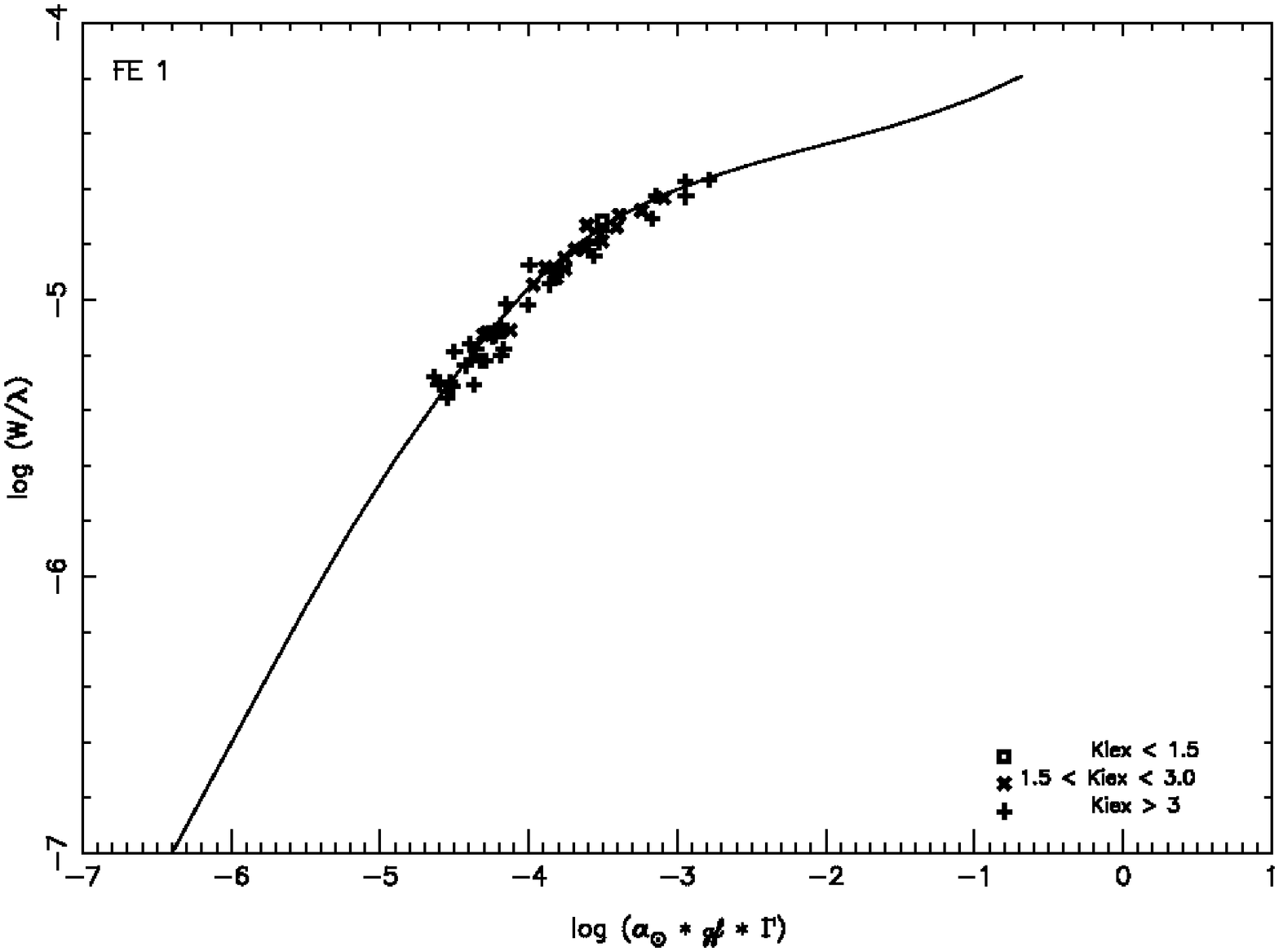}}
  \subfloat[][]{\label{CoG1328:3}\includegraphics[angle=-90,width=0.33\textwidth]{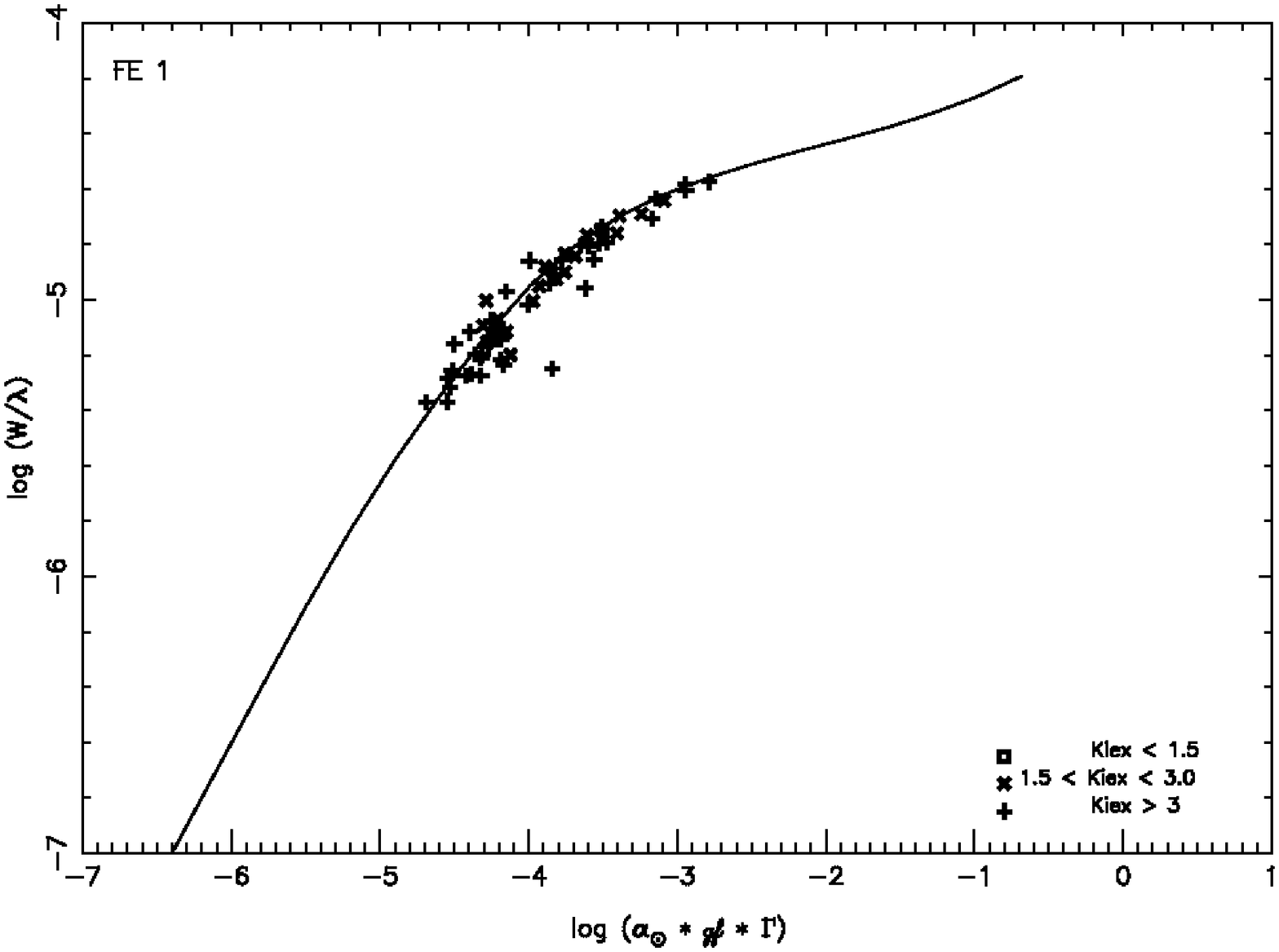}}
  \\
  \subfloat[][]{\label{CoG1328:4}\includegraphics[angle=-90,width=0.33\textwidth]{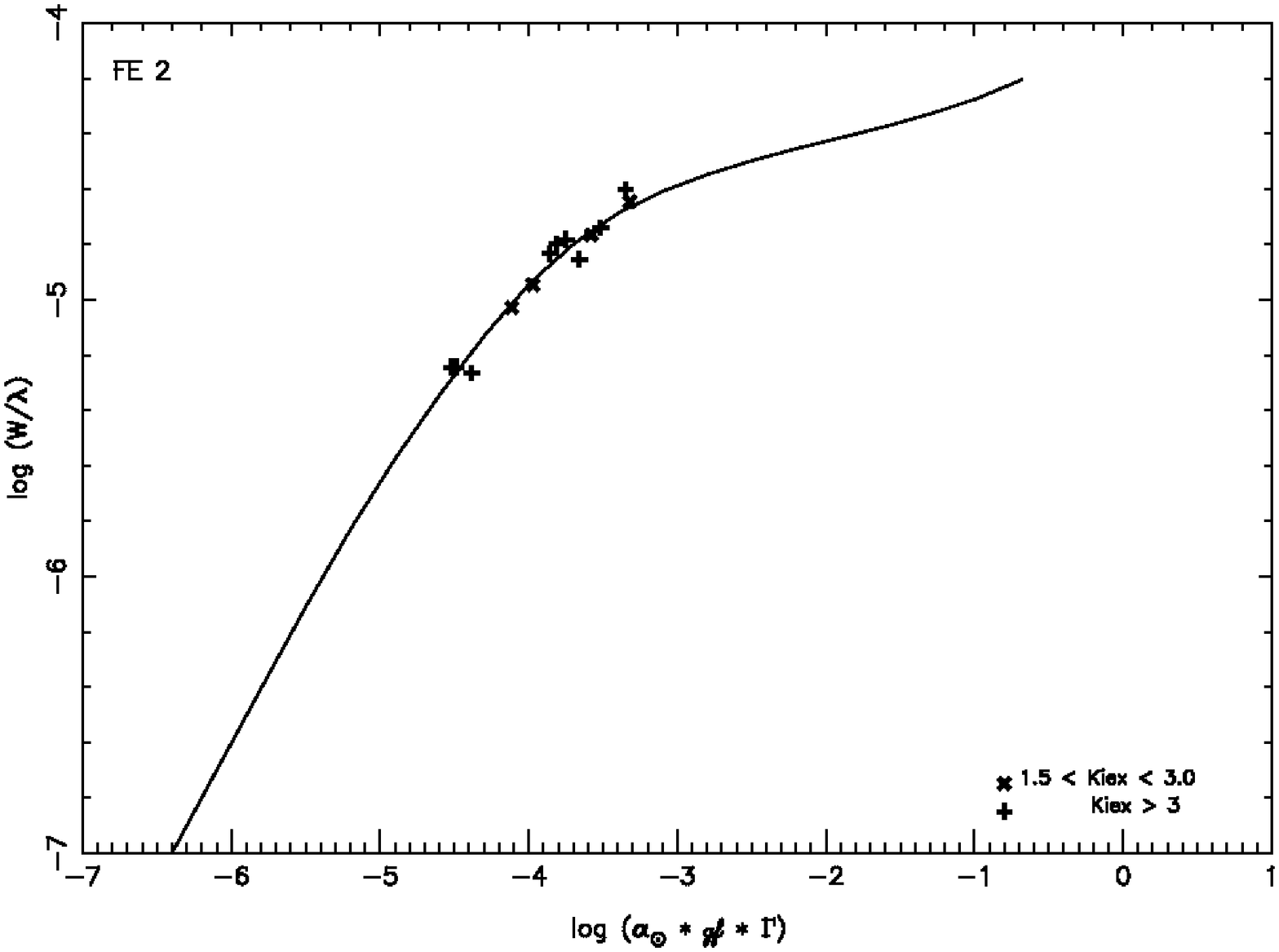}}
  \subfloat[][]{\label{CoG1328:5}\includegraphics[angle=-90,width=0.33\textwidth]{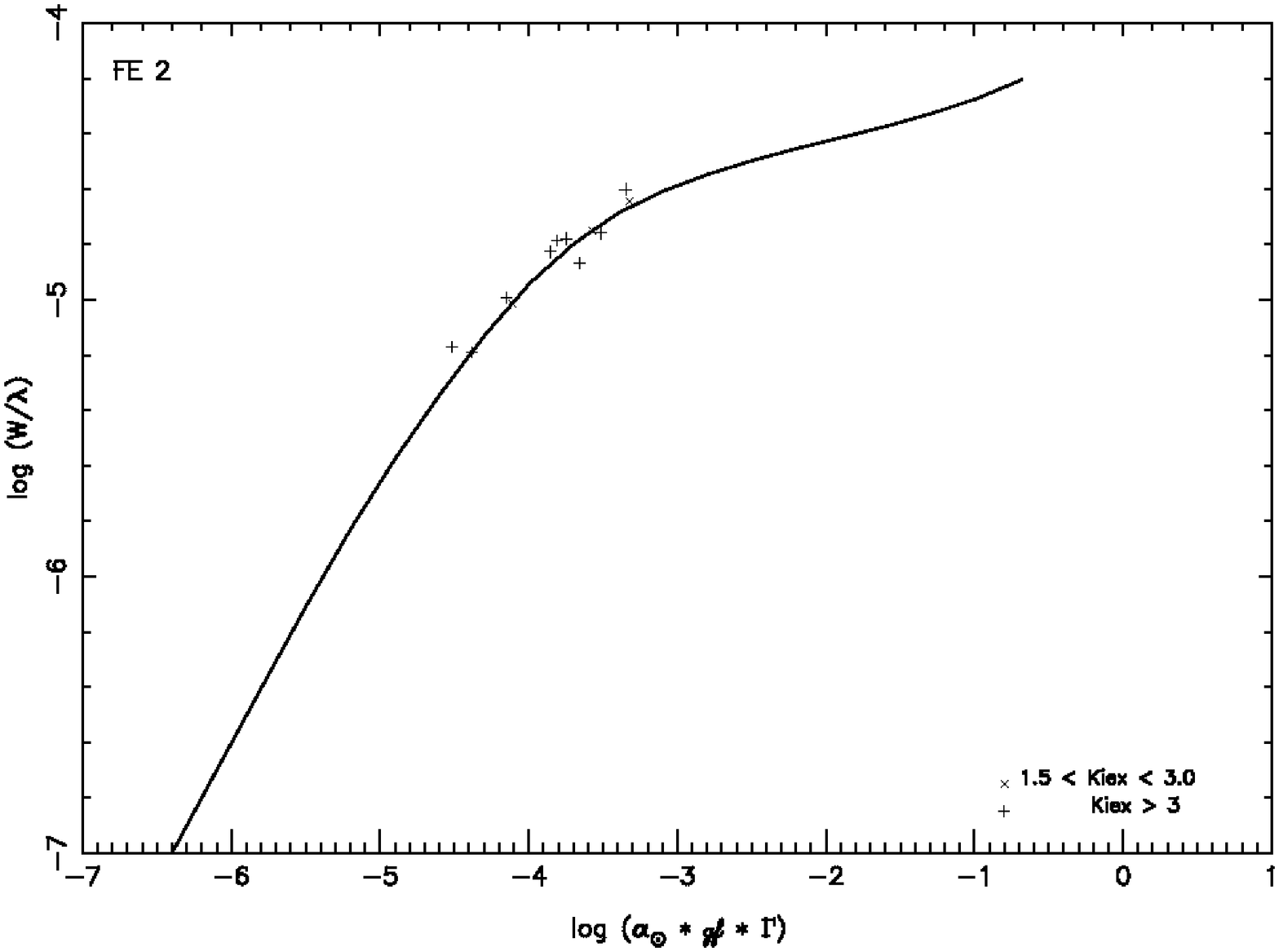}}
  \subfloat[][]{\label{CoG1328:6}\includegraphics[angle=-90,width=0.33\textwidth]{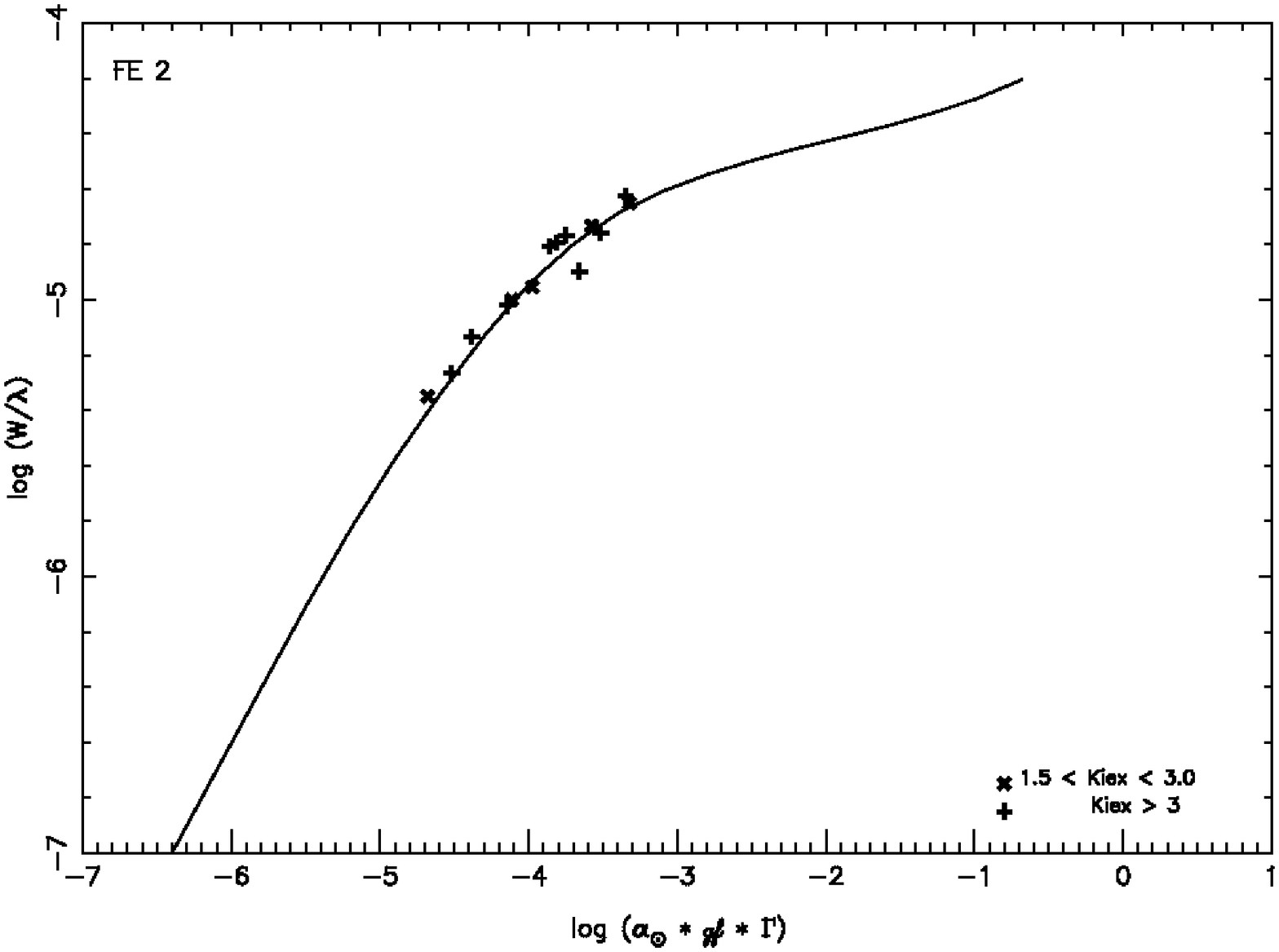}}
\end{sidewaysfigure}


\begin{sidewaysfigure}
  \centering
  \label{CoG1333}\caption{Curves of growth (Fe~I, Fe~II) for the HV~1333 spectra. From left to right, MJD=54785.07912217, MJD=54785.09358357, 54785.10804394
}
  \subfloat[][]{\label{CoG1333:1}\includegraphics[angle=-90,width=0.33\textwidth]{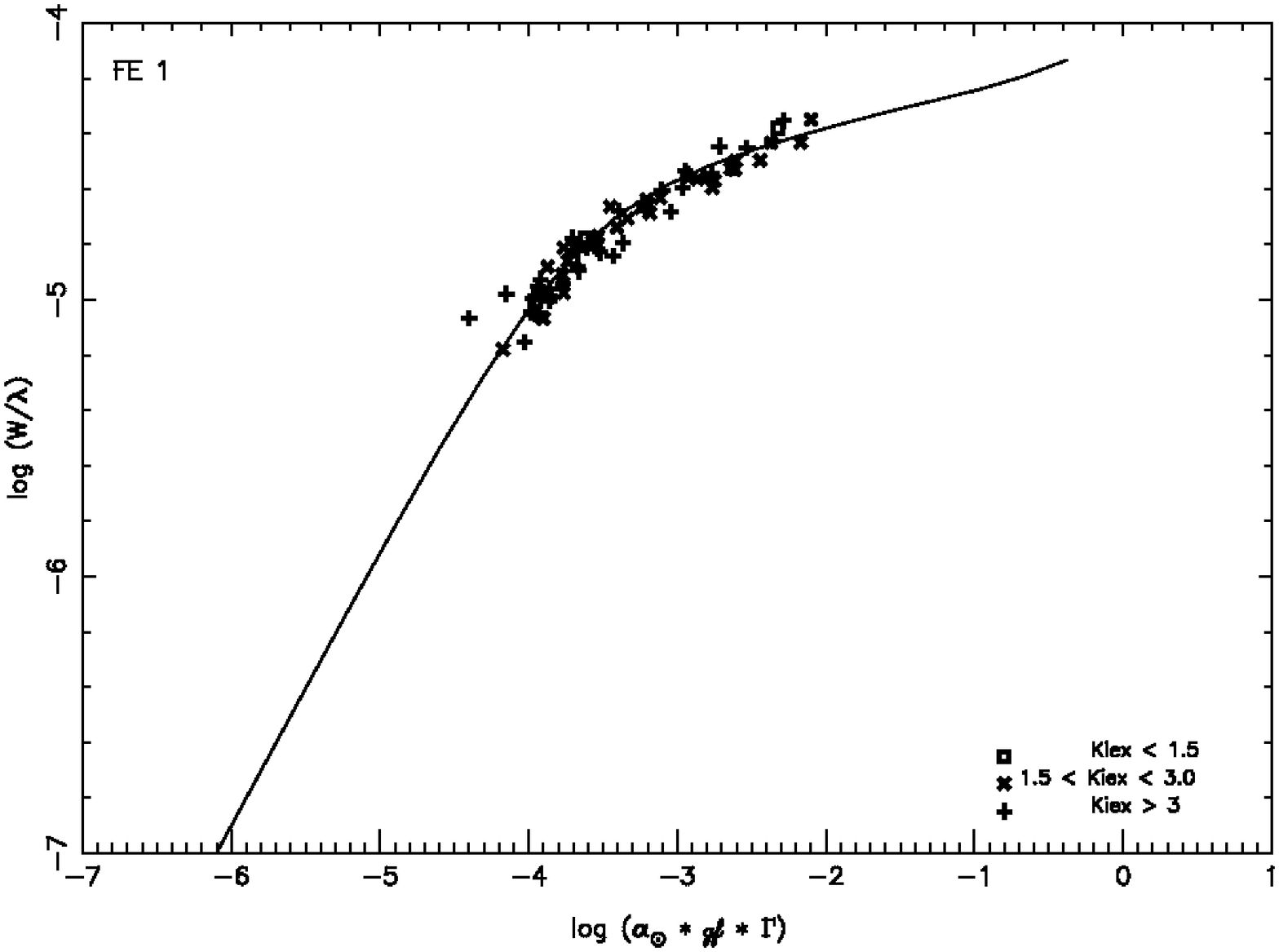}}
  \subfloat[][]{\label{CoG1333:2}\includegraphics[angle=-90,width=0.33\textwidth]{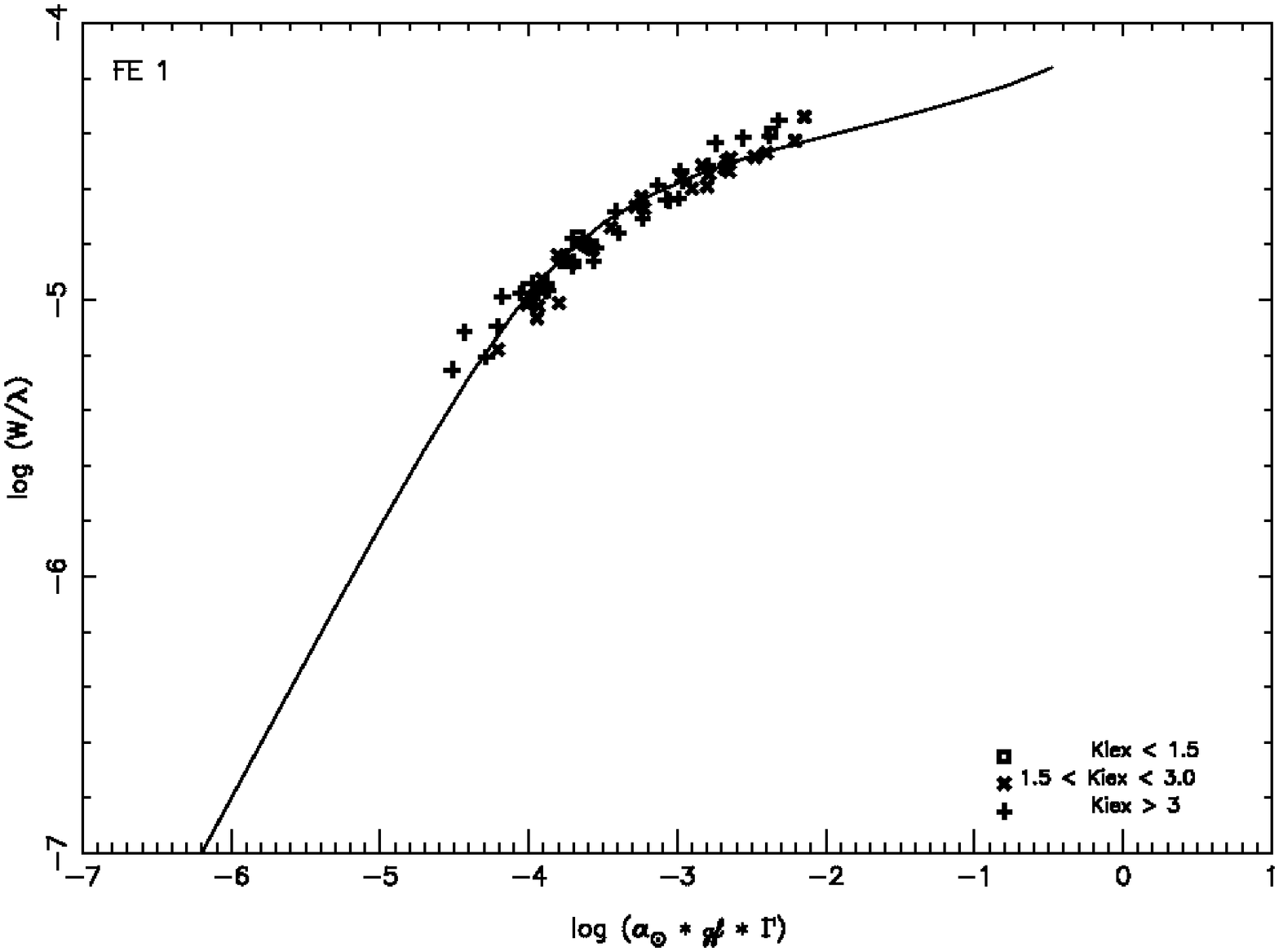}}
  \subfloat[][]{\label{CoG1333:3}\includegraphics[angle=-90,width=0.33\textwidth]{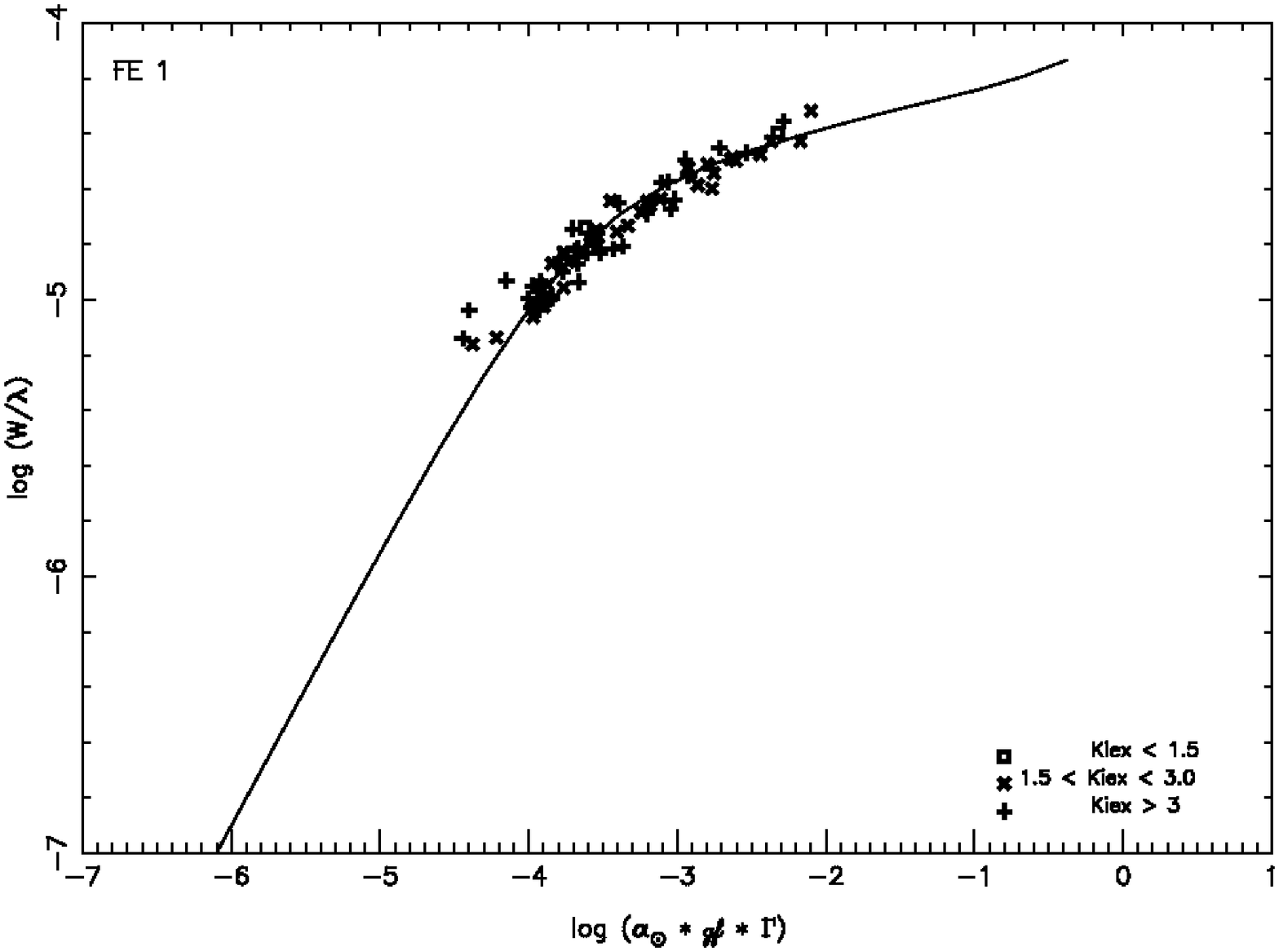}}
  \\
  \subfloat[][]{\label{CoG1333:4}\includegraphics[angle=-90,width=0.33\textwidth]{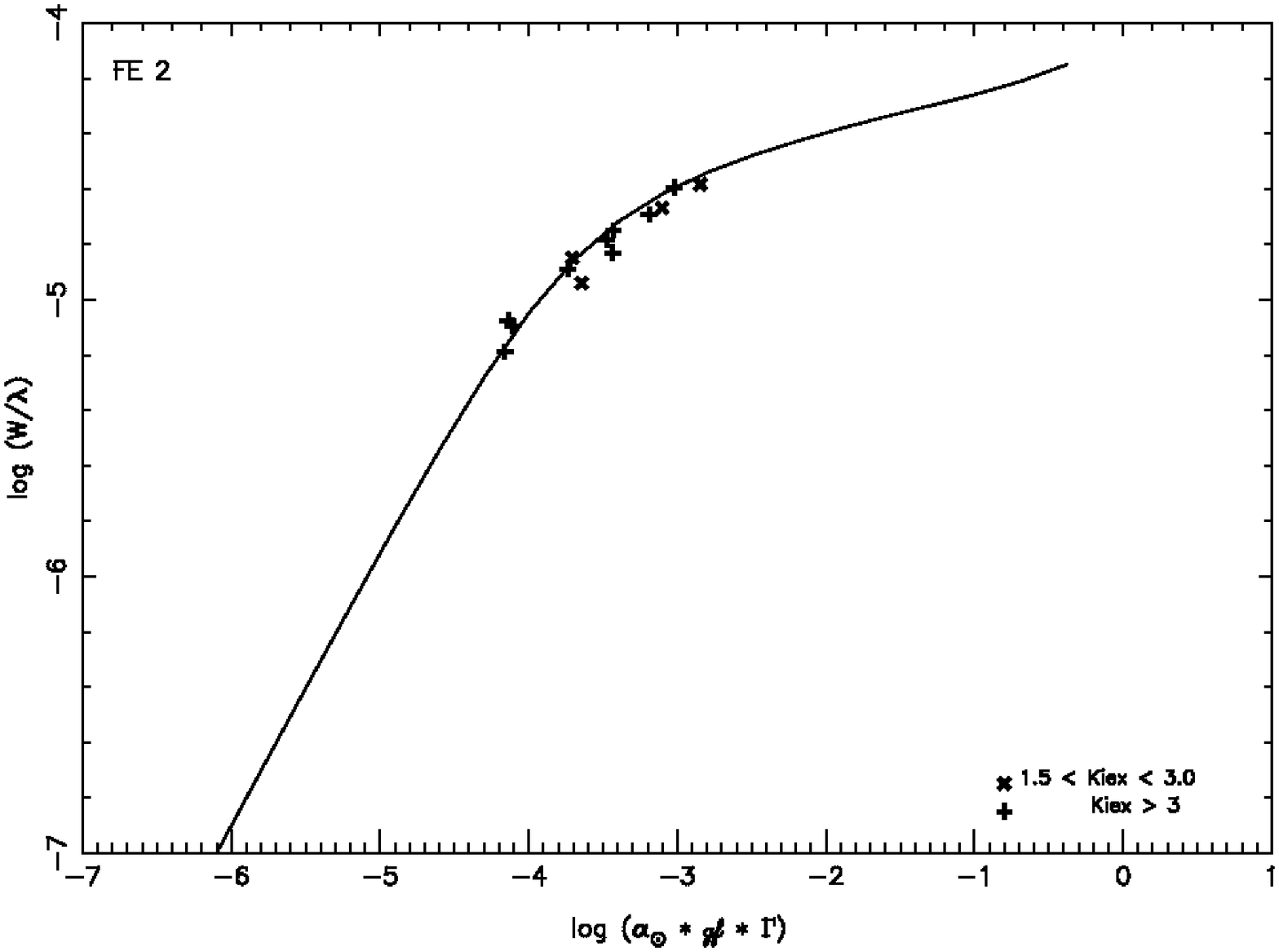}}
  \subfloat[][]{\label{CoG1333:5}\includegraphics[angle=-90,width=0.33\textwidth]{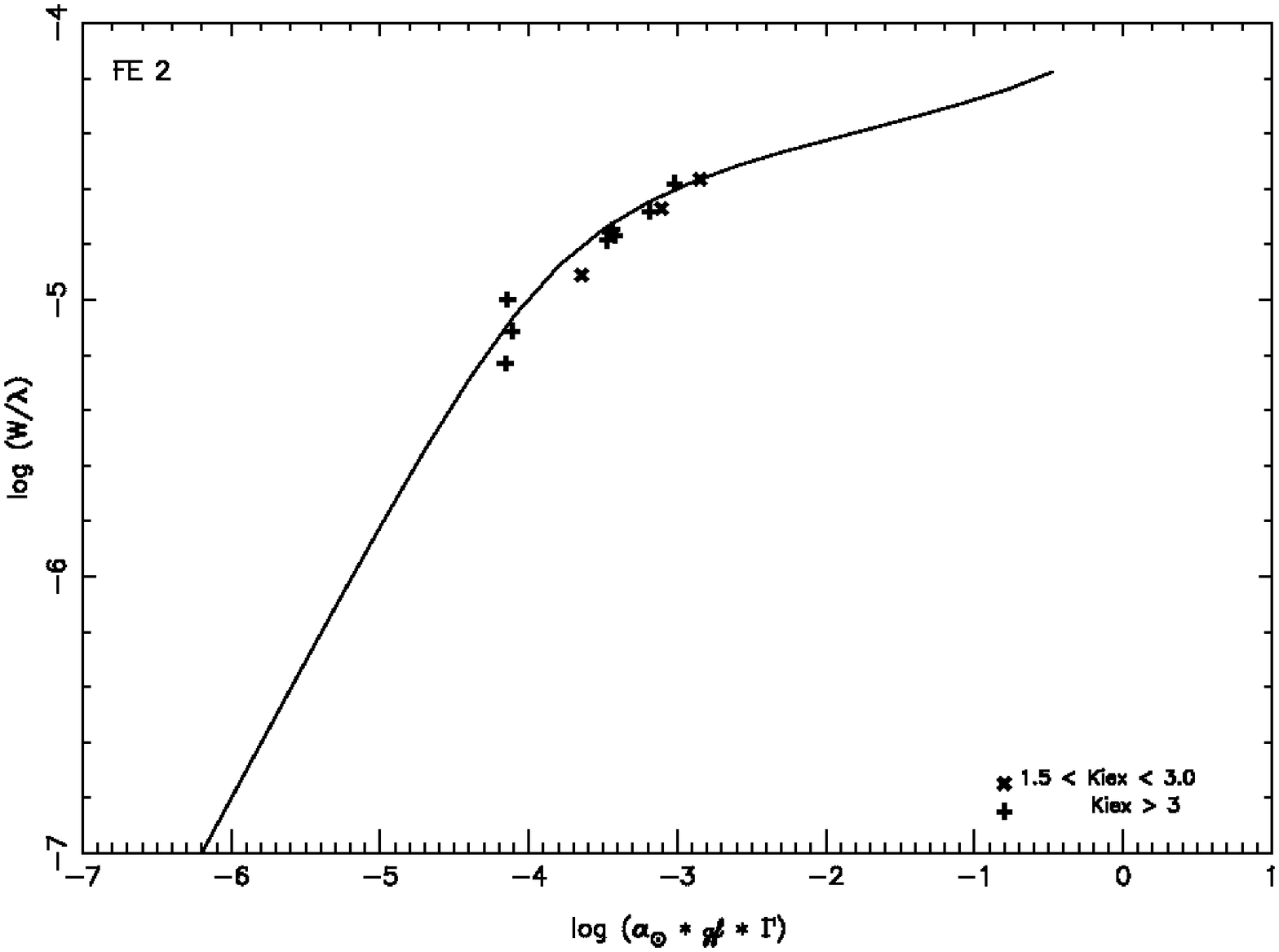}}
  \subfloat[][]{\label{CoG1333:6}\includegraphics[angle=-90,width=0.33\textwidth]{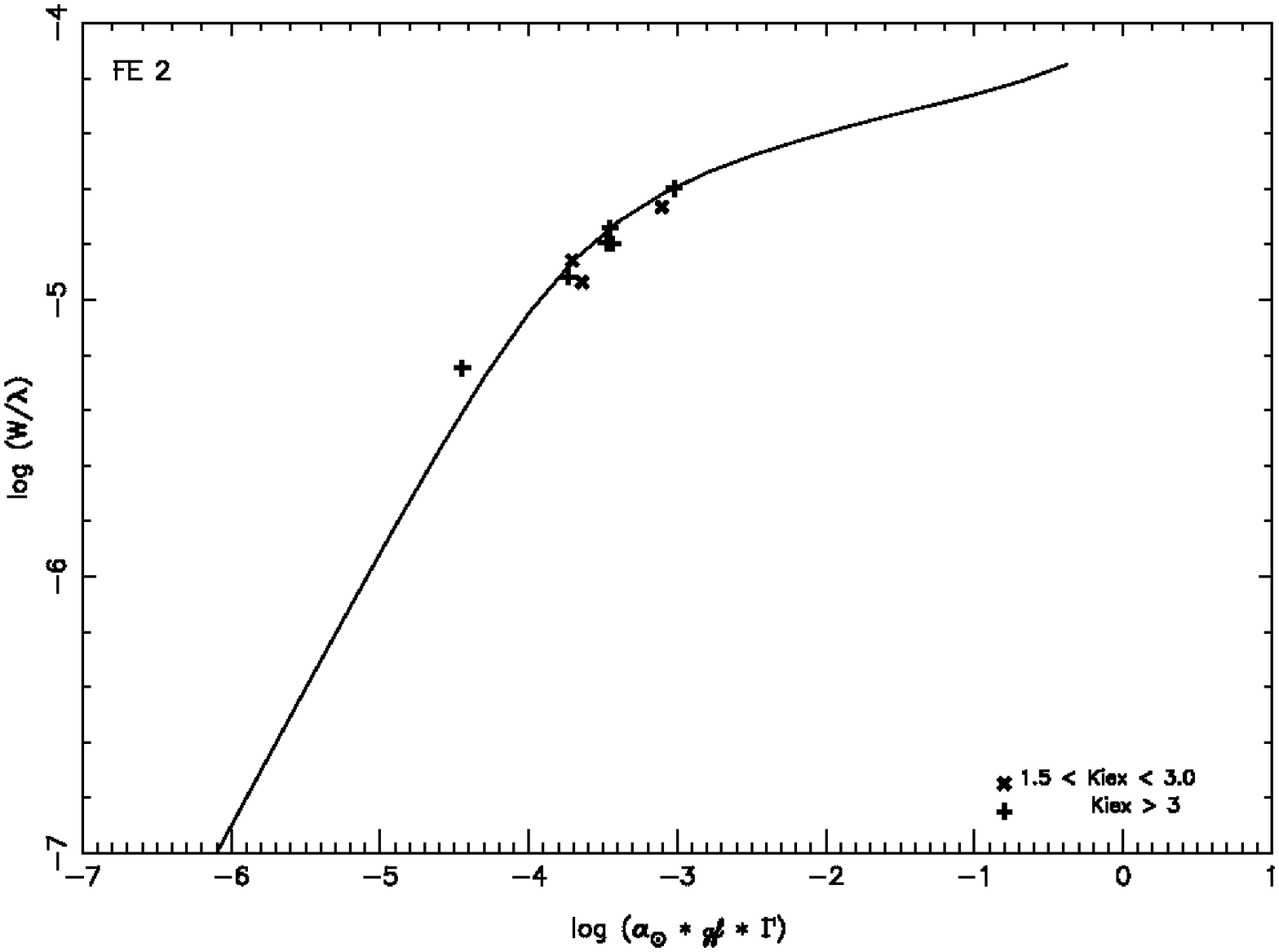}}
\end{sidewaysfigure}


\begin{sidewaysfigure}
  \centering
  \label{CoG1335}\caption{Curves of growth (Fe~I, Fe~II) for the HV~1335 spectra. From left to right, MJD=54785.12765152, 54785.14326987, 54785.15889887
}
  \subfloat[][]{\label{CoG1335:1}\includegraphics[angle=-90,width=0.33\textwidth]{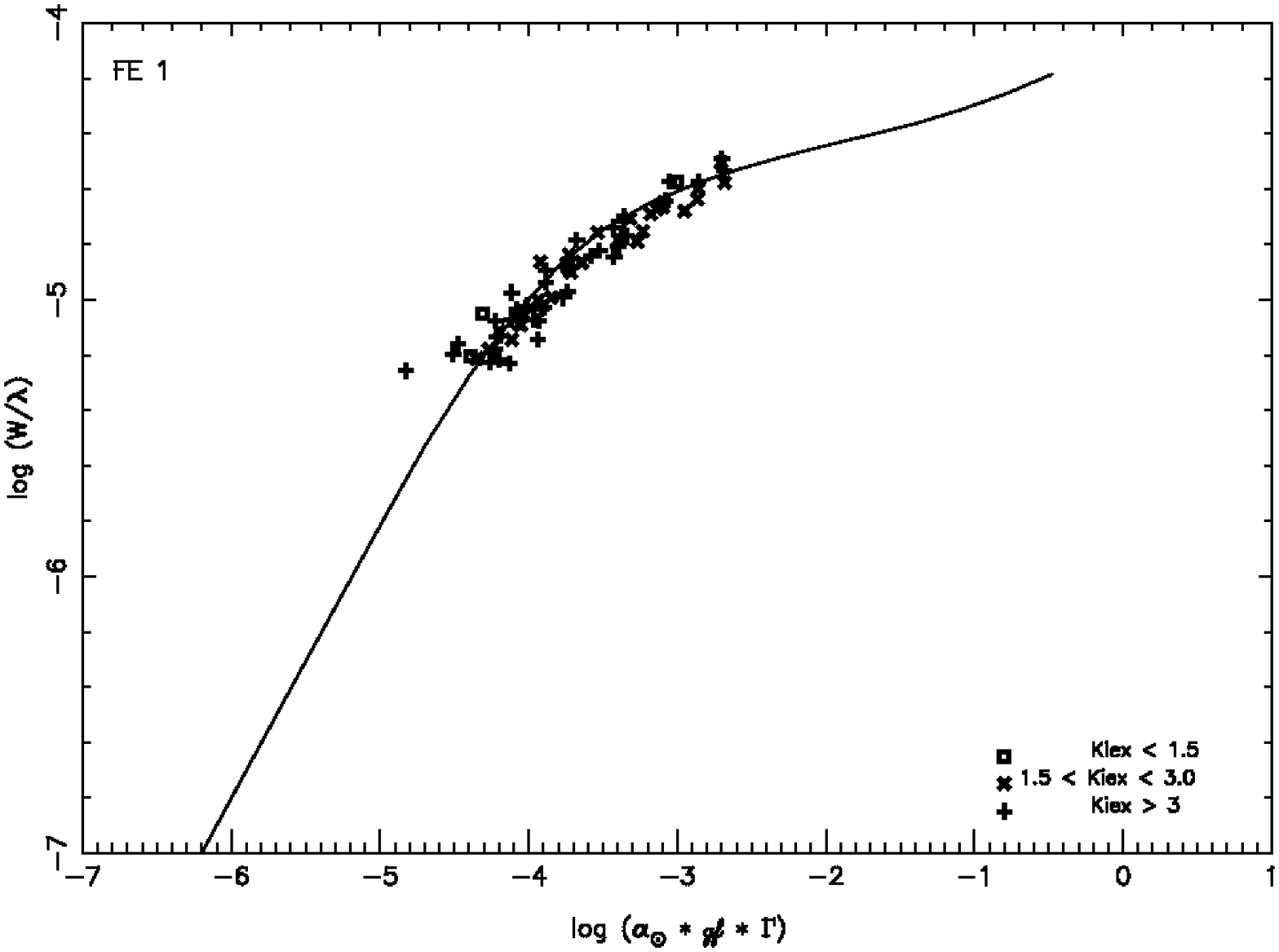}}
  \subfloat[][]{\label{CoG1335:2}\includegraphics[angle=-90,width=0.33\textwidth]{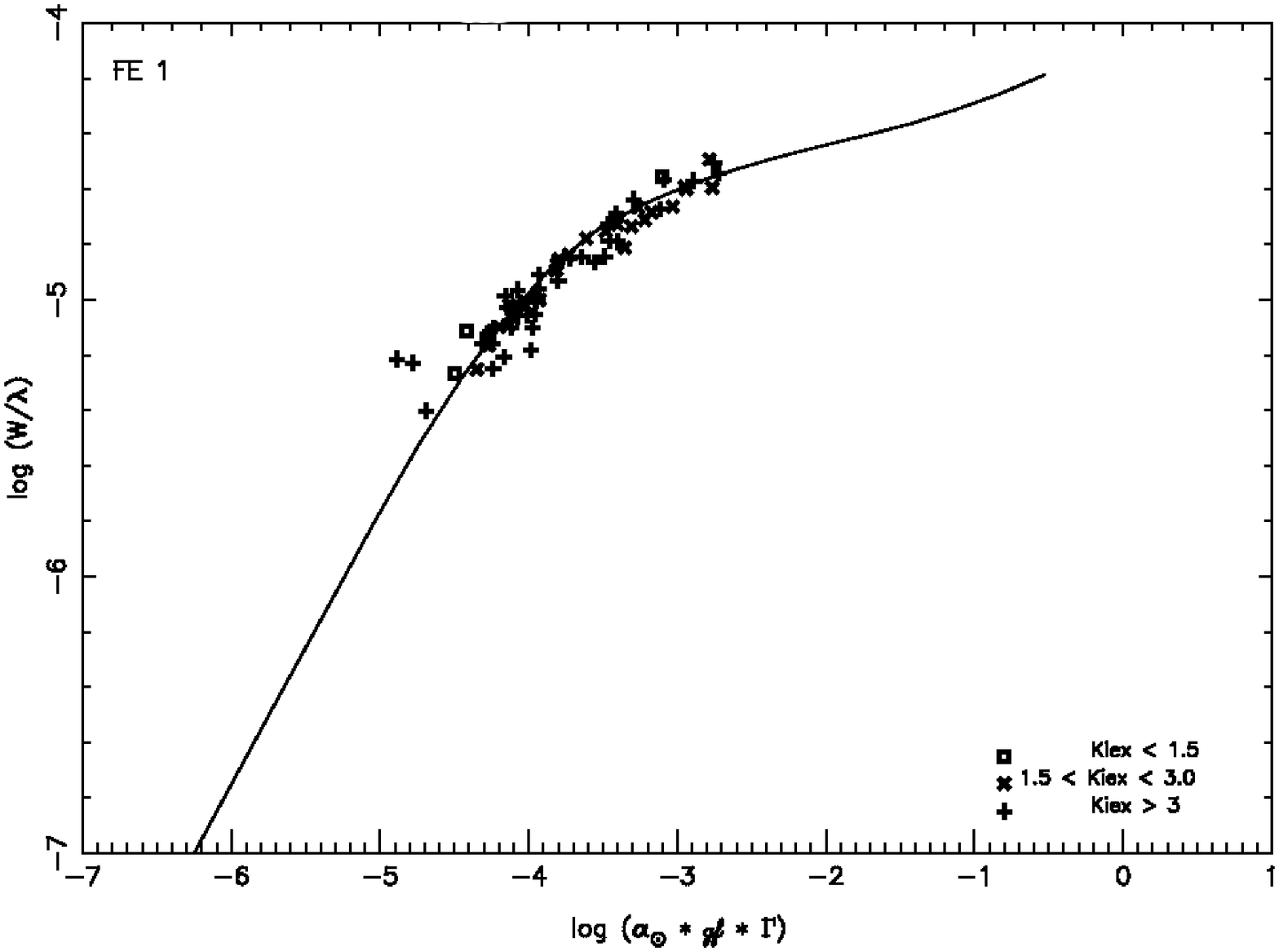}}
  \subfloat[][]{\label{CoG1335:3}\includegraphics[angle=-90,width=0.33\textwidth]{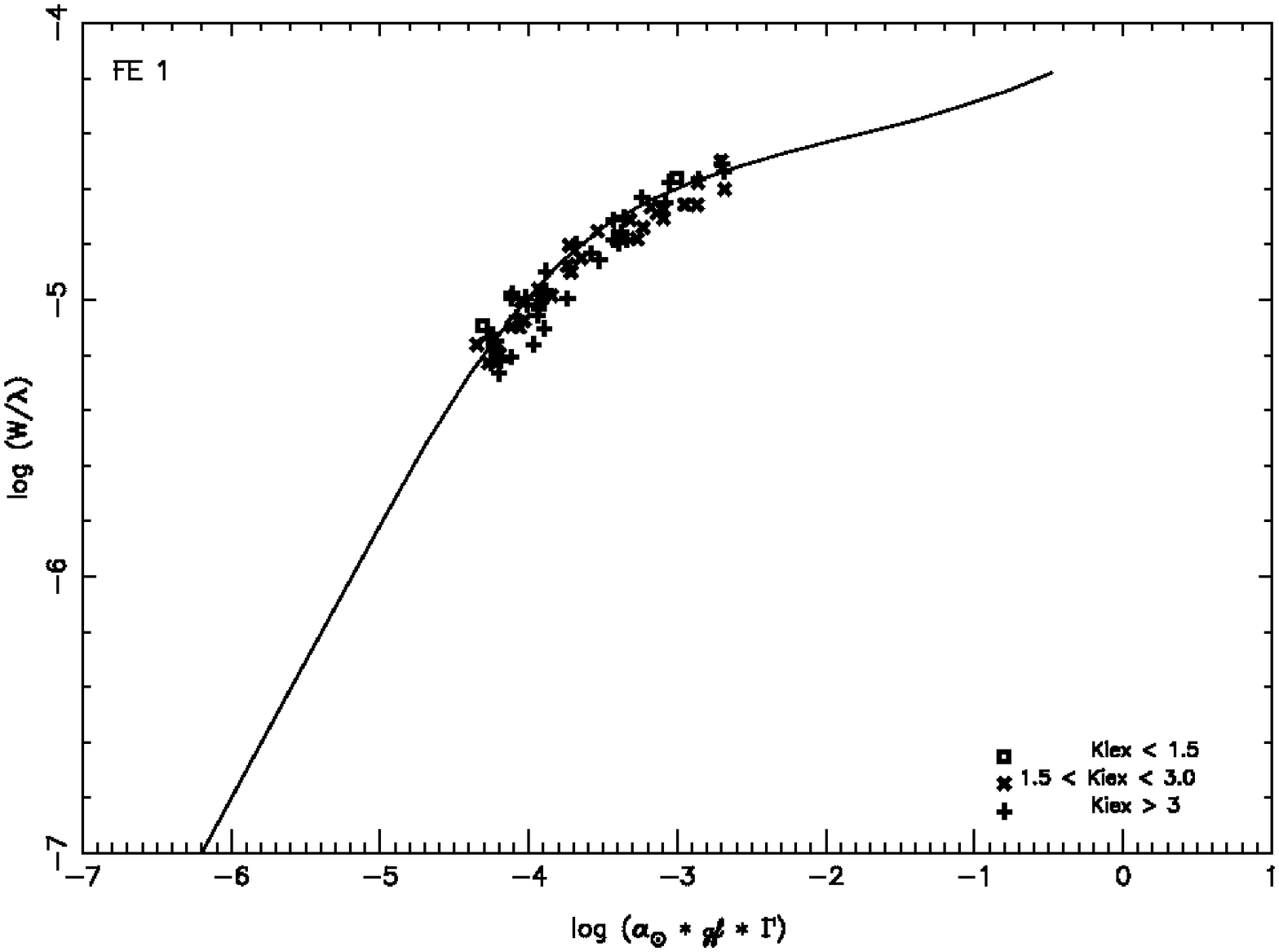}}
  \\
  \subfloat[][]{\label{CoG1335:4}\includegraphics[angle=-90,width=0.33\textwidth]{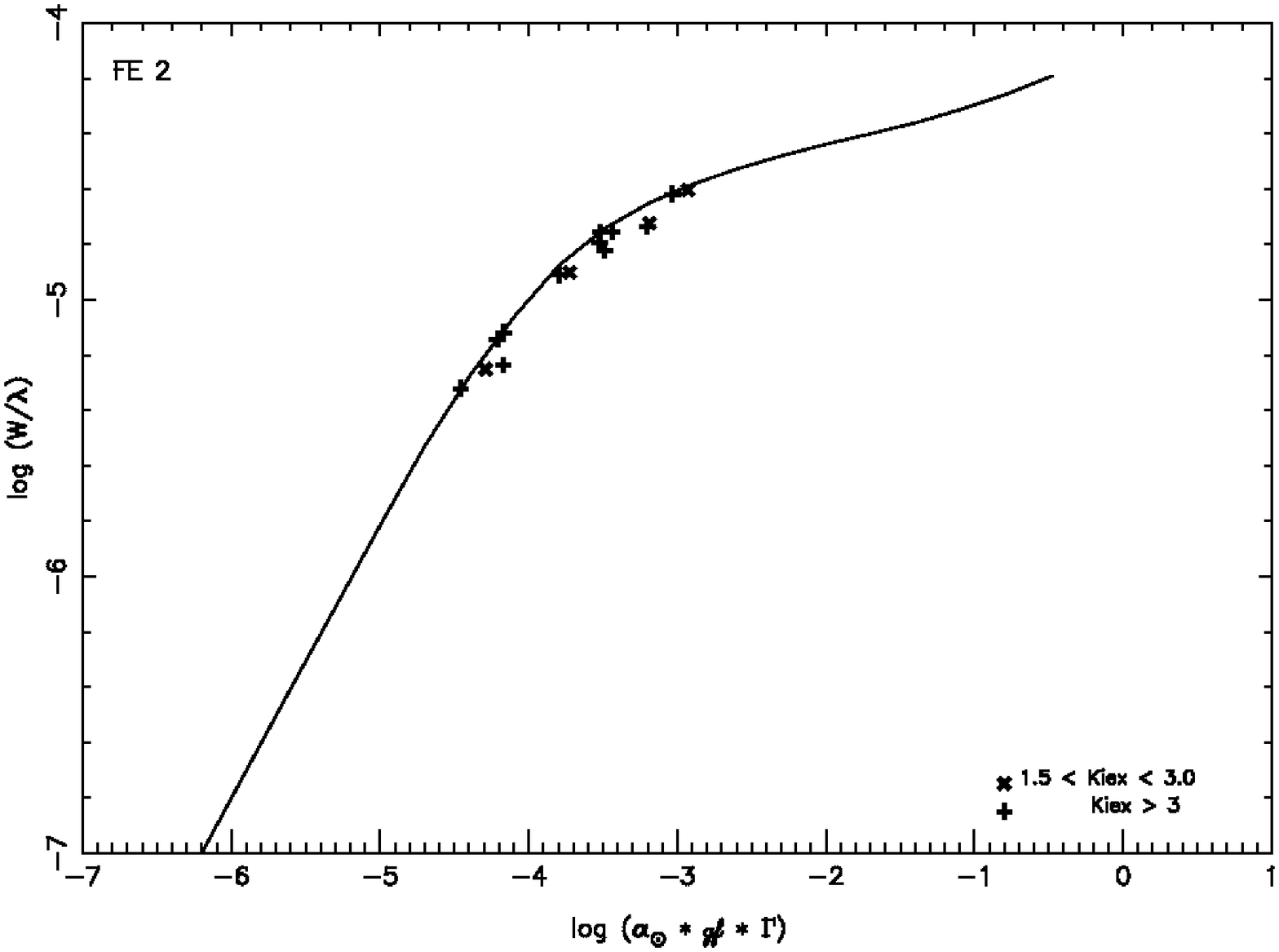}}
  \subfloat[][]{\label{CoG1335:5}\includegraphics[angle=-90,width=0.33\textwidth]{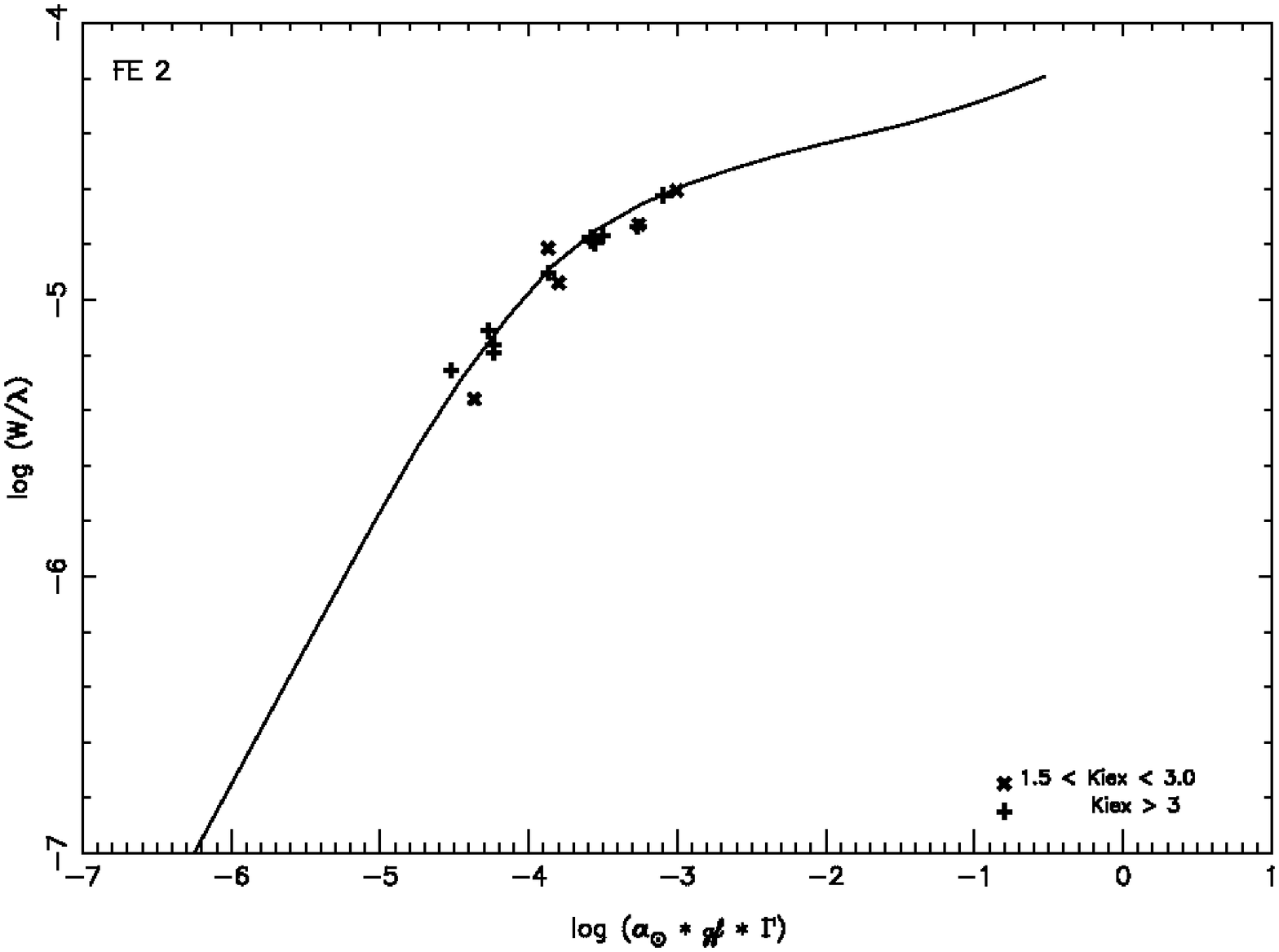}}
  \subfloat[][]{\label{CoG1335:6}\includegraphics[angle=-90,width=0.33\textwidth]{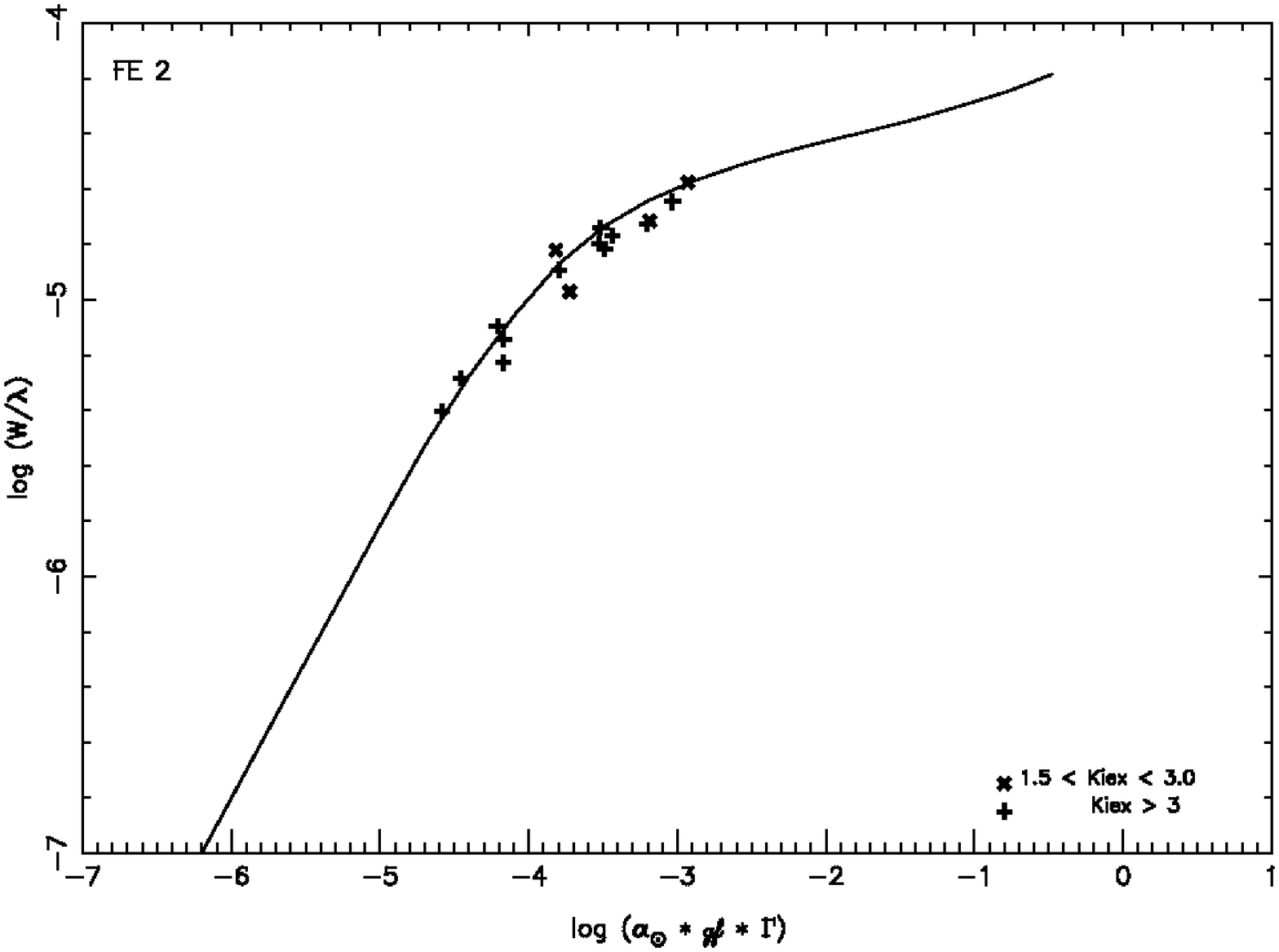}}
\end{sidewaysfigure}


\end{appendix}

\end{document}